\documentclass[aps,preprint,nofootinbib,floatfix]{revtex4}
\usepackage{epsfig}
\usepackage{amsmath}
\usepackage{amssymb}
\usepackage{amsbsy}
\usepackage{amsfonts}
\usepackage{graphics}
\usepackage{latexsym}
\usepackage{dcolumn}
\usepackage{wasysym}
\usepackage{graphicx}
\usepackage[figuresright]{rotating}

\newcommand{\Tr}{\ensuremath{\mathrm{Tr}}}

\begin{document}

\title{
\leftline{\parbox{3 cm}{\small \rm DESY 07-055\\[-2ex]HU-EP-07/08\\[-2ex]LMU-ASC 79/06 \\[-2ex]MKPH-T-07-05}}
\vskip 1cm
Exploring the structure of the quenched QCD vacuum \\
with overlap fermions}

\author{E.-M.~Ilgenfritz$^1$, K.~Koller$^2$, Y.~Koma$^3$, G.~Schierholz$^{4,5}$,
T.~Streuer$^6$, V.~Weinberg$^{7,8}$}

\affiliation{\vspace*{0.3cm} 
$^1$ Institut f\"ur Physik, Humboldt Universit\"at zu Berlin, 12489 Berlin, 
Germany \\
$^2$ Sektion Physik, Universit\"at M\"unchen, 80333 M\"unchen, Germany \\
$^3$ Institut fuer Kernphysik, Johannes-Gutenberg Universit\"at Mainz, 55099 Mainz, Germany \\
$^4$ Deutsches Elektronen-Synchrotron DESY, 22603 Hamburg, Germany \\
$^5$ John von Neumann-Institut f\"ur Computing NIC, 15738 Zeuthen, Germany \\
$^6$ Department of Physics and Astronomy, University of Kentucky, Lexington, 
KY 40506-0055, USA \\ 
$^7$ Deutsches Elektronen-Synchrotron DESY, 15738 Zeuthen, Germany   \\
$^8$ Institut f\"ur Theoretische Physik, Freie Universit\"at Berlin, 14196 Berlin, Germany }
                                         
\author{- QCDSF Collaboration -}
\affiliation{ }

\begin{abstract}
Overlap fermions have an exact chiral symmetry on the lattice and are thus 
an appropriate tool for investigating the chiral and topological structure of 
the QCD vacuum. We study various chiral and topological aspects of quenched 
gauge field configurations.  This includes the localization and chiral properties 
of the eigenmodes, the local structure of the ultraviolet filtered field strength tensor, 
as well as the structure of topological charge fluctuations. We conclude that the 
vacuum has a multifractal structure.
\end{abstract}


\maketitle

\section{Introduction}
\label{sec:introduction}

Overlap fermions~\cite{Neuberger:1997fp,Neuberger:1998wv} 
possess an exact chiral symmetry on the lattice~\cite{Luscher:1998pq}
and realize the Atiyah-Singer index theorem at finite 
cut-off~\cite{Hasenfratz:1998ri}. 
Furthermore, they 
lend themselves to a local definition of the topological charge 
density~\cite{Niedermayer:1998bi}.
Altogether, this makes overlap fermions a powerful tool for investigating 
the chiral and topological structure of the QCD vacuum.
In this paper we shall address the vacuum of the quenched theory at zero 
temperature. Preliminary results have been published 
in~\cite{Koma:2005sw,Ilgenfritz:2005hh,Weinberg:2006ju,Ilgenfritz:2006gc}.
In a forthcoming paper we shall extend the investigation to dynamical QCD at zero
and finite temperature. 
First results obtained for full QCD in the vicinity of the 
chiral phase transition have been reported in ~\cite{Weinberg:2005dh}. 

\begin{figure}[t]
\begin{center}
\hspace*{-1.0cm}\epsfig{file=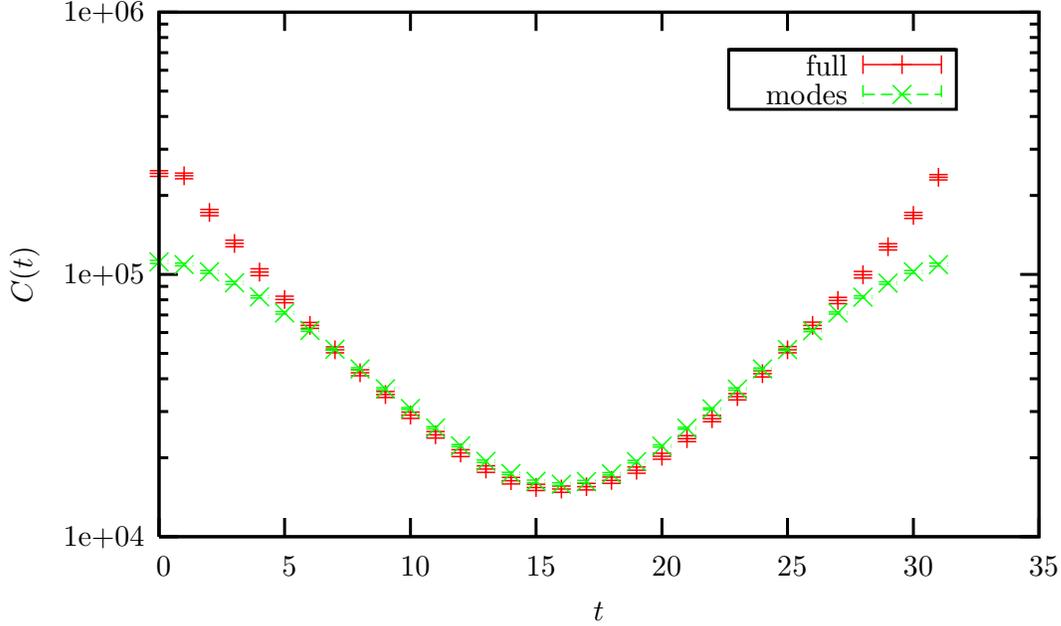,width=14cm, clip=}
\end{center}
\caption{The full pion propagator compared with the truncated pion propagator
constructed as an all-to-all correlator summing over the 40 lowest modes only
with a quark mass $\frac{am}{2\rho}=0.01$ (see Eq.~(\ref{eq:massive_overlap})). 
This calculation uses 250 configurations on the $16^3\times32$ lattice generated
at $\beta=8.45$.}
\label{fig:pionpropagator}
\end{figure}

It is known for some time~\cite{DeGrand:2000gq,DeGrand:2001tm} that the long 
distance properties of QCD are well described by the low-lying eigenmodes of the
overlap operator. The question of low-mode dominance has been raised earlier in
Ref.~\cite{Ivanenko:1997nb,Neff:2001zr}. In Fig.~\ref{fig:pionpropagator} we 
compare the full 
pion propagator with the truncated one spanned only by the 40 lowest eigenmodes.

We see that both propagators tend to each other at distances $\gtrsim 0.5$ fm. 
A suitable truncation
of the overlap operator thus acts as an infrared filter, which allows us
to separate the truly nonperturbative degrees of freedom of the QCD vacuum
from the ultraviolet noise. This is in accord with conventional wisdom,
namely that chiral symmetry breaking is encoded in the low-lying modes of
the Dirac operator and both, chiral symmetry breaking and confinement have 
the same dynamical origin. To make this more precise, 
the vacuum structure, 
seen changing with an increasing number of fermionic modes, 
is the general theme of this paper. Thus, we shall come back also to the early 
saturation of the pion propagator in more detail later in this paper.

Earlier lattice investigations of QCD vacuum structure as reviewed in 
Refs.~\cite{Haymaker:1998cw,Greensite:2003bk,DiGiacomo:2005yq} 
partly relied on gauge fixing and subsequent projection onto appropriate 
subgroups of color $SU(3)$. The topics under discussion there are mainly related
to the confinement of heavy quarks. In contrast, with the gauge invariant 
overlap approach, we come closer to the confinement issue of light quarks. 
As a particular advantage, it establishes a direct link between topological 
excitations and light quark propagation, which was missing so far. 
After all, we believe that light fermions are a major element of the low-energy 
effective action of QCD~\cite{Wetterich:2004kg}.

In the instanton model (as a prototype of other 
semiclassically 
motivated models), with the almost-zero-mode band approximation, this link is realized 
in a very economic way. We summarize these models by saying that they all are 
based on classical, selfdual or anti\-self\-dual solutions of the Euclidean 
equations of motion carrying zero modes of definite chirality.
While the vacuum fields are formed in these models as superpositions of such extended 
solutions, in the almost-zero-mode band approximation 
the spectrum of the lowest quark eigenstates
in the complex vacuum is analogously formed by linear combinations of the 
corresponding zero modes.
Examples are the instanton liquid model (for a review see
Ref.~\cite{Schafer:1996wv}) where this program has been carried through, 
or the caloron gas model~\cite{Gerhold:2006sk} where this has still to be done.
Both models are based on gauge fields piecewise coherent over $O(0.5~\rm{fm})$ 
but with decaying field strength correlations beyond that distance. 
In order to justify these models, a good deal of lattice vacuum 
studies has been devoted to instanton and caloron searches.

In the past, cooling~\cite{Ilgenfritz:1985dz,Teper:1985ek}, underrelaxed 
cooling~\cite{Smith:1998wt}, restricted improved cooling~\cite{GarciaPerez:1998ru}
and other smoothing techniques including inverse 
blocking~\cite{DeGrand:1996ih,DeGrand:1996zb,Feurstein:1996cf}, 
renormalization group cycling~\cite{DeGrand:1997gu}, finally rationalized as 
smearing~\cite{DeGrand:1997ss}, have been used for this purpose.   
Since these methods contain a bias towards classical solutions, the results 
have been taken with certain reservations. 
In particular, the number density of lumps (``instantons and antiinstantons'') 
was strongly influenced by systematical effects~\cite{Negele:1998ev}.
Only recently, when valence overlap fermions have become a viable option, 
a local definition of the topological charge density, 
$q(x) \propto \mathrm{Tr}~(F_{\mu\nu} \Tilde{F}_{\mu\nu})$, in terms of the
overlap Dirac operator (see Eq. (\ref{eq:eq-q})) became practicable
(see {\it e.g.} Ref.~\cite{Horvath:2002yn}).
Surprisingly, for Monte Carlo configurations 
this density shows a three-dimensional, laminar and highly
singular structure which seems to rule out the instanton and similar pictures.

Easier accessible are mode-truncated, {\it i.e.} ultraviolet filtered
versions of the topological charge density (see Eq. (\ref{eq:qlambdacut})).
Still with large computational effort, such evaluations of the topological
charge density are now superseding cooling or smearing techniques. 
The latter techniques were necessary to be applied in the past to the gauge 
field before the topological charge density could be calculated according to 
a gluonic definition (see for example Ref.~\cite{Bilson-Thompson:2002jk}). 
The number of fermionic modes (ultraviolet filtering) now replaces  
the number of smearing iterations as the control parameter of smoothness. 
The overlap definition of the topological charge  density, 
if recast~\cite{Horvath:2002yn} into its spectral representation 
(\ref{eq:qlambdacut}), can be evaluated with any desired degree 
of ultraviolet smearing represented by a cut-off $\lambda_{\rm cut}$. 
In this paper the two aspects (the infrared and the ultraviolet one) 
of the topological structure will be described. Whereas the infrared 
structure is definitely associated with chiral symmetry breaking, 
the relation of the surprising divergent structure in the 
ultraviolet~\cite{Horvath:2002yn,Horvath:2003yj,Horvath:2005rv}
to the phenomenology of confinement, as stressed by~\cite{Witten:1978bc}, 
is still hypothetic.

The {\it all-scale topological charge density} (\ref{eq:eq-q}), containing 
fluctuations of all scales, from the lattice spacing $a$ to global structures 
percolating through the full lattice volume, is calculable~\cite{Horvath:2002yn} 
directly in terms of the overlap operator.
This does not require to know its full spectrum and all eigenmodes. 
It is calculable, although  
computationally very demanding, as a local trace due to the form of the Neuberger 
overlap Dirac operator. In contrast to this, the {\it ultraviolet filtered 
topological charge density} can be quite easily calculated. 
Moreover, the eigenmodes can be used for many 
applications~\cite{Koma:2005sw,Ilgenfritz:2005hh,Weinberg:2006ju,Ilgenfritz:2006gc}. 
In this paper, for instance, we will elucidate the low-mode dominance of the pion 
propagator in more detail.

In the past, selfdual objects have been searched for on the lattice primarily 
by looking for coherent lumps of topological 
charge~\cite{Ilgenfritz:1985dz,Teper:1985ek,DeGrand:1996ih,DeGrand:1996zb,Feurstein:1996cf,DeGrand:1997gu,DeGrand:1997ss,Smith:1998wt,GarciaPerez:1998ru} which play a
prominent role in models of vacuum structure. 
When the overlap-based topological charge became calculable, these models were 
heavily challenged~\cite{Horvath:2001ir,Horvath:2002gk}. 
Early arguments~\cite{Witten:1978bc} against the instanton dominance of the 
functional integral have been raised again. The striking new argument against 
the instanton structure was that the four-dimensional extendedness typical for 
semiclassical lumps cannot be 
reconciled with the newly discovered three-dimensional sign-coherent 
global structures~\cite{Horvath:2003yj} 
which are infinitely thin in the co-dimension (equal 
to one~\cite{Horvath:2005rv}). This is the picture that has emerged for the 
all-scale topological density. All what can be said in defense of instantons 
etc. is that it is unknown how the effect of quantum fluctuations modifies 
the topological charge density profile of a classical instanton or caloron
which is described by overlap fermions in full agreement with the classical
profile.

In this paper, concerned with fully quantum lattice configurations, we will
further develop this picture. It is pointed out that the topological charge 
density, in regions of higher density, possesses  
lower-dimensional connected structures. In the full four-dimensional landscape 
we find all substructures, from zero-dimensional peaks over one-dimensional ridges,
to two-dimensional walls. They are all nested inside the three-dimensional sign 
coherent structures mentioned above. On the other hand, it will become clear 
in which sense a model based on a dilute gas of (anti-)selfdual domains can 
also be supported by an overlap-fermion based analysis. 
Evidently, for this an ultraviolet filtered version of the overlap-based 
field strength tensor~\cite{Gattringer:2002gn,Liu:2006wa,Liu:2007hq} is necessary, 
supplemented by the mode-truncated 
topological charge density~\cite{Horvath:2002yn}. Both can be projected out by 
a certain small number of true low-lying overlap modes.  

This latter point of view is qualitatively attractive because the low-lying 
fermionic modes are known to encode the most important phenomenological properties
of the QCD vacuum relevant for the physics of light hadrons. 
The case for concentrating on the lowest part of the Dirac spectrum is illustrated
by the fact already mentioned 
that a modest number of eigenmodes is sufficient to reproduce the 
propagator of the lightest (pseudoscalar)
hadrons~\cite{Ivanenko:1997nb,Neff:2001zr,DeGrand:2000gq,DeGrand:2001tm}
(see also Fig.~\ref{fig:pionpropagator}).  The lowest-mode dominance allows for 
a systematic improvement of the propagator by the so-called low-mode averaging 
method~\cite{DeGrand:2004qw,Giusti:2005sx,Galletly:2006hq}. 
In this and a following paper this will be investigated more carefully.

Recently, the localization 
properties~\cite{Aubin:2004mp,Bernard:2005nv,Gubarev:2005az,Polikarpov:2005ey,Koma:2005sw,Polikarpov:2006ki}, 
and the effective dimension of the lower eigenmodes, and their local chiral 
properties~\cite{DeGrand:2001pj,Edwards:2001nd,Horvath:2001ir,Horvath:2002gk} 
have received strong interest. The results partly contradict each other.
Concerning the localization the situation has been reviewed in 
Ref.~\cite{deForcrand:2006my}. There, de Forcrand points out the basic 
difference between $SU(2)$ fields simulated with Wilson action, when scale $a$ 
dislocations have too low action~\cite{Pugh:1989ek} to compensate for their 
entropy $\sim \log a^{-4}$, and $SU(3)$ gauge fields simulated with L\"uscher-Weisz 
action with dislocations suppressed. 
The general motivation behind this high attention for issues like localization and
(fractal) dimensionality was the hope~\cite{Zakharov:2004jh,Zakharov:2006vt} 
that the lowest modes are bound to certain singularities of the gauge field 
that in turn would explain confinement.  Here, ``singular'' is understood
in a sense opposite to semiclassical lumps, just in the spirit of 
Witten's~\cite{Witten:1978bc} criticism of instantons. 
This is discussed from a more recent point of view in 
Refs.~\cite{Zakharov:2006te,Zakharov:2006tf}. 
Uncorrelated instantons, indeed, are found 
unrelated to confinement. It should be mentioned in passing, however, that 
other semiclassical configurations with non-trivial asymptotic holonomy 
can be related to confinement, as was recently 
demonstrated in Ref.~\cite{Gerhold:2006sk}.

Another source of inspiration was the hypothetical analogy drawn between the 
finite-temperature chiral transition to the Anderson transition (metal insulator 
transition)~\cite{Garcia-Garcia:2005dp,Garcia-Garcia:2005vj,Garcia-Garcia:2006gr}
in condensed matter physics. As precursor of the chiral symmetry restoring 
transition, before the spectral gap opens, the spectrum in the gap region is 
expected to become critical, exposing both spectral criticality and multifractality 
of modes. We will come back to these aspects in a forthcoming 
publication dealing with finite temperature full QCD.
In the present case of quenched QCD the whole low-lying spectrum seems to be 
critical and multifractal.

This paper is organized as follows.
In Section~\ref{sec:tools} the basic tools and settings of the quenched 
simulation and for handling the eigenmodes of the overlap operator are explained. 
In Section~\ref{sec:spectralresults} the properties of individual modes are discussed.
This begins with the distribution of the topological charges $Q$ 
(given by the number and chirality of zero modes)
and the spectral density of non-zero modes.
From the latter, the chiral condensate is extracted by fitting our finite-volume 
data to quenched chiral perturbation theory~\cite{Damgaard:2001xr}.  
Then the localization behavior, the dimensionality 
and the local chiral properties (chirality $X(x)$) of individual modes 
are discussed. 
In Section~\ref{sec:topdensity} we describe our experience with the 
topological charge density defined~\cite{Niedermayer:1998bi} 
through the overlap Dirac operator 
and its ultraviolet filtered (mode-truncated) variant.  
The existence of a three-dimensional, singular and sign-coherent 
global structure revealed by the all-scale topological charge density,
first pointed out in Ref.~\cite{Horvath:2003yj},
is confirmed and described in more detail.
The negativity of the two-point function of the 
topological charge density~\cite{Horvath:2005cv,Koma:2005sw} 
is found to be realized for the all-scale topological charge density for high 
enough $\beta$, {\it i.e.} at lattice spacing $a \lesssim 0.1$ fm. 
In the same Section~\ref{sec:topdensity} 
we also discuss the topological structure in terms of what we 
call $q$-clusters. The multiplicity, size, distance and the percolation 
behavior of $q$-clusters is compared for various levels of ultraviolet filtering 
($\lambda_{\rm cut}$) and for the all-scale topological charge density. 
With an appropriate $\lambda_{\rm cut}$  
the $q$-cluster structure of the mode-truncated density $q_{\lambda_{\rm cut}}(x)$ 
can be made agreeing with the $R$-cluster structure defined later in 
Section~\ref{sec:UVfilterF}.
Finally, estimating the fractal dimension of $q$-clusters related to the all-scale 
topological charge density at various threshold values $q_{\rm cut}$ , 
we find that the three-dimensional structure of sign-coherent topological charge 
is supplemented by nested, lower-dimensional substructures 
at higher density $|q(x)|$. 
Obviously these features have no analog for the mode-truncated topological charge 
density. 
In Section~\ref{sec:UVfilterF} an ultraviolet filtering technique 
for the field strength tensor is introduced. It was inspired by a similar 
work by Gattringer~\cite{Gattringer:2002gn}. The representation of the 
field strength tensor in terms of the overlap Dirac operator was later 
discussed by Liu {\it et al.}~\cite{Liu:2006wa,Liu:2007hq}.
An infrared field strength can be obtained that allows one to assign at each
point in space-time  
the degree $R(x)$ of selfduality or antiselfduality.
Connected clusters (called $R$-clusters) are found such that in
the interior (anti-)selfduality is approximately satisfied.
These clusters start to percolate and form a dilute network 
if slight deviations from (anti-)selfduality are tolerated. 
Thus, locally perfectly selfdual and antiselfdual vacuum fields are 
embedded in a vacuum that is globally neither selfdual nor antiselfdual.
In Section~\ref{sec:conclusions} we summarize our findings and
discuss from this point of view the saturation of the pion propagator.
We draw conclusions and point out routes for further research.

\section{Basic tools}
\label{sec:tools}

\subsection{Lattice ensembles}
\label{sec:2.1}

Topological studies using the Wilson one-plaquette gauge field action suffer
from dislocations~\cite{Gockeler:1989qg} and should be treated with caution.
For this paper the L\"uscher-Weisz gauge field action~\cite{Luscher:1984xn}
is used which is known to suppress dislocations and to greatly reduce the 
number of unphysical zero modes of the Wilson-Dirac operator. The L\"uscher-Weisz 
gauge field action is given by
\begin{eqnarray}
S[U]&\!\!\!=\!\!\!&\beta\!\!\!\!
\sum_{\rm plaquette}\frac{1}{3}\,
\mbox{Re}\,
\mbox{Tr}\, (1-U_{\rm plaquette}) \nonumber \\
&\!\!\!+\!\!\!& \beta_R\!\!\!\! \sum_{\rm rectangle}\frac{1}{3}\, \mbox{Re}\,
\mbox{Tr}\, (1-U_{\rm rectangle}) \label{ImpAct} \\
&\!\!\!+\!\!\!& \beta_P\!\!\!\!\!\!\!\! \sum_{\rm parallelogram}\frac{1}{3}\,
\mbox{Re}\, \mbox{Tr}\, (1-U_{\rm parallelogram})   \nonumber
\end{eqnarray}
in terms of the standard plaquette, the planar rectangle and the parallelogram
loop terms. The latter are closed along the diagonally opposite parallel links 
on the surface of a 3D cube.

The coefficients $\beta_R$ and $\beta_P$
are taken from tadpole improved perturbation 
theory~\cite{Luscher:1985zq,Lepage:1992xa,Alford:1995hw,Snippe:1997ru} 
\begin{eqnarray}
\beta_R & = & - \frac{\beta}{20~u_0^2}~\left( 1 + 0.4805~\alpha \right) \; ,\\
\beta_P & = & - \frac{\beta}{u_0^2}~0.03325~\alpha  \; ,
\end{eqnarray}
where
\begin{equation}
u_0 = \left( \frac{1}{3}~\langle~\mbox{Re}~\mbox{Tr}~U_{\rm plaquette}~\rangle \right)^{1/4} \qquad , \qquad 
\alpha = - \frac{\log \left( u_0^4 \right)}{3.06839} \; .
\end{equation}
The couplings $\beta_R$ and $\beta_P$ have been selfconsistently determined
through the calculation of the average plaquette 
in Ref.~\cite{Gattringer:2001jf} for a set of $\beta$ values.
We can write $\beta=c_0 \frac{6}{g^2}$, $\beta_R=c_1 \frac{6}{g^2}$ and
$\beta_P=c_2 \frac{6}{g^2}$ with $c_0+8c_1+8c_2=1$, which fixes the relation
between $\beta$ and the bare coupling $g^2$.

To investigate the volume dependence of our data we have simulated at
three different volumes at fixed coupling $\beta=8.45$.
To explore the $a$ dependence of the results also a $12^3\times 24$ lattice at 
$\beta=8.10$ has been employed with approximately the same physical volume as 
the $16^3\times 32$ lattice at $\beta=8.45$.
Finally, a large ensemble of rather coarse configurations on a $16^3\times 32$ 
lattice at $\beta=8.00$ became available in the QCDSF collaboration. A first
physics analysis on this basis has been 
presented in Ref.~\cite{Galletly:2006hq}. 
The physical volume of this ensemble almost equals the 
biggest physical volume studied on $24^3\times 48$ at $\beta=8.45$.

In Table~I the statistics of lattices used in our investigation is listed. 
The lattice spacing for $\beta=8.00$ and $\beta=8.45$ was determined in \cite{Galletly:2006hq}, where we used the pion decay constant of $f_{\pi}=92.4$ MeV to set the scale.
The value of the lattice spacing at $\beta=8.10$ is interpolated by fitting our measured pion decay constants at  $\beta=8.00$ and $\beta=8.45$  to   equation (6) of \cite{Gattringer:2001jf}.
 
It is in this scale that the spatial linear extent $a~L_s$ of the 
lattices, the 4D volumes $V$ and the topological susceptibilities $\chi_{\rm top}$ 
are obtained. 
The topological susceptibilities will be discussed in Section~\ref{sec:3.1}
\begin{table}
\vspace*{1cm}
\begin{center}
\begin{tabular}{|c|c|c|c|c|c|c|c|}
\hline
$\beta$ & $a$ [fm] & $L_s^3 \times L_t$ & $a~L_s$ [fm] & $V$ [fm$^4$] & $\chi_{\rm top}$ & \# of config. & \# of modes \\
\hline
\hline
8.45    & 0.105(2) & $12^3\times 24$ & 1.3 &  5 & $[167(3) \rm{MeV}]^4$ &  437 & $O(50)$ \\
8.45    & 0.105(2) & $16^3\times 32$ & 1.7 & 16 & $[169(3) \rm{MeV}]^4$ &  400 & $O(150)$ \\
8.45    & 0.105(2) & $24^3\times 48$ & 2.5 & 81 & $[168(4) \rm{MeV}]^4$ &  250 & $O(150)$ \\
\hline
8.10    & 0.142(2) & $12^3\times 24$ & 1.7 & 15 & $[171(1) \rm{MeV}]^4$ &  251 & $O(150)$ \\
\hline
8.00    & 0.157(3) & $16^3\times 32$ & 2.5 & 74 & $[172(4) \rm{MeV}]^4$ &  2156 & $O(170)$ \\
\hline
\end{tabular}
\vspace{0.3cm}
\caption{Details of the quenched ensembles used in this study: 
couplings $\beta$, lattice spacings $a$ determined from the pion decay constant, 
lattice sizes ($L_s$ and $L_t$), physical lattice sizes ($a~L_s$ and the volume 
$V$), topological susceptibilities $\chi_{\rm top}$, 
the statistics of configurations and available overlap fermion modes.} 
\end{center}
\label{tab:ensembles}
\end{table}

\subsection{Implementation of Neuberger overlap fermions}
\label{sec:2.2}

Overlap fermions~\cite{Neuberger:1997fp,Neuberger:1998wv}
have an exact chiral symmetry on the lattice~\cite{Luscher:1998pq}
and provide the cleanest known theoretical description of lattice fermions.
Their implementation of chiral symmetry and the possibility to exactly
define the index theorem on the lattice at finite lattice spacing 
allow one to investigate the relationship of topological properties of 
gauge fields and the dynamics of fermions.
A further advantage of overlap fermions, in contrast to Wilson fermions,
is that they are automatically $O(a)$ improved~\cite{Capitani:1999uz}.

\begin{figure}[t]
\begin{center}
\hspace*{-1.0cm}\epsfig{file=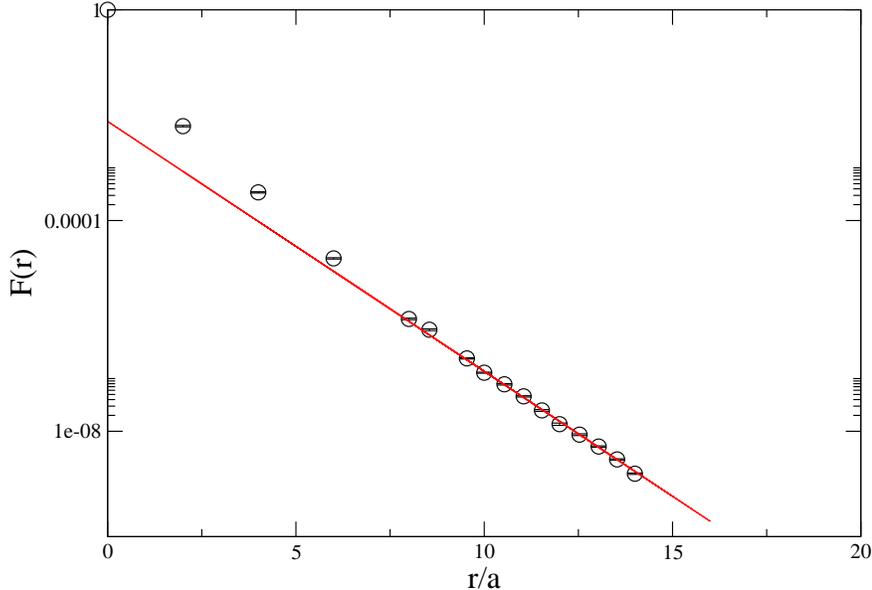,angle=270,width=12cm, clip=}
\end{center}
\caption{The effective range function $F(r)$ as a function of $r/a$ on the
$16^3\times32$ lattice at $\beta=8.45$ for $\rho=1.4$, together with an 
exponential fit.}
\label{fig:locality}
\end{figure}

The massless Neuberger overlap operator is defined by
\begin{equation}
D(0) = \frac{\rho}{a}\left(1+\frac{D_W}{\sqrt{D_W^\dagger D_W}}\right),
\quad D_W = M -\frac{\rho}{a},
\label{DN}
\end{equation}
where we use the Wilson-Dirac operator $D_W$ as input. $M$ is Wilson's
hopping term with $r=1$. 
The negative mass parameter $\rho$ is chosen to be $1.4$, which 
represents a reasonable compromise between the physical requirement of 
good locality properties~\cite{Hernandez:1998et,Galletly:2006hq} of the 
overlap operator and a performance requirement,  
demanding a small condition number of $H_W^2$,  where $H_W=\gamma_5 D_W$ is
the Hermitean Wilson-Dirac operator. 
In Fig.~\ref{fig:locality} the effective range of $D(0)$, represented by
the decay of
\begin{equation}
F(r)= \left\langle \left\langle \max_x |D(0;x,y)|\,\Big|_{|x-y|=r}
\right\rangle_y  \,\right\rangle_U \,,
\label{eq:locality_of_D}
\end{equation}
with respect to the Euclidean distance
\begin{equation}
 |x| = \left(\sum_{\mu=1}^4 x_\mu^2 \right)^{\frac{1}{2}} \; ,
\end{equation}
is shown, as obtained on a $16^3\times 32$ lattice at $\beta=8.45$ for $\rho=1.4$ ,

Asymptotically, $F(r) \propto \exp \left(-\mu~r/a \right)$, where $\mu$ depends on 
$\rho$ (and {\it a priori} also on the roughness of the configurations, {\it i.e.} $\beta$).
It turns out, however, that the fitted slope $\mu = 1.11(1)$ is practically 
independent of $\beta$. Thus, the overlap Dirac operator is not ultralocal. 
It has a range that shrinks together with the lattice spacing towards the 
continuum limit. This is sufficient for any Dirac operator $D$ to be called local.
    
To compute the sign function
\begin{equation}
{\rm sgn}(D_W) = \frac{D_W}{\sqrt{D_W^\dagger D_W}} \equiv
\gamma_5\, {\rm sgn}(H_W) \; ,  \quad H_W = \gamma_5~D_W \; , 
\end{equation}
we write  
\begin{equation}
{\rm sgn}(H_W) = \sum_{i=1}^N {\rm sgn}(\mu_i) \chi_i \chi_i^{\dagger}
+ P_{\perp}^N {\rm sgn}(H_W) \; .
\end{equation}
The first part is exactly handled in the subspace of eigenvectors
$\chi_i$ of $H_W$, the rest acting in the orthogonal subspace is approximated 
by a minmax polynomial~\cite{Giusti:2002sm} satisfying 
\begin{equation}
|P_{\rm minmax}(x) - \frac{1}{\sqrt{x}}| < \epsilon
\qquad {\rm for} \qquad \mu_{N+1}^2 < x < \mu_{\rm max}^2 \; , 
\end{equation}
such that
\begin{equation}
{\rm sgn}(H_W) = \sum_i^N {\rm sgn}(\mu_i) \chi_i \chi_i^{\dagger}
+ P_{\perp}^N~H_W~P_{\rm minmax}(H_W^2) \; .
\end{equation}
The lowest $N = O(50)$ eigenvalues $\mu_i$ of $H_W$ and the corresponding 
eigenvectors $\chi_i$ have to be computed.  
The degree of the polynomial $P_{\rm minmax}$ depends on $\epsilon$ and 
is proportional to the condition number $\kappa = \mu^2_{\rm max}/\mu^2_{N+1}$ 
of $H_W^2$ in the orthogonal subspace spanned by the modes 
$\{\chi_i\,|\,(1- P_{\perp}^N) \chi_i = 0\}$.
For configurations generated with the L\"uscher-Weisz action, the condition number 
of $H_W$ is essentially smaller in comparison with the Wilson action 
and only slightly depends on $\rho$.   

$D(0)$ as described above is a massless Dirac operator. 
The massive overlap Dirac operator
is easily obtained from the massless one, as~\cite{Capitani:1999uz}
\begin{equation}
D(m) = \left( 1 - \frac{a~m}{2~\rho} \right) D(0) + m \; .
\label{eq:massive_overlap}
\end{equation}

Later on, in our spectral analysis we use the improved massless Neuberger 
operator~\cite{Chiu:1998aa,Chiu:1998gp},
\begin{equation}
D^{\rm imp}(0) = \left( 1 - \frac{a}{2~\rho}~D(0) \right)^{-1} D(0) \; .\
\label{eq:improved_overlap}
\end{equation}
The resulting eigenvalues $\lambda_{\rm imp}$ of $D^{\rm imp}(0)$ are the 
stereographical projection of the eigenvalues $\lambda_N$ of $D(0)$ from the 
Ginsparg-Wilson circle to the imaginary axis. 
The non-zero eigenvalues appear in pairs $\lambda_{\rm imp} \equiv \pm i~\lambda$. 
In terms of these imaginary parts $\lambda$, the spectral density 
and the characteristic properties of the eigenmodes are discussed. 
Also any mode cut-off $\lambda_{\rm cut}$ refers to this $\lambda$.

Finally, the improved massive overlap Dirac operator~\cite{Galletly:2006hq}, 
which is used for example for the pion propagator in this paper, is 
\begin{equation}
D^{\rm imp}(m) = D^{\rm imp}(0) + m \; .
\label{eq:improved_massive_overlap}
\end{equation}

\subsection{Eigenmodes of the Neuberger operator}
\label{sec:2.3}

In principle, the emerging Neuberger overlap operator $D(0)$ could have
$n_-$ plus $n_+$ exact zero modes, $D(0) \psi_n = 0$, with $n_-$ ($n_+$) being 
the number of modes with negative (positive) chirality, 
$\gamma_5 \psi_n = - \psi_n$ ($\gamma_5 \psi_n = + \psi_n$).
The index of $D(0)$ agrees with the topological charge, $Q = n_- - n_+$.
In practice, however, there are always {\it only} negative {\it or} positive 
chirality eigenmodes for a given lattice configuration.
The non-zero eigenvalues $\lambda_N$ and their non-chiral eigenvectors with
$D(0) \psi_{\lambda_N} = \lambda_N \psi_{\lambda_N}$ appear in complex
conjugate pairs $\lambda_N$ and $\lambda_N^*$ and satisfy
$\sum_x \left(\psi_{\lambda_N}^\dagger(x),\gamma_5 \psi_{\lambda_N}^{}(x) \right) = 0$.

Locally, the eigenmodes $\psi_{\lambda}$ of the Neuberger operator 
can be characterized by the scalar density
\begin{equation}
p_{\lambda}(x) =  |\psi_{\lambda}(x)|^2 = \sum_{\sigma,c} \psi_{\lambda}^{\sigma~c~*}(x)
\psi_{\lambda}^{\sigma~c}(x) \;, \quad \sum_x p_{\lambda}(x) = 1 \; ,
\end{equation}
where $\sigma$ and $c$ denote spinor and color indices. 

Besides the scalar density, another important density needed 
is the pseudoscalar density
\begin{equation}
p_{\lambda~5}(x) = \sum_{\sigma,\sigma^{\prime},c} \psi_{\lambda}^{\sigma~c~*}(x)
\gamma_5^{\sigma~\sigma^{\prime}} \psi_{\lambda}^{\sigma^{\prime}~c}(x) \;,
\quad \sum_x p_{\lambda5}(x) = \pm 1 \quad {\rm or} \quad 0
\label{eq:eq25}
\end{equation}
for chiral or non-chiral modes, respectively. While the zero modes are
globally chiral, the non-chiral non-zero modes may still have a 
local chirality,
\begin{equation}
p_{\lambda \pm}(x) = \sum_{\sigma,c} \psi_{\lambda}^{\sigma~c~*}(x)
P_{\pm}^{\sigma~\sigma^{\prime}} \psi_{\lambda}^{\sigma^{\prime}~c}(x) \;,
\end{equation}
with the projectors $P_{\pm} = \left( 1 \pm \gamma_5 \right)/2$ onto positive
and negative chirality. 

\subsection{Overlap definition of the all-scale and the 
mode-truncated topological density}
\label{sec:2.4}

As for any $\gamma_5$-Hermitean Dirac operator
satisfying the Ginsparg-Wilson relation, 
the topological charge density for the Neuberger operator $D(0)$
can be expressed as~\cite{Hasenfratz:1998ri}:
\begin{equation}
q(x)=-~\mathrm{tr}~\left[ \gamma_5~\left( 1 - \frac{a}{2}~D(0;x,x) \right) \right] \; , \quad Q=\sum_x q(x) \; ,
\label{eq:eq-q}
\end{equation}
where the trace $\mathrm{tr}$ is taken over color and spinor indices. 

To compute the topological charge density, we use two different
approaches \cite{Horvath:2002yn,Horvath:2003yj}.
In the first approach, the trace of the overlap
operator is directly evaluated according to Eq.~(\ref{eq:eq-q}).
This is computationally very demanding  and is
therefore performed on only 53 (5) configurations in the case of a 
$12^3\times 24$ lattice ($16^3\times 32$ lattice, respectively) at $\beta=8.45$.
The all-scale density $q(x)$ computed in this way includes charge fluctuations 
at all scales. For this density, interesting anisotropic, global structures have been
detected and discussed by Horvath {\it et al.}~\cite{Horvath:2005rv}.
In this paper we will give more details on the multifractal properties
of $q(x)$.

The second technique involves the computation of the topological charge
density based only on the low-lying modes of the overlap Dirac operator.
This approach is a gauge invariant filtering that leaves the lattice configuration 
unchanged (as well as the effective lattice spacing as the physical scale). 
Using the spectral representation of the Dirac operator, the truncated
eigenmode expansion of the topological charge density reads
\begin{eqnarray}
q_{\lambda_{\rm cut}}(x)&=&-\sum_{|\lambda|<\lambda_{\rm cut}}(1-\frac{\lambda}{2})\; p_{\lambda5}(x) \; ,
\label{eq:qlambdacut}
\end{eqnarray}
with $p_{\lambda5}(x)$ defined in Eq. (\ref{eq:eq25}).
Truncating the expansion at $\lambda_{\rm cut}$ acts like an ultraviolet 
filter by 
removing the short-distance fluctuations from $q(x)$. We will study 
in more detail how 
the properties of this density depend on the choice of $\lambda_{\rm cut}$. 
With a suitable cut-off the clusters of this density will coincide with
the $R$-clusters of (anti-)selfdual domains 
to be introduced in Section~\ref{sec:5.3}.
Note that the total topological charge $Q=\sum_x q_{\lambda_{\rm cut}}(x)$ 
is not affected by the choice of $\lambda_{\rm cut}$, the level of 
truncation.

\subsection{Cluster analysis} 
\label{sec:2.5}

In this paper we will describe the properties of different 
densities in terms of their cluster properties. This is the appropriate
place to describe the cluster algorithm in general terms.
As an example we discuss this for the topological charge density $q(x)$.

\begin{enumerate}
\item The cluster analysis first requires tagging the lattice sites that 
will form the clusters. In the present case these are all sites with 
$|q(x)| > q_{\rm cut}$. 
\item Next these tagged sites are assigned to a set of {\it link-connected} 
clusters. Two sites $x$ and $y$ are said to be link-connected if they are 
immediate neighbors on the lattice connected by a link, $x=y\pm\hat{\mu}$. 
Two link-connected and tagged sites $x$ and $y$ then belong to the same 
cluster unless $q(x)$ and $q(y)$ have opposite sign. Obviously,
the latter veto is ineffective in the case of a positive density like the 
scalar density of eigenmodes.
\item Cluster percolation is defined by the cluster correlation function 
which is given as the following ensemble average: 
\begin{equation}
f(r) = 
\frac{\sum_{x,y}~\langle~\sum_c~\Theta_{c}(x)~\Theta_{c}(y)~\rangle~\delta(r-|x-y|)}
{\sum_{x,y} \delta(r-|x-y|)} \; ,
\label{eq:clusterdef} 
\end{equation}
where $\Theta_c(x)$ is the characteristic function of a cluster $c$,
{\it i.e.} $\Theta_{c}(x)=1$ if $x \in c$ and $\Theta_{c}(x)=0$ otherwise.
In the definition (\ref{eq:clusterdef}) 
$r$ is the Euclidean distance. 
If $r_{\rm max}$ is the largest distance possible on 
the periodic lattice, we call the value $f(r_{\rm max})$ ``connectivity''.
The onset of percolation is defined as the value of the cluster-defining 
quantity, {\it i.e.} of the cut-off $q_{\rm cut}=q_{\rm perc}$, such that 
the connectivity $f(r_{\rm max}) \ne 0$ for $q_{\rm cut} < q_{\rm perc}$.
\item An important quantity describing the system of clusters is the 
Euclidean distance between two clusters, for example the two biggest clusters. 
The distance $d(c,c^{\prime})$ between two clusters labeled $c$ and $c^{\prime}$ 
can be defined as the maximum over sites $x \in c$ of the minimal 
Euclidean distance between $x$ and {\it any} site $y \in c^{\prime}$,
\begin{equation}
d(c,c^{\prime}) = \max_{x \in c} \left( \min_{y \in c^{\prime}} (|x-y|) \right) \; .
\end{equation}
\item Other quantities describing the system of clusters in a given
configuration are the fraction of occupied volume, the fraction of the largest 
cluster to the total occupied volume, 
the size and charge distributions of clusters and the total multiplicity 
of clusters.
\end{enumerate}

We will use this terminology when we shall discuss qualitative features 
of certain densities.

\subsection{Random walkers} 
\label{sec:2.6}

For the characterization of a $q$-cluster with respect to its fractal 
dimension, we use the random walker method briefly described below.
We consider a collection of independent random 
walkers starting from the maximum $q_{\rm max}$ of the topological charge density 
inside a cluster among a collection of clusters defined by a cut-off $q_{\rm cut}$.
Depending on the ratio $q_{\rm cut}/q_{\rm max}$, the cluster is more or 
less extended.  The cluster determines where the random walkers are 
permitted to go.
The random walk is defined by the condition that at each time step 
$\tau \to \tau + 1$ the walker 
is required to jump with equal probability to one of the neighboring sites 
that also belong to the cluster. If $q_{\rm cut}/q_{\rm max}$ is large, 
the cluster is small, such that the random walk soon arrives at a stationary regime. 
If $q_{\rm cut}/q_{\rm max}$ is smaller, the cluster is extended in one or more 
directions, the number of which one want to explore. The fractal dimension $d^{*}$
which is open for the random walkers will then be reflected by the return 
probability to the starting point, 
\begin{equation}
P({\vec 0},\tau)=\left(2 \pi \tau \right)^{-\frac{d^{*}}{2}} \; ,
\label{eq:returnprob}
\end{equation}
following a power-like decay with the number $\tau$ of time-steps. 
This can be defined for one cluster or as average over all clusters.

Also the distribution of all walkers in one cluster at a given time-step 
can be useful to characterize the shape of the cluster. This method is applicable
to fractal clusters of any kind, for example also those of the scalar density of
the individual fermionic modes.

\section{Spectral results based on the lowest $\mathbf{O(150)}$ eigenmodes}
\label{sec:spectralresults}

\subsection{The Ginsparg-Wilson circle and the topological charge 
from the index of $D(0)$}
\label{sec:3.1}

\begin{figure}[t]
\begin{center}
\epsfig{file=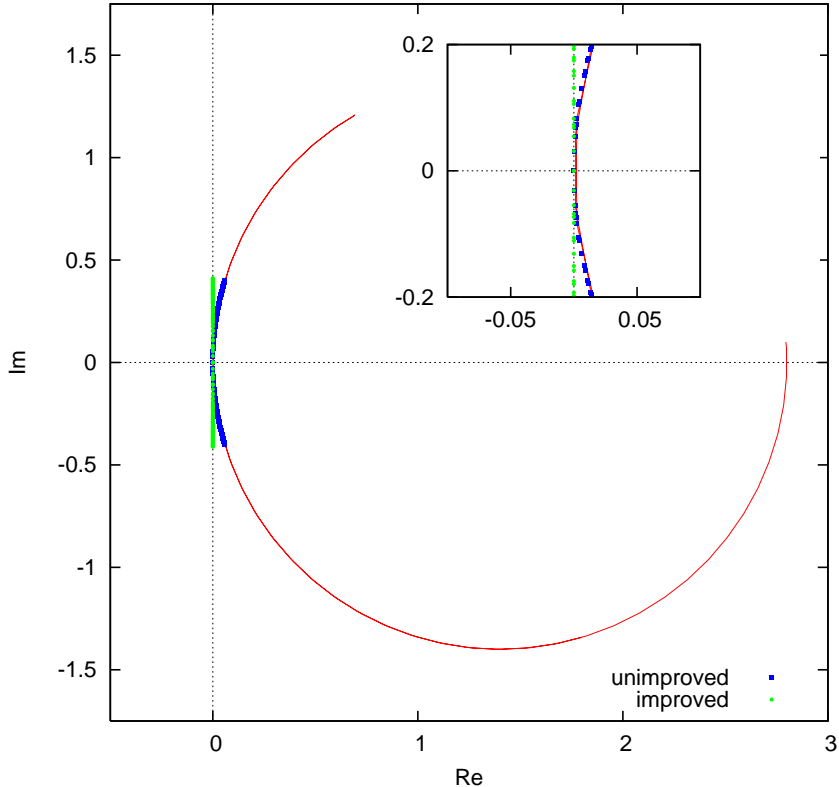,angle=270,width=12cm}
\end{center}
\caption{The Ginsparg-Wilson circle and the analyzed part of the unimproved 
and improved spectrum for a $Q=3$ configuration on the $16^3 \times 32$
lattice generated at $\beta=8.45$. The insert shows the lowest part of the 
spectrum magnified. The zero eigenvalue is threefold degenerate.}
\label{fig:GWcircle}
\end{figure}

Using the above construction of the Neuberger overlap operator, we 
perform its diagonalization by a variant of the implicitly restarted 
Arnoldi algorithm. Mostly an amount of $O(150)$ eigenvalues and 
eigenvectors per configuration has been computed and stored. 
The complex-valued eigenvalues $\lambda_N$ are located on the 
Ginsparg-Wilson circle of radius 
$\rho$ around the point $(\rho,0)$ in the complex plane. This is shown in 
Fig.~\ref{fig:GWcircle}.

\begin{figure}[t]
\begin{center}
\epsfig{file=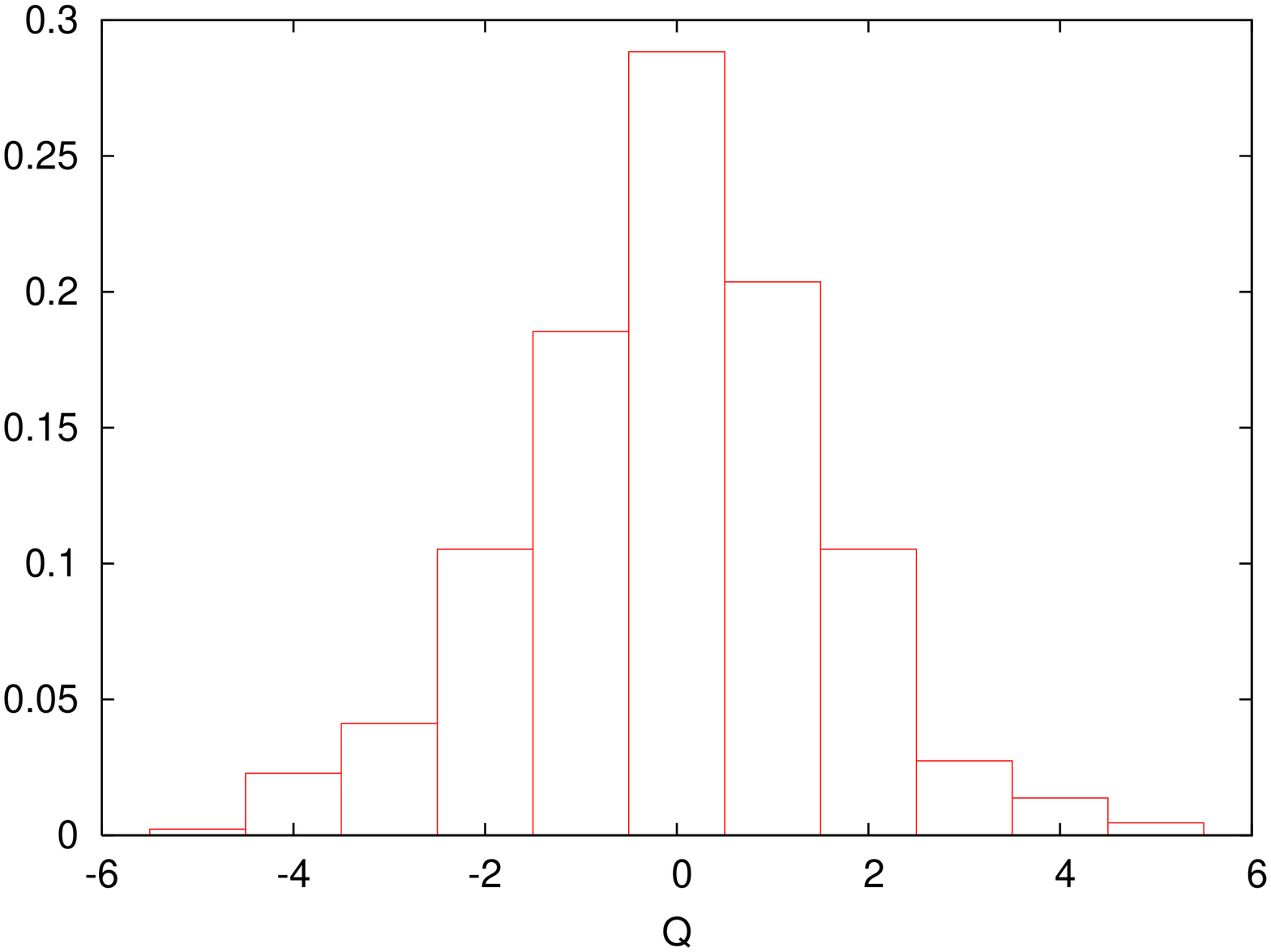,angle=270,width=8cm}\\ 
(a) \\
\begin{tabular}{cc}
\epsfig{file=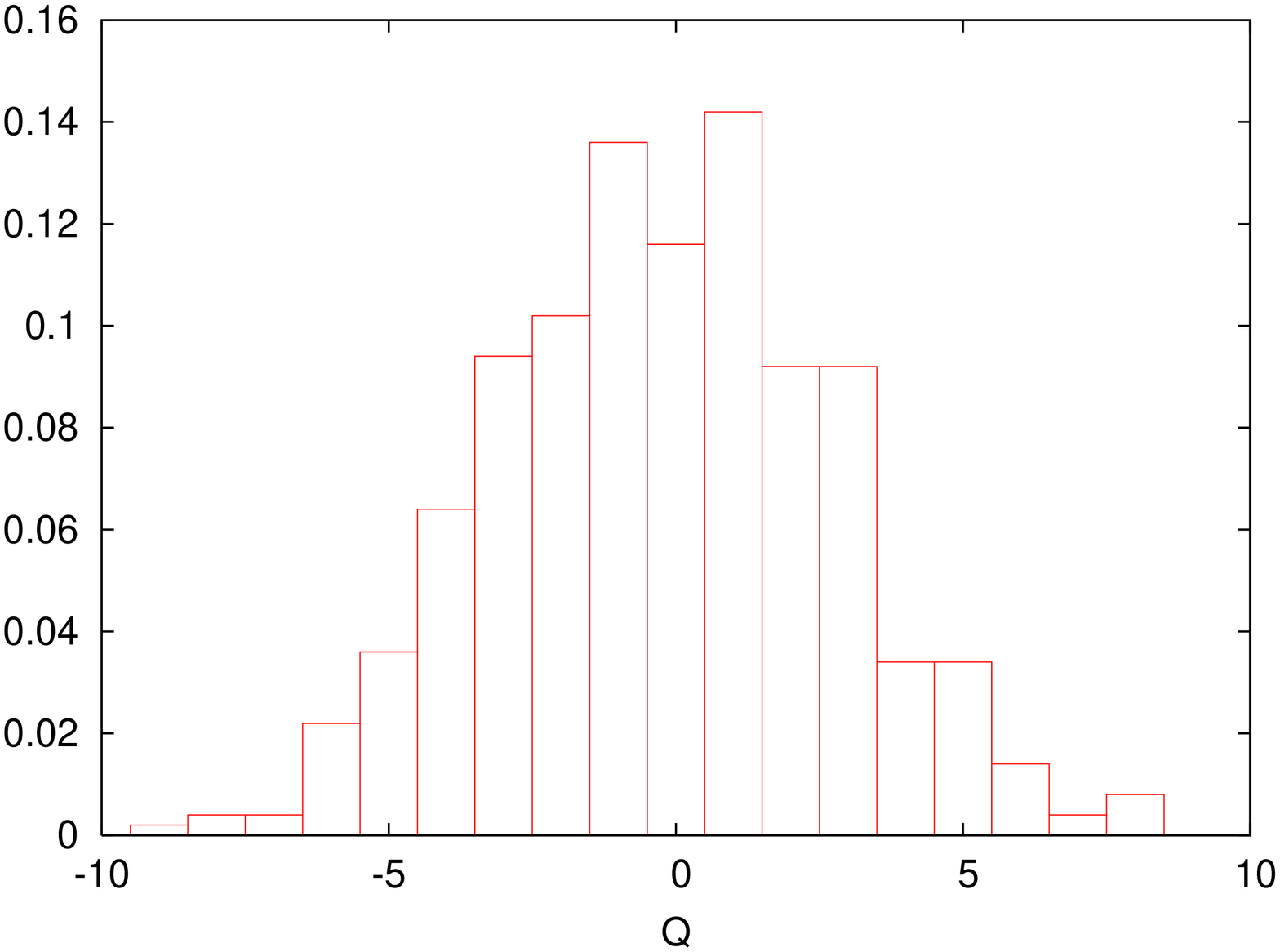,angle=270,width=8cm}&
\epsfig{file=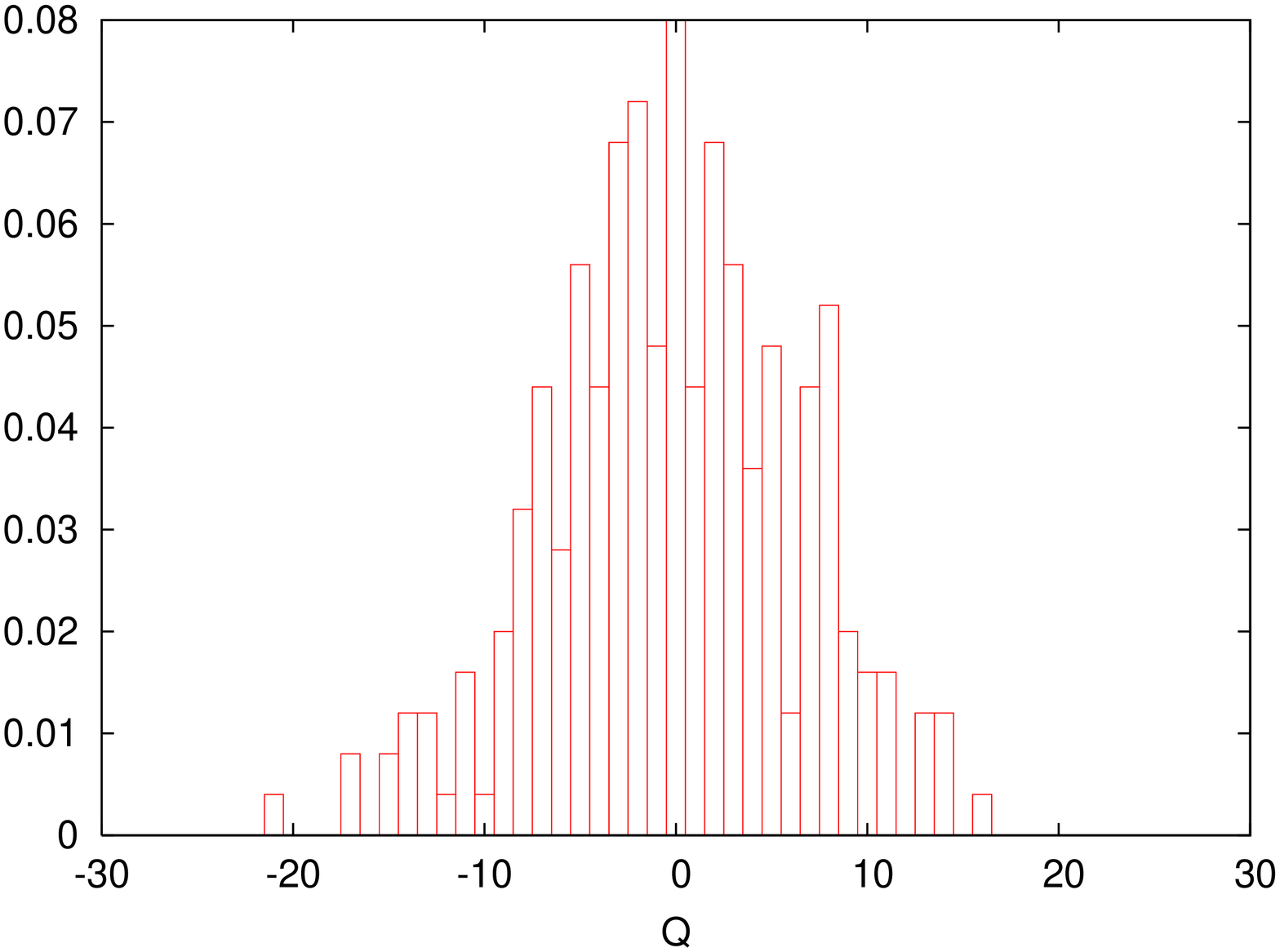,angle=270,width=8cm}\\
(b) & (c)\\
\epsfig{file=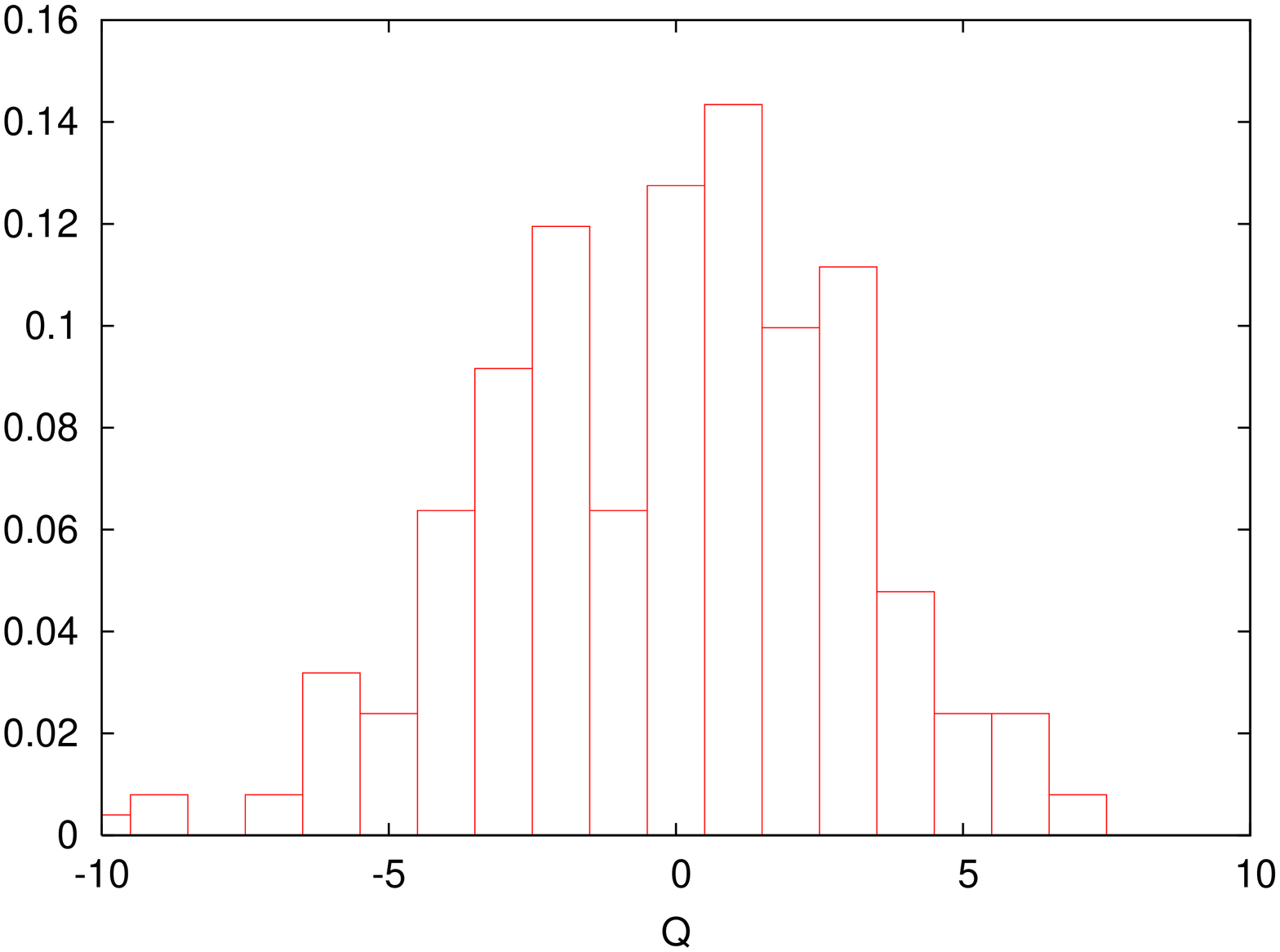,angle=270,width=8cm}&
\epsfig{file=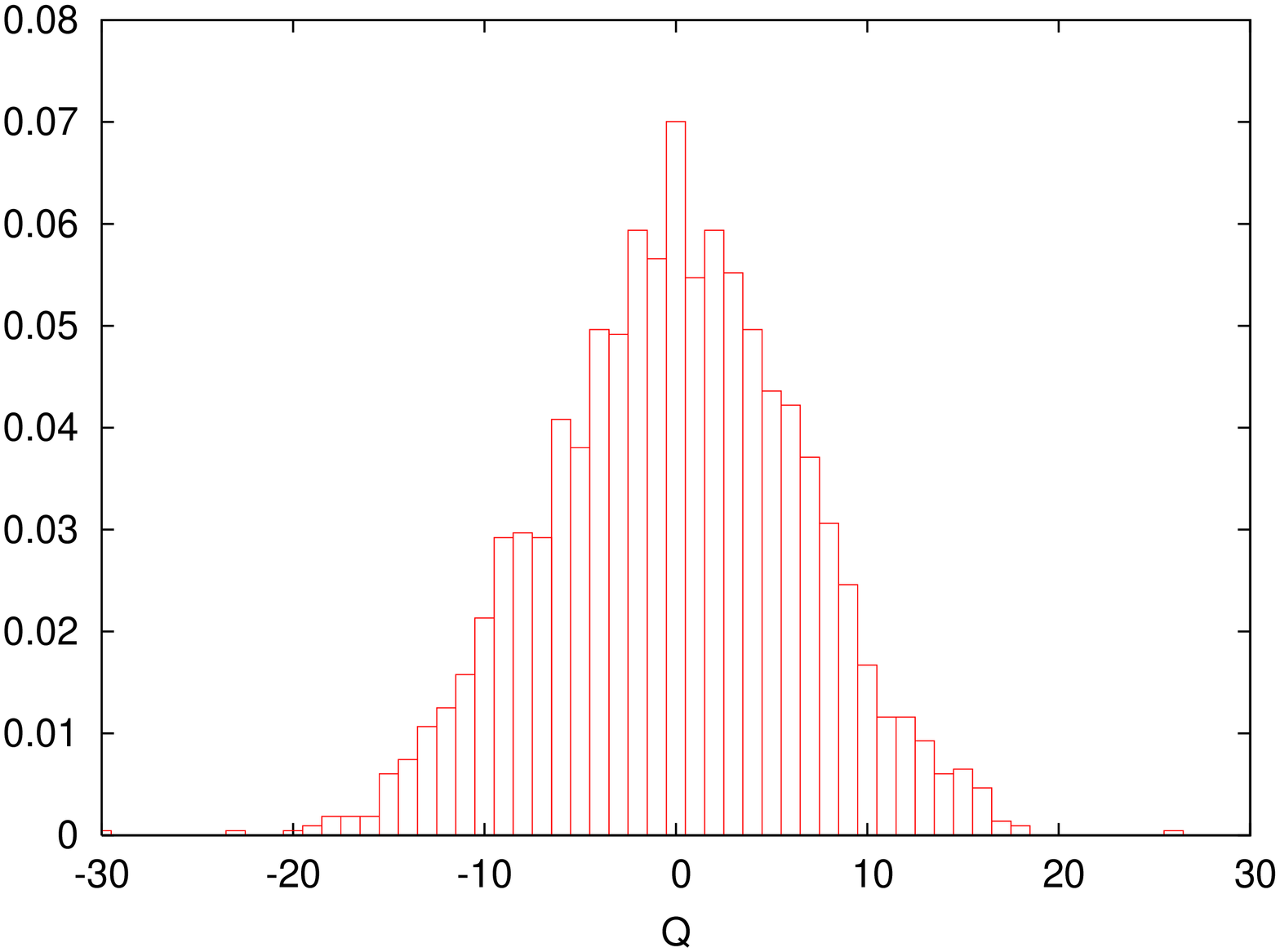,angle=270,width=8cm}\\
(d) & (e)
\end{tabular}
\end{center}
\caption{The normalized distributions of topological charge $Q$, 
in subfigures ordered according to descending $\beta$, 
for (a) $12^3\times24$ at $\beta=8.45$,
(b) $16^3\times32$ at $\beta=8.45$,
(c) $24^3\times48$ at $\beta=8.45$,
(d) $12^3\times24$ at $\beta=8.10$,
(e) $16^3\times32$ at $\beta=8.00$.
Note that the physical volumes corresponding to (b) and (d) are roughly equal,
and similarly for (c) and (e).}
\label{fig:Q-distribution}
\end{figure}

In Fig.~\ref{fig:Q-distribution} (a) to (e) the distribution of topological
charge $Q=n_{-}-n_{+}$ over the various ensembles is shown as determined from 
the number and chirality of zero modes of the lattice configurations.
As has been said, we have never found zero modes with different chirality
{\it simultaneously} in the same configuration.

From the distribution of topological charges shown in Figs.~\ref{fig:Q-distribution}
the topological susceptibilities
\begin{equation}
\chi_{\rm top} = \frac{\langle Q^2 \rangle}{V} \; ,
\end{equation}
have been evaluated. They are also given in Table~I.
It is remarkable that all physical lattice sizes realized with $\beta=8.45$ are 
big enough to get the topological susceptibility equal within error bars. 
At the same time, the topological susceptibilities of the coarser lattices 
are only slightly larger, in accordance with the fact that other quantities 
are found to scale well with $\beta$~\cite{Galletly:2006hq}. 
The topological susceptibilities at $\beta=8.1$ and
$\beta=8.0$ are equal to each other within error bars although the
physical volumes differ by a factor five. All results are smaller 
than the topological susceptibility of quenched $SU(3)$ gauge theory reported
in Ref.~\cite{DelDebbio:2004ns} to be $(191 \pm 5 \mathrm{MeV})^4$. 
This value was given using $f_K=160(2) \mathrm{MeV}$ to set the scale.

\subsection{Spectral density of non-zero modes}
\label{sec:3.3}

The spontaneous breaking of chiral symmetry by the dynamical creation of a
nonvanishing chiral condensate $\langle\bar\Psi\Psi\rangle$ is related to
the spectral density $\rho(\lambda)$ of non-zero modes of the Dirac operator
near zero by the Banks-Casher relation \cite{Banks:1979yr}
$\langle \bar\Psi \Psi \rangle = - \pi~\rho(0)$. 

The spectral density of 
the continuous modes at finite volume is formally given by
\begin{equation}
\rho(\lambda,V)=\frac{1}{V} \big\langle\sum_{\bar\lambda} \delta(\lambda
 -\bar{\lambda})\big\rangle ,
\end{equation}
where the sum extends only over positive (non-zero) values of 
$\bar{\lambda}= \mbox{Im}~\lambda_{\rm imp}$ of the eigenvalues of the improved 
massless overlap operator $D^{\rm imp}(0)$. 
The average $\langle \ldots \rangle$ is taken over the ensemble of  
gauge field configurations.

In the finite volume and for small eigenvalues the spectral 
density can be computed from the chiral low-energy effective theory. 
For $\lambda < E_T$,  $E_T$ being the Thouless energy $E_T \approx 
f_{\pi}^2/\Sigma \sqrt{V}$, the low-energy effective partition function is 
dominated by the zero momentum modes, and the zero-momentum approximation of 
the chiral low-energy effective theory is equivalent to chiral random matrix 
theory. 

In random matrix theory the spectral density is given as 

\begin{equation}
\rho(\lambda,V)=\Sigma_{\rm eff}~\sum_Q~w(Q)~\rho_Q(\Sigma_{\rm eff}V\lambda) \; ,
\label{eq:rho}
\end{equation}
where $\Sigma_{\rm eff}$ is an effective value of the chiral condensate, and
\begin{equation}
\rho_Q(x)=\frac{x}{2}(J_{|Q|}^2(x)-J_{|Q|+1}(x)J_{|Q|-1}(x))\;,
\end{equation}
is the microscopic spectral density \cite{Wilke:1997gf} in the sector of fixed
topological charge $Q$, expressed in terms of Bessel functions $J_n(x)$.

We take the weights $w(Q)$ of the sectors of topological charge $Q$ 
(normalized to $\sum_Q w(Q) = 1$) from our measured charge distributions
presented in Fig.~\ref{fig:Q-distribution}. 
To take into account the effects of higher orders in chiral perturbation 
theory we also add a term $a_1 \lambda + a_2 \lambda^2$  to the fitting 
formula (\ref{eq:rho}).

The spectral densities for the three different volumes available 
at the same $\beta=8.45$, and the coarser ensembles 
of $\beta=8.10$ ($12^3\times24$) and $\beta=8.00$ ($16^3\times32$), 
together with our fits are presented in Fig. 5.

\begin{figure}[t]
\begin{center}
\hspace*{-1.0cm}\epsfig{file=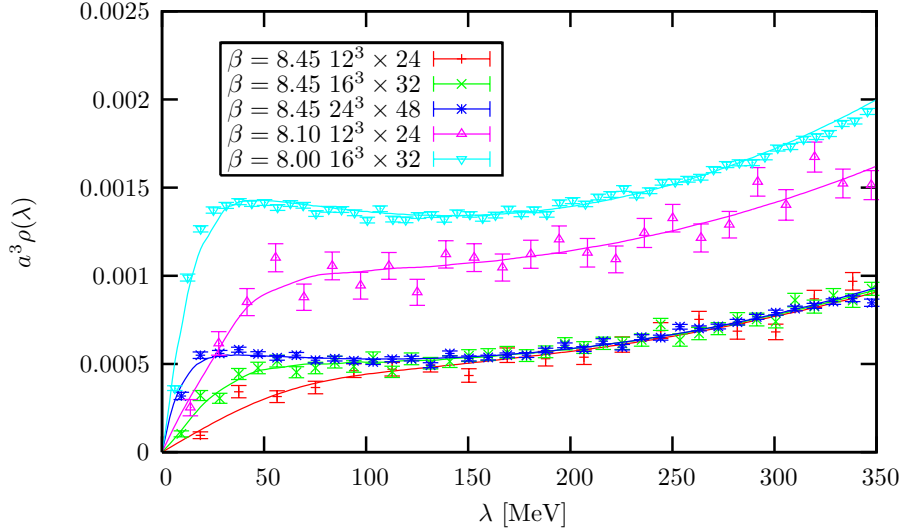,width=12.0cm}
\caption{The spectral densities $a^{3}~\rho(\lambda,V)$ 
at $\beta=8.45$ for the three lattice sizes 
under study ($12^3\times24$, $16^3\times32$ and $24^3\times48$) 
and the spectral densities at $\beta=8.1$ and $\beta=8,0$ for
lattice sizes ($12^3\times24$ and $16^3\times32$, resp.),  
together with fits using random matrix theory predictions.}
\end{center}
\label{fig:spectraldensity}
\end{figure}

In the region of small eigenvalues ($\lambda < 100$ MeV) one can see a strong 
volume dependence of the spectral density. This is in agreement with predictions 
from quenched chiral perturbation theory, 
where it has been shown \cite{Damgaard:2001xr} that the effective value of the 
chiral condensate  $\Sigma_{\rm eff}$  used in (\ref{eq:rho}) diverges 
logarithmically as the volume is sent to infinity. 

\subsection{Localization and fractal dimension of the eigenmodes}
\label{sec:3.4}

In this paper, we concentrate on details of localization and fractal 
dimension of the overlap eigenmodes. Later on, in Section~\ref{sec:topdensity}, 
similar aspects will be discussed for the topological density as derived from 
the overlap operator.

A useful measure to quantify
the localization of eigenmodes~\cite{Aubin:2004mp,Polikarpov:2005ey} is the 
inverse 
par\-ti\-ci\-pa\-tion ratio (IPR)
\begin{equation}
I(\lambda) = L_s^3~L_t~I_2(\lambda) \equiv L_s^3~L_t~\sum_x p_{\lambda}(x)^2 \; ,
\end{equation}
with a scalar density $p_{\lambda}(x)$ that is normalized for all eigenfunctions 
to $\sum_x p_{\lambda}(x)=1$. While one would have $I = L_s^3~L_t$ 
if the scalar density would 
have support only on one lattice point, the IPR decreases as the density becomes 
more delocalized, reaching $I=1$ when the scalar density is maximally spread over 
all lattice sites. For independent Gaussian 
distributed $\psi^{\sigma~c}(x)$ at each site 
(subject to an overall normalization) the density is {\it not} maximally spread, 
but still delocalized. This case corresponds to $I = \pi/2$.

\begin{figure}[t!]
\begin{center}
\begin{tabular}{cc}
\epsfig{file=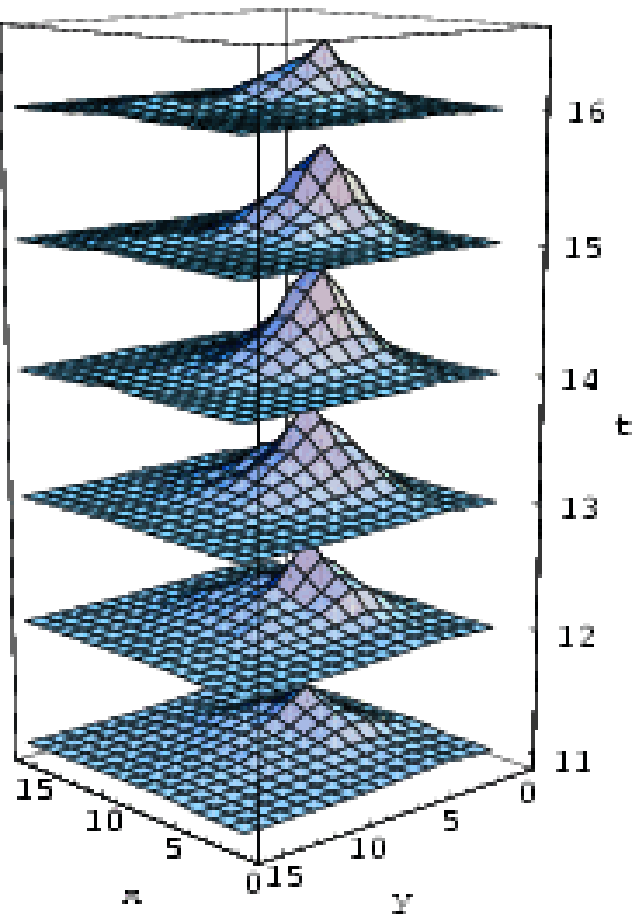,width=8cm}&
\epsfig{file=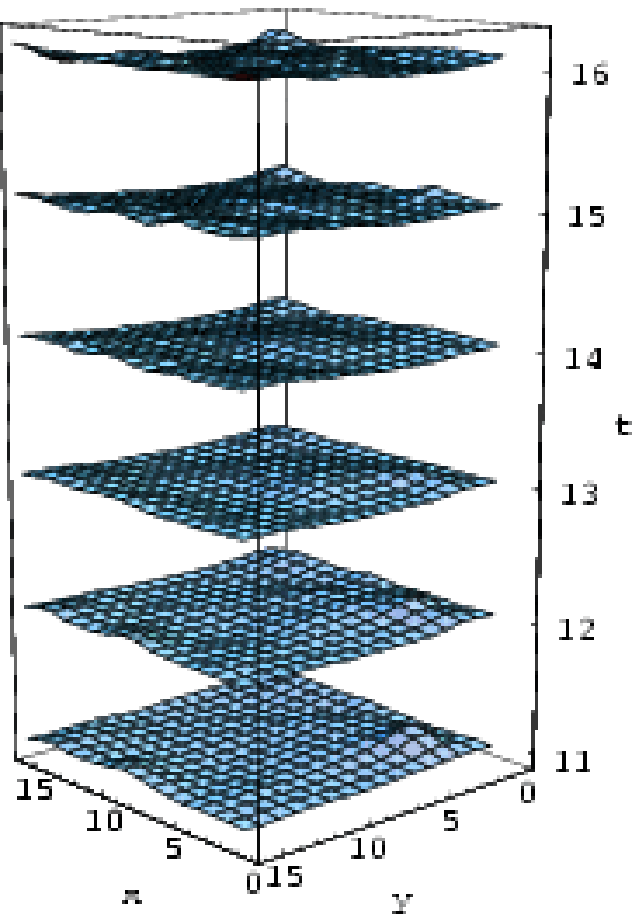,width=8cm} \\
(a) & (b)
\end{tabular}
\end{center}
\caption{The scalar density of two typical non-zero eigenmodes of the overlap 
operator, shown in two-dimensional profile for a series of subsequent timeslices 
of a $Q=0$ configuration on the $16^3\times32$ lattice generated at $\beta=8.45$. 
(a) shows the first non-zero mode, (b) the highest (144th) analyzed mode.}
\label{fig:mode-visualization}
\end{figure}
The zero modes, that do not contribute to the chiral condensate,
are always highly localized compared to the bulk of non-zero modes.
But it turns out that the lowest non-zero modes may be localized as well.
In Fig.~\ref{fig:mode-visualization}, in a series of subsequent timeslices, 
the scalar densities of the lowest (a) and of the highest analyzed non-zero mode (b) 
are shown for one $Q=0$ configuration~\footnote{This configuration serves as 
an example for many other structural observations in this paper.} 
of the $16^3\times32$ ensemble generated at $\beta=8.45$. 
The lowest non-zero mode is clearly localized in a few timeslices. But the scalar 
density of the highest analyzed mode is also slightly inhomogeneous.

\begin{figure}[t]
\begin{center}
\epsfig{file=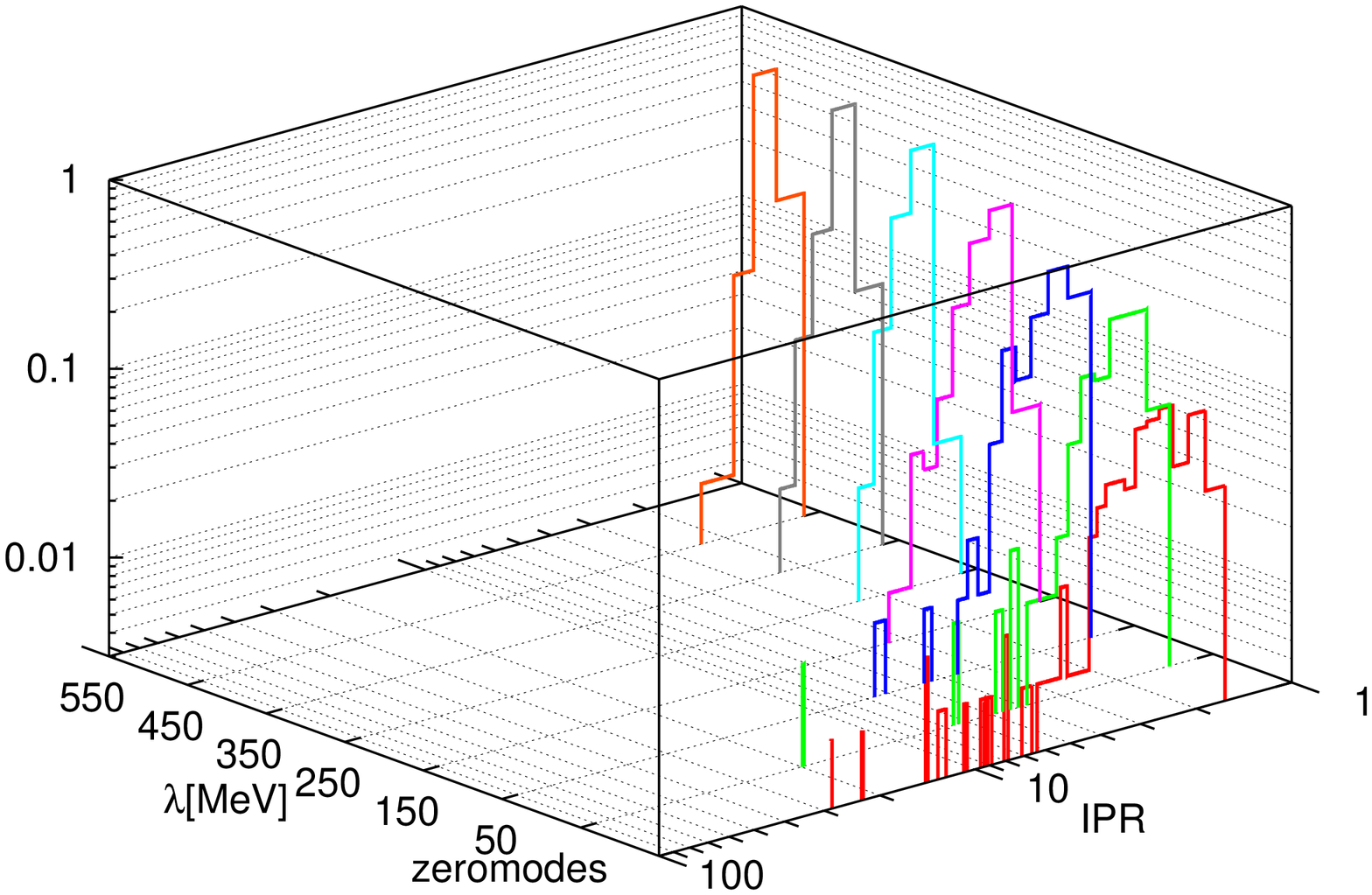,angle=270,width=8cm}\\ 
(a) \\
\begin{tabular}{cc}
\epsfig{file=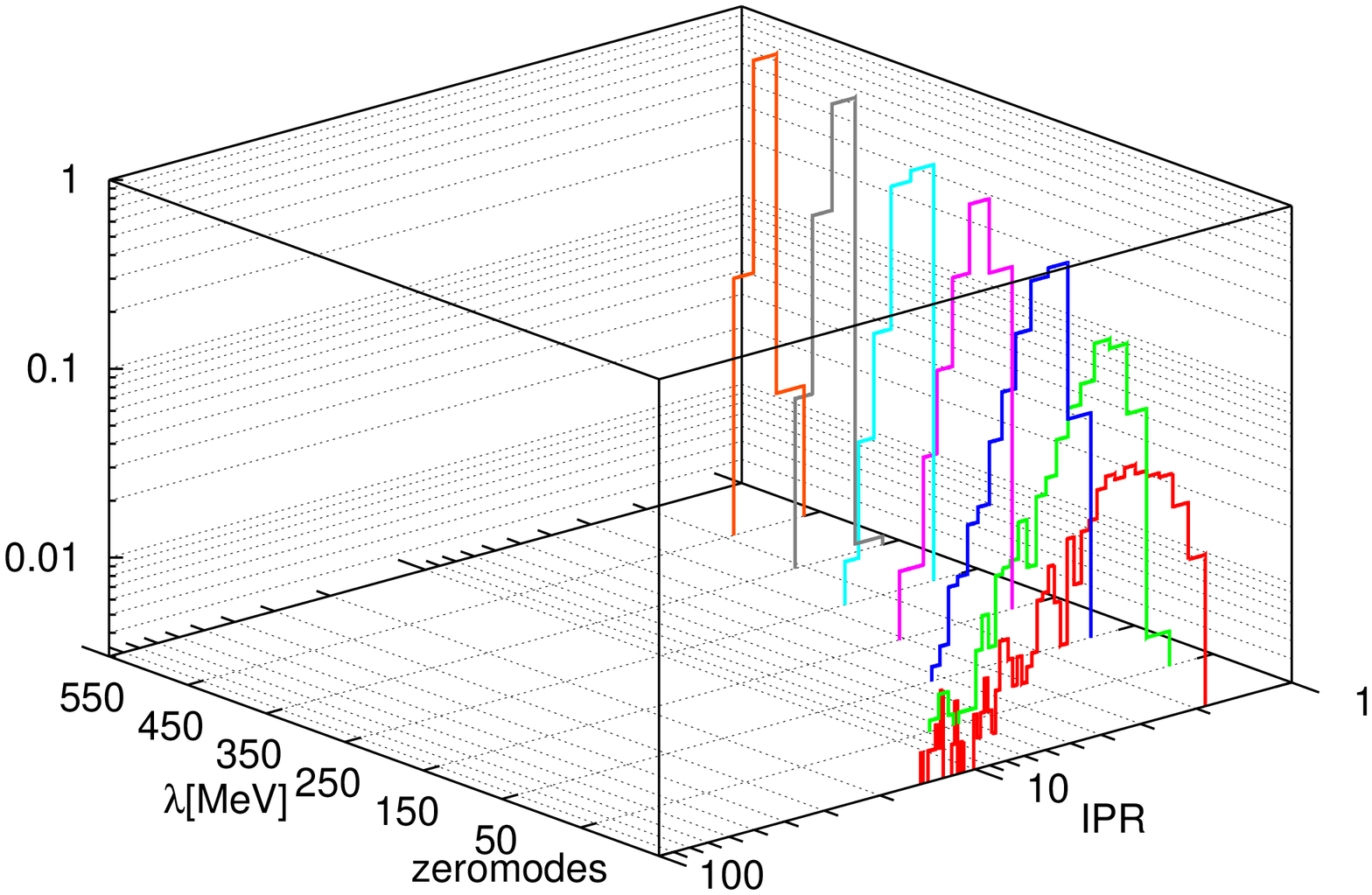,angle=270,width=8cm}&
\epsfig{file=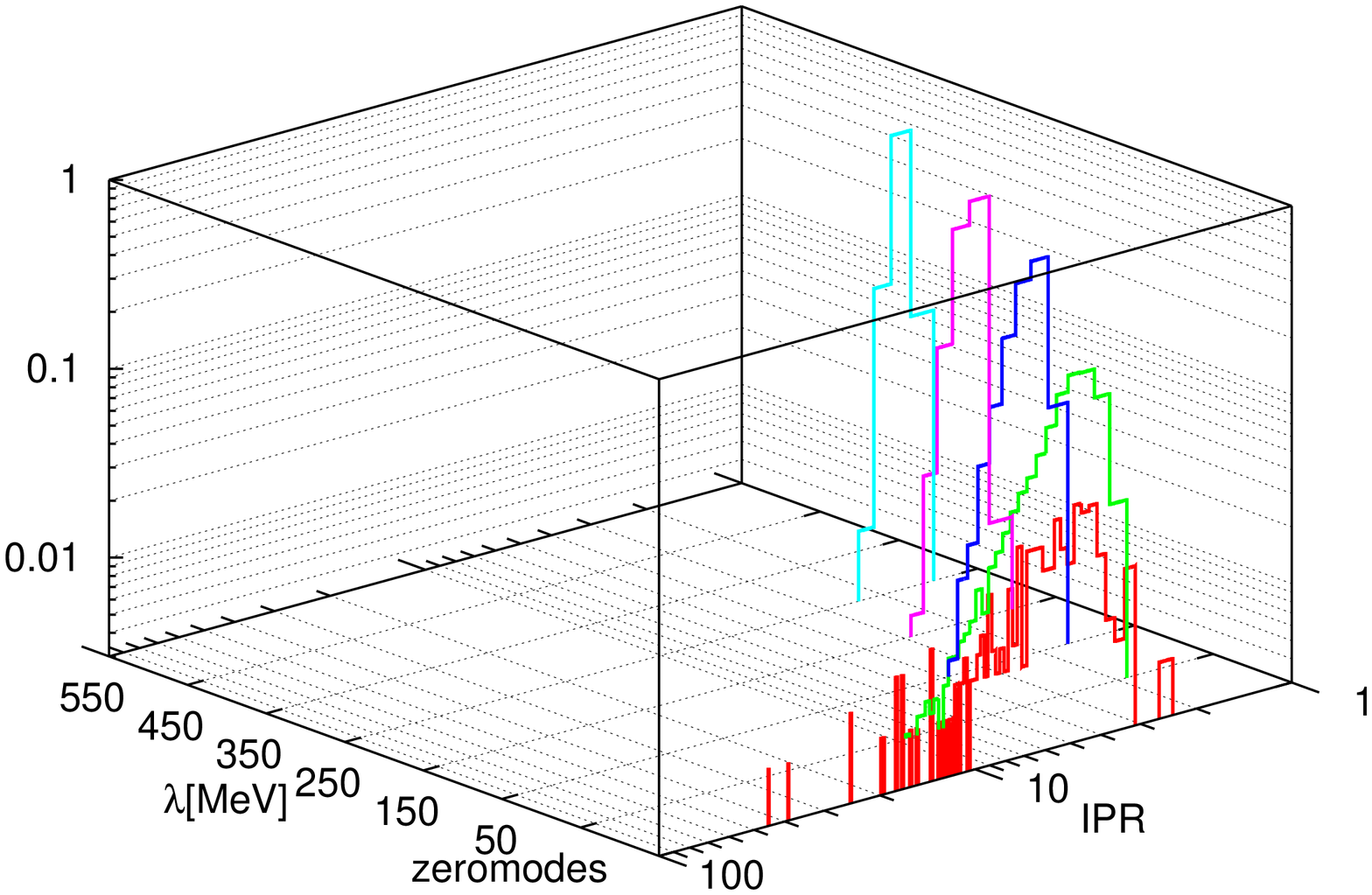,angle=270,width=8cm}\\
(b) & (c)\\
\epsfig{file=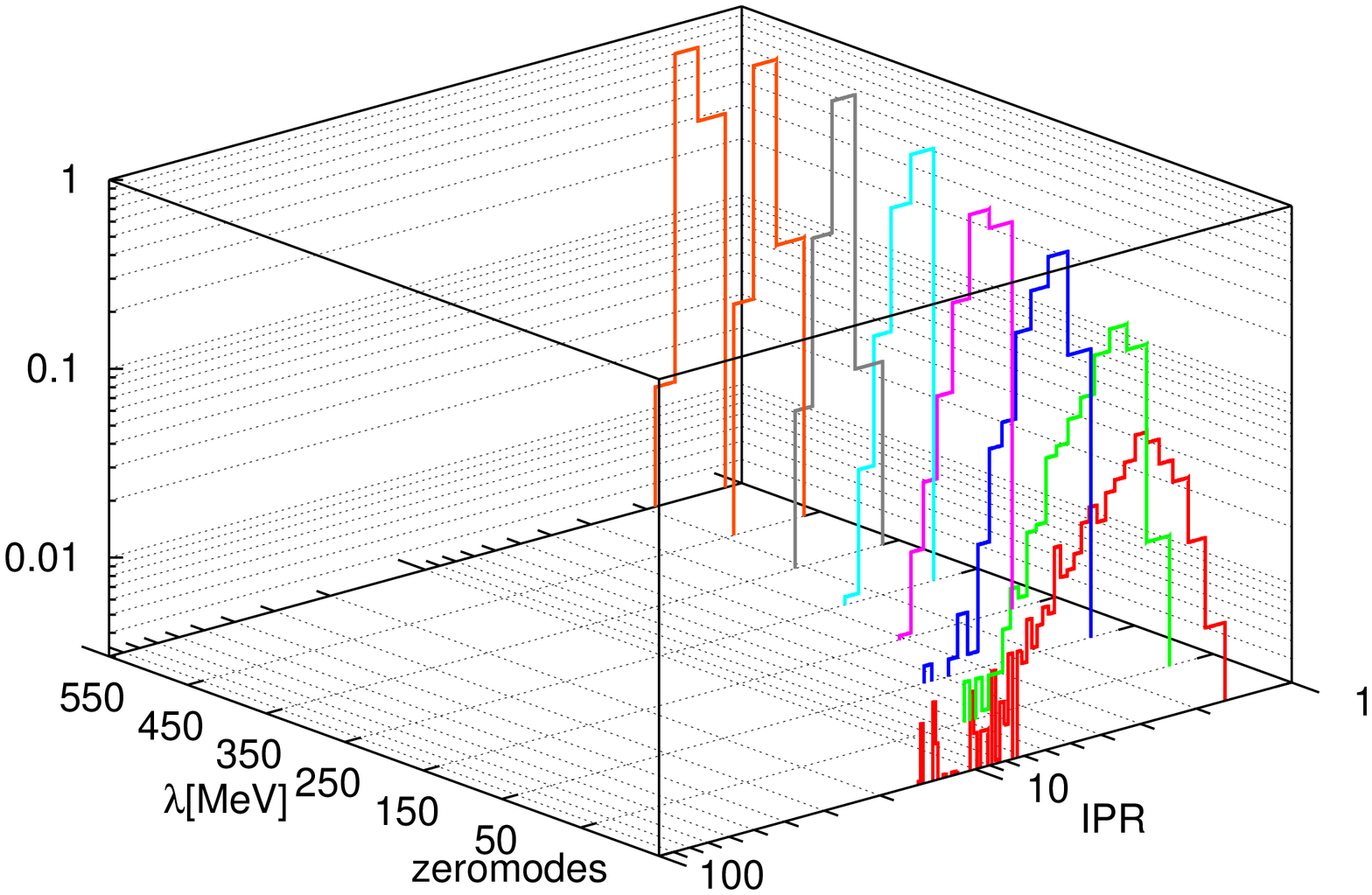,angle=270,width=8cm}&
\epsfig{file=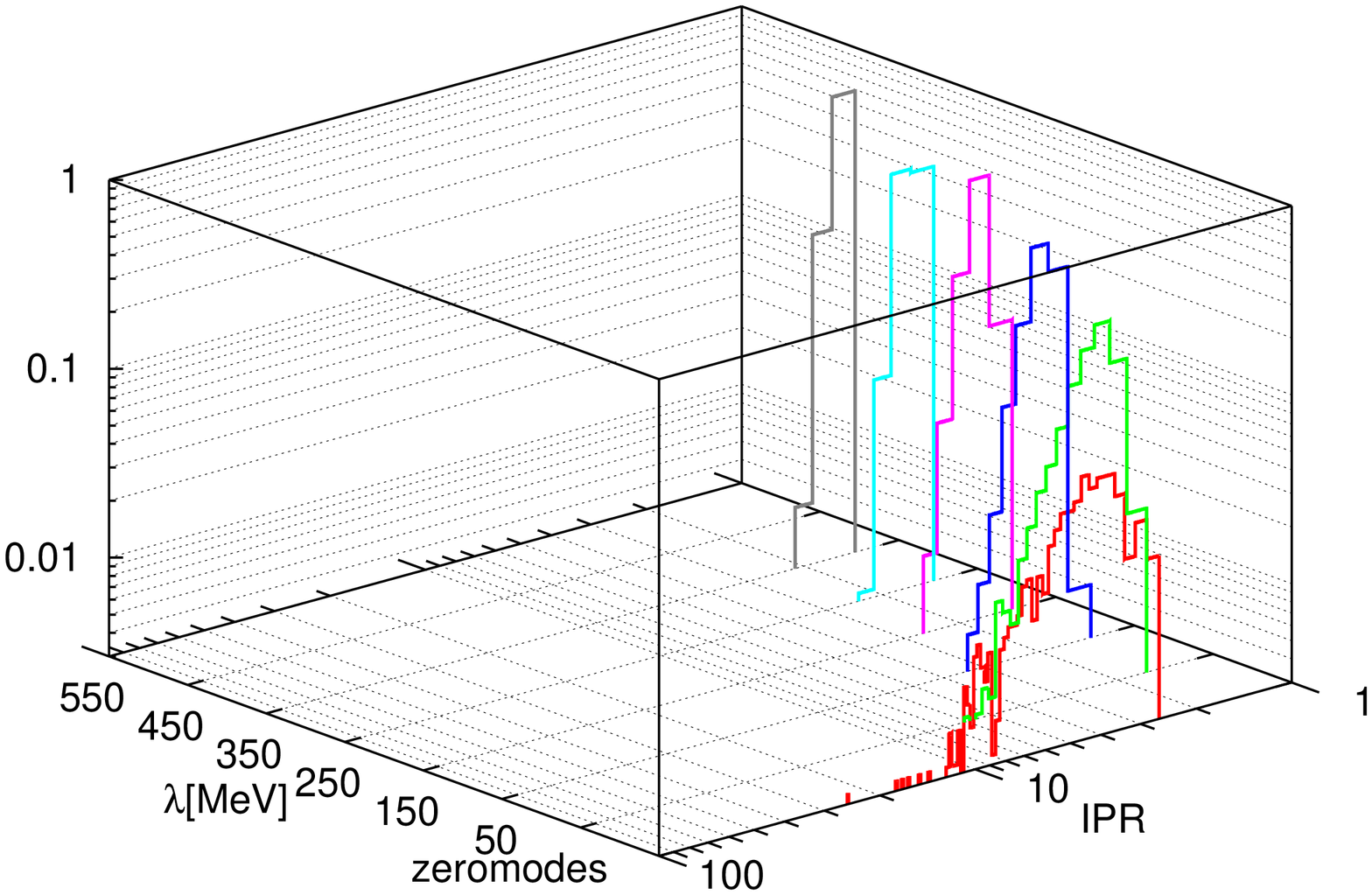,angle=270,width=8cm}\\
(d) & (e)
\end{tabular}
\end{center}
\caption{Normalized histograms of IPRs as functions of $\lambda$,
in subfigures ordered according to descending $\beta$: 
(a) $12^3\times24$ at $\beta=8.45$, 
(b) $16^3\times32$ at $\beta=8.45$, 
(c) $24^3\times48$ at $\beta=8.45$, 
(d) $12^3\times24$ at $\beta=8.10$, 
and (e) $16^3\times32$ at $\beta=8.00$, for zero modes and non-zero modes
in $\lambda$ bins with width 100 MeV.}
\label{fig:ipr-1}
\end{figure}

We study now more systematically the values of the IPR that occur in various 
parts of the spectrum. Also the zero modes are included in this discussion.
In Fig.~\ref{fig:ipr-1} 
histograms are shown with respect to 
the IPR for the complete set of 
all analyzed eigenmodes on the considered  configurations 
for the five lattice ensembles, separately in bins of $\lambda$.
One can see again that the zero and lower modes 
are remarkably more localized, 
whereas the bulk of the higher modes is delocalized. 

\begin{figure}[t]
\begin{center}
\epsfig{file=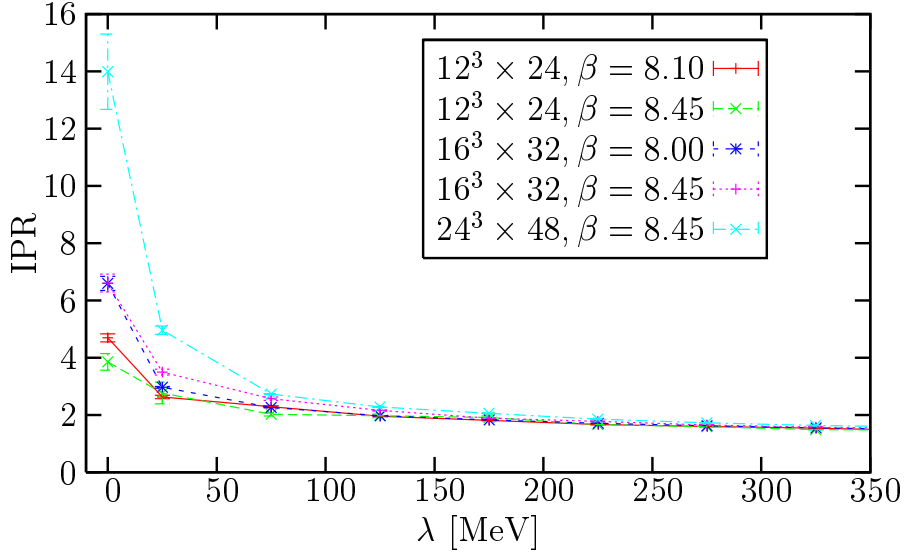,width=12cm}
\end{center}
\caption{The average IPR for zero modes and for non-zero modes in $\lambda$ 
bins of width 50 MeV for the five ensembles.} 
\label{fig:ipr-2}
\end{figure}

In Fig.~\ref{fig:ipr-2} the IPR {\it averaged over bins} with
a bin width $\Delta \lambda = 50$ MeV and for the zero modes (which are 
considered separately) is plotted. The average IPR shows a dependence on $L_s$ 
and on the lattice spacing $a$ 
for zero modes and non-zero modes only in the range $\lambda < 150$ MeV 
(for the first three bins). Beyond that interval the average IPR is practically 
independent of $L_s$ and $a$. There one finds $\langle I \rangle \leq 2.0$ .  
It would be difficult to define an exact ``mobility edge'' through the localization. 
A proper definition would require to tell the minimal number of modes  
above which the pion starts propagating.~\footnote{Although leading to a
quantitatively wrong propagator, we will see that even the zero mode 
contribution alone would allow the pion to propagate.}

\begin{figure}[t!]
\begin{center}
\epsfig{file=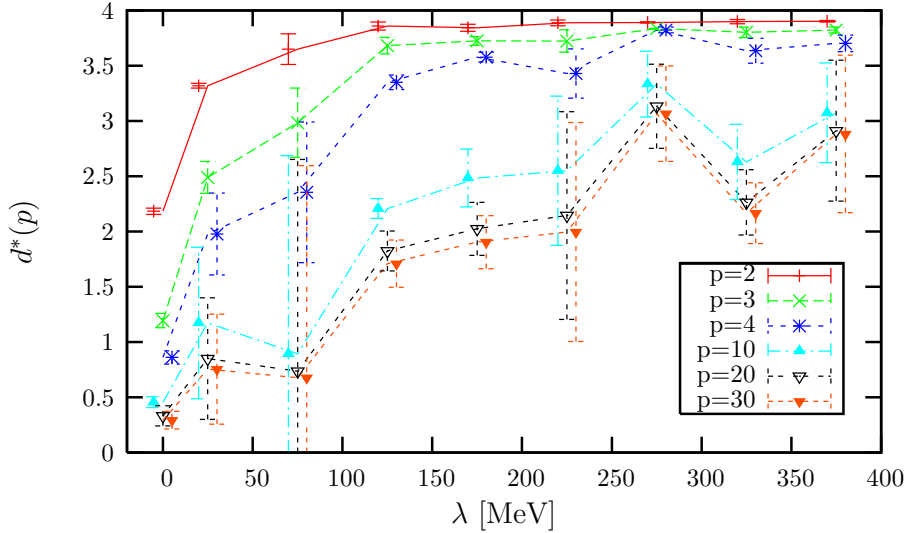,width=12cm}
\end{center}
\caption{The fractal dimension $d^{*}(p)$ obtained from fits of 
the volume dependence of the averages of the generalized $I_p$,  
presented for zero modes and for non-zero modes in $\lambda$ bins of 
width 50 MeV for the three ensembles with different volume and common 
$\beta=8.45$ .} 
\label{fig:dimension-from-general-IPR}
\end{figure}

The volume and $a$ dependence for the zero modes and the first three 
bins of non-zero modes allows some conclusions concerning the (fractal) 
dimension and the special localization properties of these modes.
In view of a generalization of the IPR that will be made 
in a next step, we denote the effective 
dimension of an eigenmode based on the {\it usual} IPR as $d^{*}(2)$   
(to remind us of the second moment of the scalar density $p(x)$).
From the volume dependence of the average usual IPR, following
$\langle I \rangle \propto (V/a^4)^{1-d^{*}(2)/4}$~\cite{Aubin:2004mp}, 
one gets this dimension as an upper effective dimension.
Similarly, the lattice spacing $a$ dependence can be described by 
$\langle I \rangle = b_0 + b_1~a^{d^{*}(2)-4}$. 
From the first bin in Fig.~\ref{fig:ipr-2} it is clearly seen that the 
IPR grows with the physical volume $V$ at fixed $a$ and with $a^{-1}$ at 
fixed physical volume $V$. 
We fit the zero modes and the lowest non-zero modes in a couple of bins 
with a width $\Delta \lambda = 50$ MeV. 
The results of the fit are presented in the uppermost curve of 
Fig.~\ref{fig:dimension-from-general-IPR}. For the zero modes an
effective dimension of $d^{*}(2) = 2.2(1)$ and for the first 
interval of non-zero modes with $\lambda \le 50$ MeV an effective dimension
$d^{*}(2) = 3.3(1)$ are found.

Gubarev {\it et al.}~\cite{Gubarev:2005az,Polikarpov:2005ey} came to somewhat 
different conclusions
guided by the $a$ dependence of the usual IPR 
$\langle I \rangle = V~I_2$ 
in pure $SU(2)$ Yang-Mills theory, simulated with the Wilson action. 
Their fits for the $a$ 
dependence of the IPR, with $d^{*}(2)$ restricted to integers, gave 
the lowest $\chi^2$ per degree of freedom for the choice $d^{*}(2) = 1$ for the 
zero modes and choosing $d^{*}(2) = 0$ for the lowest non-zero modes. 
This discrepancy is most probably to blame to the Wilson action used 
to create the quenched configurations resulting in the proliferation of
dislocations~\cite{Pugh:1989ek}.  Our results concerning the usual 
IPR are similar to the findings of Aubin {\it et al.}~\cite{Aubin:2004mp}.

While the usual IPR of the low modes strongly depends on $V$ and $a$, 
and hence the modes are localized with an effective dimension 
between $d^{*}(2) = 2$ and $3$, the IPR of the bulk of higher modes is independent 
of $V$ and $a$. That means that these modes almost freely extend throughout 
$d^{*}(2) \lesssim d=4$ dimensions. We will see, studying the percolation behavior,
that also the higher modes are not like simple plane waves, but experience a 
kind of soft localization.
At this place it might be helpful to present in the Table~II the average number 
of non-zero modes below some frequently used cut-off values $\lambda_{\rm cut}$\footnote{In the following the cut-off values 200, 400, 600 and 800 MeV on the $12^3\times 24$ lattice at $\beta=8.1$ should be considered as approximate values. Using the interpolated value for the lattice spacing at $\beta=8.1$, the exact values are 195, 389, 584 and 778 MeV, respectively.}. 

\begin{table}[h]
\vspace*{1cm}
\begin{center}
\begin{tabular}{|c|c|c|c|c|}
\hline
$\beta$ & lattice size  & $a~\lambda_{\rm cut}$ & $\lambda_{\rm cut}$ & average \# of NZM  \\
\hline
\hline
8.45  & $16^3\times 32$ & $0.1064$    & $200$ MeV           & 13.1(2)     \\
      &                 & $0.2128$    & $400$ MeV           & 36.2(2)     \\
      &                 & $0.3374$    & $634$ MeV           & 92.3(2)     \\
\hline
8.10  & $12^3\times 24$ & $0.14$      & $200$ MeV           & 11.0(2)     \\
      &                 & $0.28$      & $400$ MeV           & 28.1(2)     \\
      &                 & $0.42$      & $600$ MeV           & 59.3(2)     \\
      &                 & $0.56$      & $800$ MeV           & 117.3(4)    \\
\hline
\end{tabular}
\caption{The average number of non-zero modes (NZM) below various spectral cut-offs
for two ensembles used in this study. The cut-offs are given both in MeV and 
in $1/a$.}
\end{center}
\label{tab:number_of_modes}
\end{table}

In the theory of the metal-insulator transition tools for a quantitative 
description of critical level statistics are given and the multifractal 
properties of the eigenfunctions are in the focus of interest. 
For this purpose, the notion of IPR has been 
generalized~\cite{Kravtsov:1997,Kravtsov:1996} to
\begin{equation}
I_p(\lambda) = \frac{\sum_{x,n}
 |\psi_{\lambda_n}(x)|^{2p} \Theta_{\epsilon}(\lambda -\lambda_n)} 
                   { \sum_n \Theta_{\epsilon}(\lambda -\lambda_n)} \; ,
\end{equation}
with a free parameter $p$ and a window function 
$\Theta_{\epsilon}(\lambda -\lambda_n)=1/\epsilon$ 
for $|\lambda-\lambda_n|\le \epsilon/2$ and vanishing elsewhere.
Depending on whether the metal phase, the insulator phase or the critical region 
is met, this quantity would scale in a different way with the volume of the 
specimen (see Table~III, results see Fig.~\ref{fig:dimension-from-general-IPR}).

\begin{table}[h]
\vspace*{1cm}
\begin{center}
\begin{tabular}{|l|l|l|}
\hline
$ I_p(\lambda) \propto L^{-d (p-1) }$ & metallic  & electrons~propagate~freely  \\
\hline
$ I_p(\lambda) \propto L^{-d^{*}(p) (p-1) } $ & critical & electrons~propagate~along  \\
&  &     low-dimensional~structures  \\
\hline
$ I_p(\lambda)= {\rm const} $ & insulator  & electrons~do~not~propagate  \\
\hline
\end{tabular}
\caption{Volume scaling of generalized IPR's according 
to the metal-insulator transition analogy.}
\end{center}
\label{tab:critical}
\end{table}
For the ``metallic'' phase, $d$ is the embedding dimension of the system, in our
case $d=4$. For the critical region the effective dimension becomes 
$d^{*}(p) < d$. In this sense, all eigenstates with $\lambda < 200$ MeV 
definitely belong 
to a ``critical region''. From the standard IPR $I=L_s^3~L_t~I_2$ ($p=2$) 
we have concluded that the dimension of the zero modes is close to two. 
In the next bins of 50 MeV each, the dimension 
quickly exceeds three, before it is finally approaching four.

For the ``critical region'' and up to $\lambda = 400$ MeV we present now the
results for the higher order $p$ fractal dimensions $d^{*}(p)$, which allows to 
recognize a varying type of localization at various levels (heights) 
of the scalar density 
$p(x)$. Notice the decrease of the multifractal dimension with increasing $p$, 
resulting from the fact that $I_p$ explores regions of higher scalar density. 
For $p=4$, for example, one sees generally a reduced dimension compared 
to $p=2$, interpolating between ``filamentary'' ($d^{*}(4) \approx 1$ 
for zero modes) and ``surface-like'' ($d^{*}(4) \approx 2$ for 
the next two bins with $\lambda < 100 \mathrm{~MeV}$), 
becoming $d^{*}(4) \approx 3.5$ at higher $\lambda$.
This indicates that the regions of higher scalar density are geometrically 
distinct from those regions where an eigenmode is present only with a tail 
of the scalar density. For $p=20$ - $30$ the analyzed
fractal dimension does not grow anymore. The envelope tells that, with respect to 
the maxima of the scalar density, zero modes and non-zero modes up to 
$\lambda = 100$ MeV are all characterized by a less than one-dimensional 
localization ({\it i.e.} isolated peaks), whereas in the following spectral region 
two-dimensional or three-dimensional localization prevails for the regions of 
highest scalar density (cf. Fig.~\ref{fig:mode-visualization}).
In contrast to this, the standard IPR based on $I_2$ alone cannot 
resolve these details and would describe the modes in the spectral region 
$\lambda > 100 \mathrm{~MeV}$ as four-dimensional. 

It is tempting to conjecture a pinning-down of the low-dimensional low-lying
modes to specific confining objects
(vortices, monopoles, close-by meron pairs etc.) and a relation to the 
localization of topological charge (see Section~\ref{sec:topdensity} 
in this context).
It is important to recognize that the difference between the zero modes 
and the lowest (say, 10) non-zero modes is less pronounced than it seems
in the result of binning.
In marked contrast to this, the highest analyzed non-zero modes are really 
qualitatively different. This could already be concluded from 
Fig.~\ref{fig:mode-visualization}. 

\begin{figure}[t]
\begin{center}
\begin{tabular}{cc}
\epsfig{file=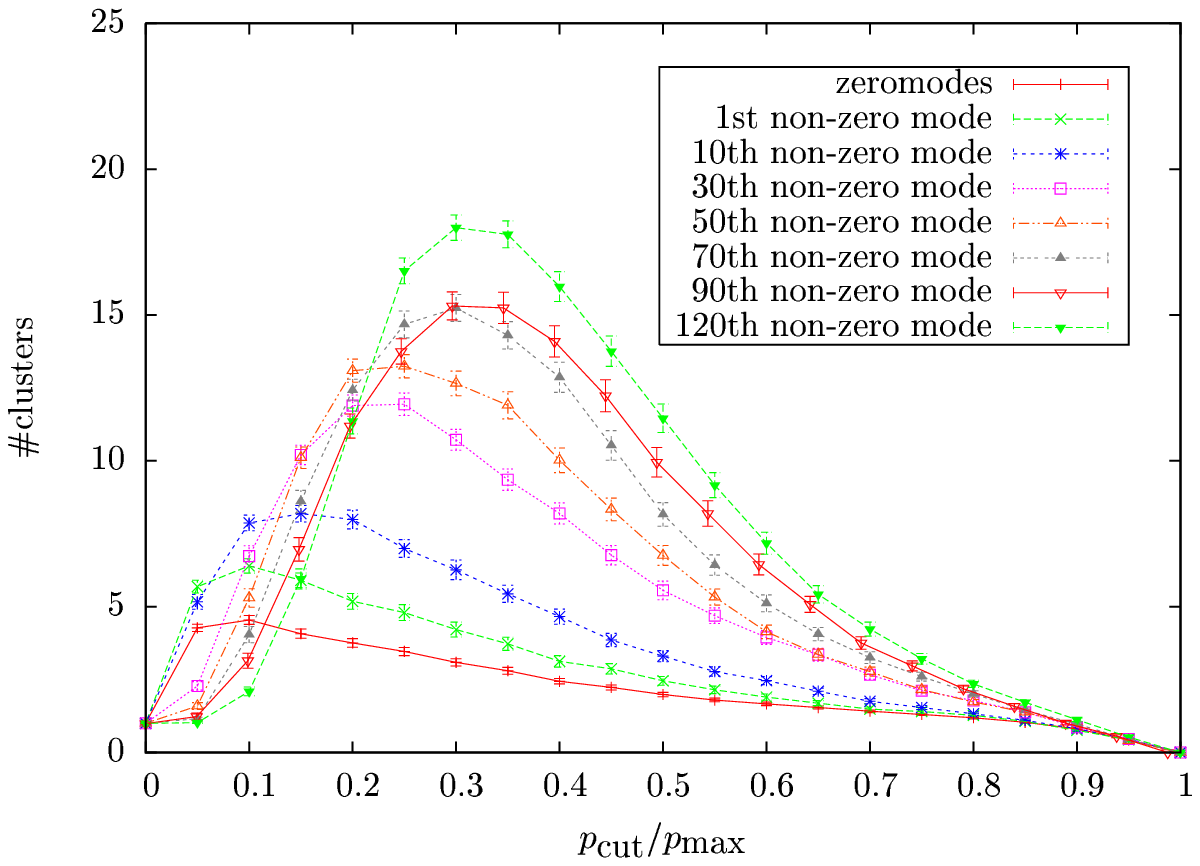,width=8cm}&
\epsfig{file=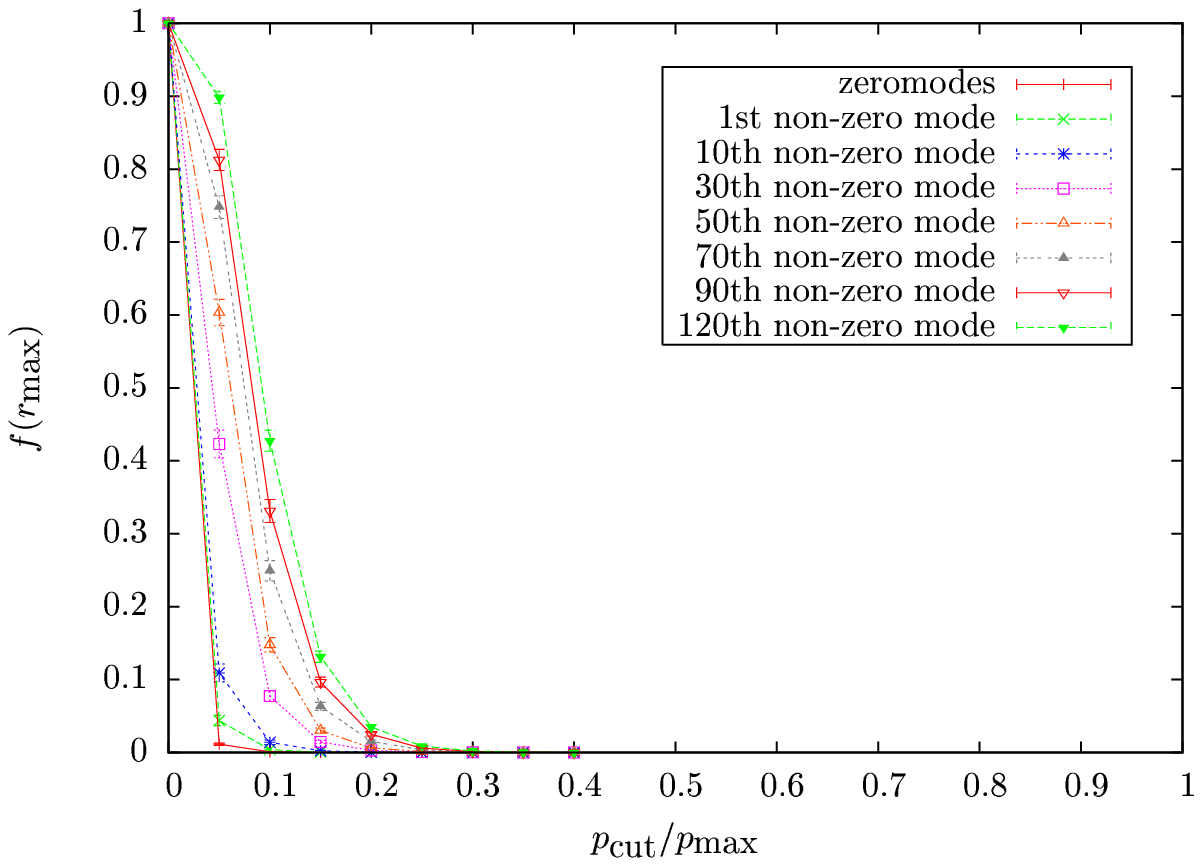,width=8cm}\\
(a) & (b)\\
\epsfig{file=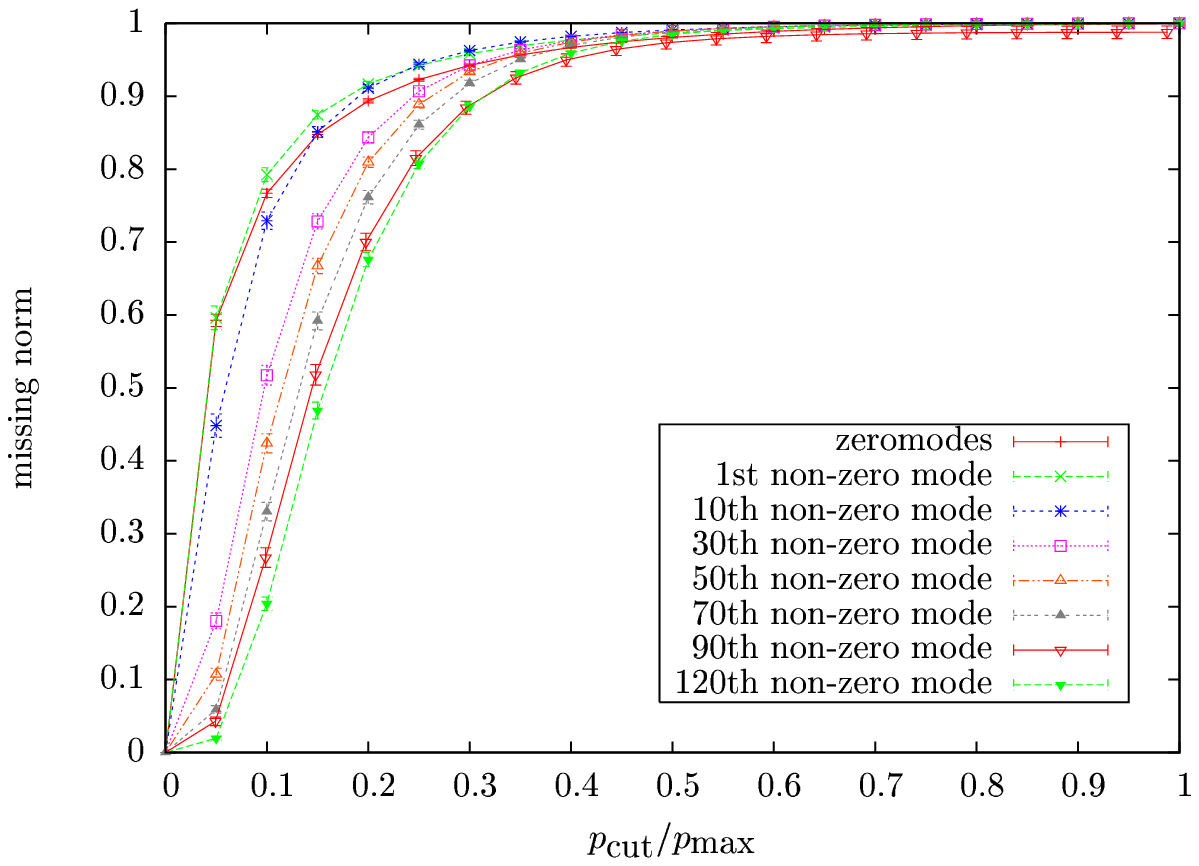,width=8cm}&
\epsfig{file=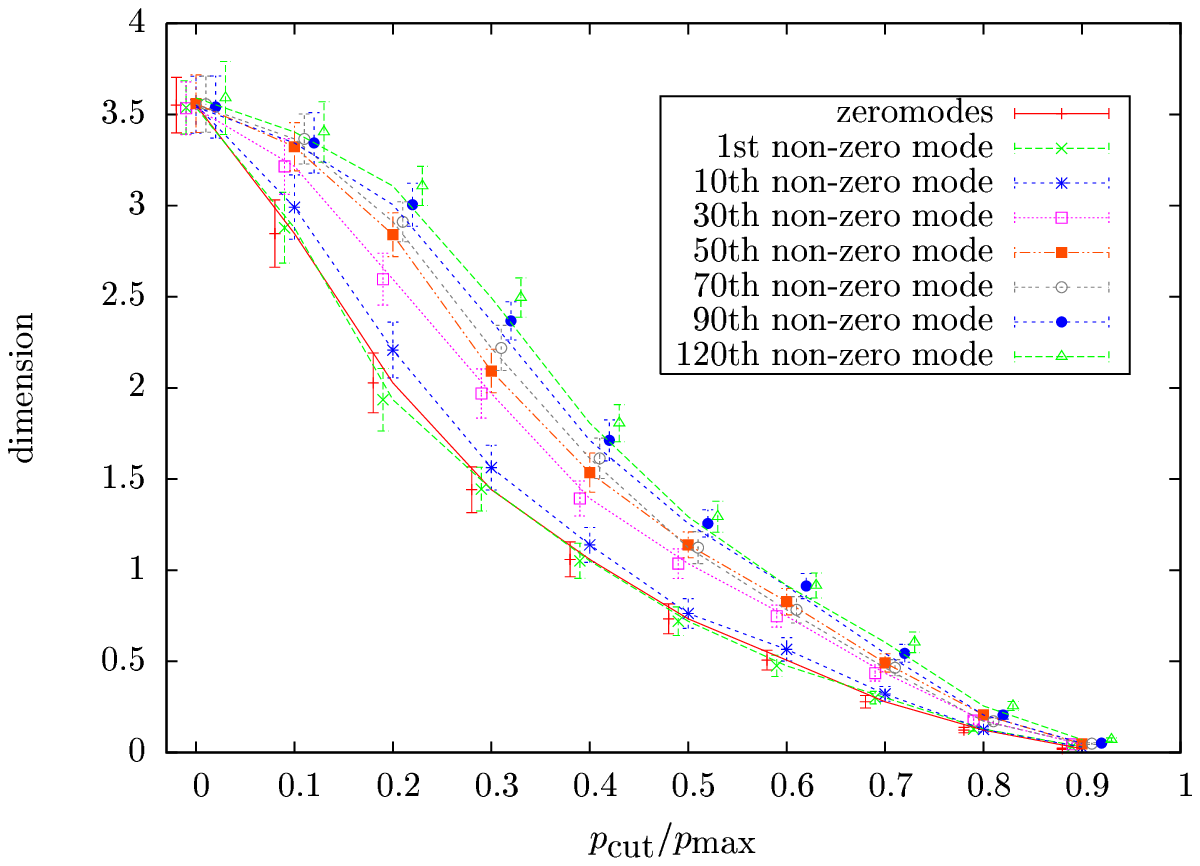,width=8cm}\\
(c) & (d)
\end{tabular}
\end{center}
\caption{Cluster analysis of some individual eigenmodes as listed in the
legend box, averaged over 170 configurations on $16^3\times32$ generated 
at $\beta=8.45$. The $p_{\rm cut}/p_{\rm max}$ dependence is shown of 
(a) the total number of separate clusters at the height $p_{\rm cut}$, 
(b) the connectivity as a signal for the percolation of the mode,
(c) the missing norm of the mode and
(d) the effective dimension of the mode at height $p_{\rm cut}$ 
determined by the return probability (\ref{eq:returnprob}) 
of random walkers.}
\label{fig:singlemodecluster}
\end{figure}

We show in Fig.~\ref{fig:singlemodecluster} (a) the result of a cluster 
analysis~\footnote{The cluster analysis with respect to some observable 
was introduced in Section~\ref{sec:2.5}.} of 170 
configurations on the $16^3\times32$ lattice generated at $\beta=8.45$. 
The average is over the zero modes in this subensemble, the 1st, 
the 10th, the 30th etc. up to the (arbitrarily chosen) 
120th non-zero mode.
We analyze the average cluster composition as function of the cut-off scalar 
density $p_{\rm cut}/p_{\rm max}$ (relative to the maximum of the scalar density 
of the respective mode $p_{\rm max} = \max_x p_{\lambda}(x)$). 
In this type of cluster analysis one attempts to decompose the set of lattice 
points with $p_{\lambda}(x) \geq p_{\rm cut}$ into one or more connected clusters, 
each enclosing
secondary maxima of $p_{\lambda}(x)$. For the highest analyzed mode, beginning at 
$p_{\rm cut} = 0.9~p_{\rm max}$, rapidly further maxima become visible, and the 
number of separate clusters grows to $\approx 20$ at
$p_{\rm cut} = 0.35~p_{\rm max}$. In contrast to this, the typical zero mode  
only slowly 
develops up to $\approx 5$ maxima at $p_{\rm cut} = 0.1~p_{\rm max}$. 
From the zero mode to the highest analyzed mode there seem to exist several
centers that ``attract'' the mode, but the mechanism might be different. 
The lowest modes might be pinned down to some lumps of topological charge
({\it \'a la} Diakonov and Petrov~\cite{Diakonov:1985eg},
whereas the localization of the highest 
(analyzed) modes might be the result of Anderson-like localization in the 
random gauge field background.

The lowest ten non-zero modes are very similar to the typical zero mode. 
The zero modes percolate, {\it i.e.} extend 
over the full lattice, only at a height 
below $p_{\rm cut} = 0.25~p_{\rm max}$. 

The highest analyzed non-zero mode percolates at $p_{\rm cut} = 0.4~p_{\rm max}$ 
(with the maximal scalar density being smaller, of course). This can be seen 
in Fig.~\ref{fig:singlemodecluster} (b). 
The difference in the percolation behavior can be concluded also from the missing 
norm~\cite{Greensite:2005yu}
\begin{equation}
R(p_{\rm cut}) = 1 - \sum_{\{x | p(x) \geq p_{\rm cut}\}} p(x)
\label{eq:missing}
\end{equation}
as function of $p_{\rm cut}/p_{\rm max}$
that is shown in Fig.~\ref{fig:singlemodecluster} (c). 
Here the similarity between the zero modes and the first 10 non-zero modes
is similarly clear.

At each height level of $p_{\lambda}$ the effective shape of the leading cluster of 
the respective mode can be explored by the random walk method described in
section~\ref{sec:2.6}. 
In the present context one studies how the return probability of the
walkers to the maximum of $p_{\lambda}(x)$ decreases with an increasing number
$\tau$ of steps. 
A fit of the power law decrease (\ref{eq:returnprob}) provides the effective dimensions $d^{*}$ shown 
in Fig.~\ref{fig:singlemodecluster} (d). 
We notice that the average dimension averaged over the 10 lowest modes 
continuously
rises from $d^{*}=0$ to $d^{*}=3.5$. It becomes three-dimensional at the same 
$p_{\rm cut}/p_{\rm max}$ when the missing norm begins the final steep 
drop from 80 \% to 0 \%.    
As for the 120th non-zero mode, the effective dimension $d^{*}$ rises more
steeply and reaches $d^{*} \approx 3$ at higher $p_{\rm cut}/p_{\rm max}$ 
when the missing norm is already dropped to 60 \% .
For a survey of dimensionalities of overlap eigenmodes see Tables VII and VIII
in the Appendix.

\subsection{Chiral properties of the non-zero modes}
\label{sec:3.5}

In this section we characterize the different parts of the low-lying
spectrum of non-zero modes by the distribution of sites over the {\it local} 
chirality. 
Later on, the local chirality of the lowest modes is shown to be locally 
correlated to the topological density.

In a given lattice point $x$, one might be interested
in the intensity ratio 
\begin{equation}
r_{\lambda}(x) = \frac{p_{\lambda +}(x)}{p_{\lambda -}(x)}
\label{eq:r_for_X}
\end{equation}
of the two chiral projections for a given mode $\psi_{\lambda}$.
For the chiral zero modes this ratio is $\infty$ or $0$ everywhere. For the
non-chiral, non-zero modes it is some function of $x$. It is useful to relate 
$r_{\lambda}(x)$ to~\cite{Edwards:2001nd} 
\begin{equation}
X_{\lambda}(x) = \frac{4}{\pi} \arctan \left(\sqrt{r_{\lambda}(x)}\right) - 1 \in [-1,+1] \; .
\label{eq:arctan}
\end{equation}
It will be seen that relatively low-lying modes have regions where the local 
chirality carries information about the (anti-)selfdual character of the 
background field, to be precise, where $0.5 < |X(x)| < 1$.

\begin{figure}[t]
\begin{center}
\begin{tabular}{cc}
\epsfig{file=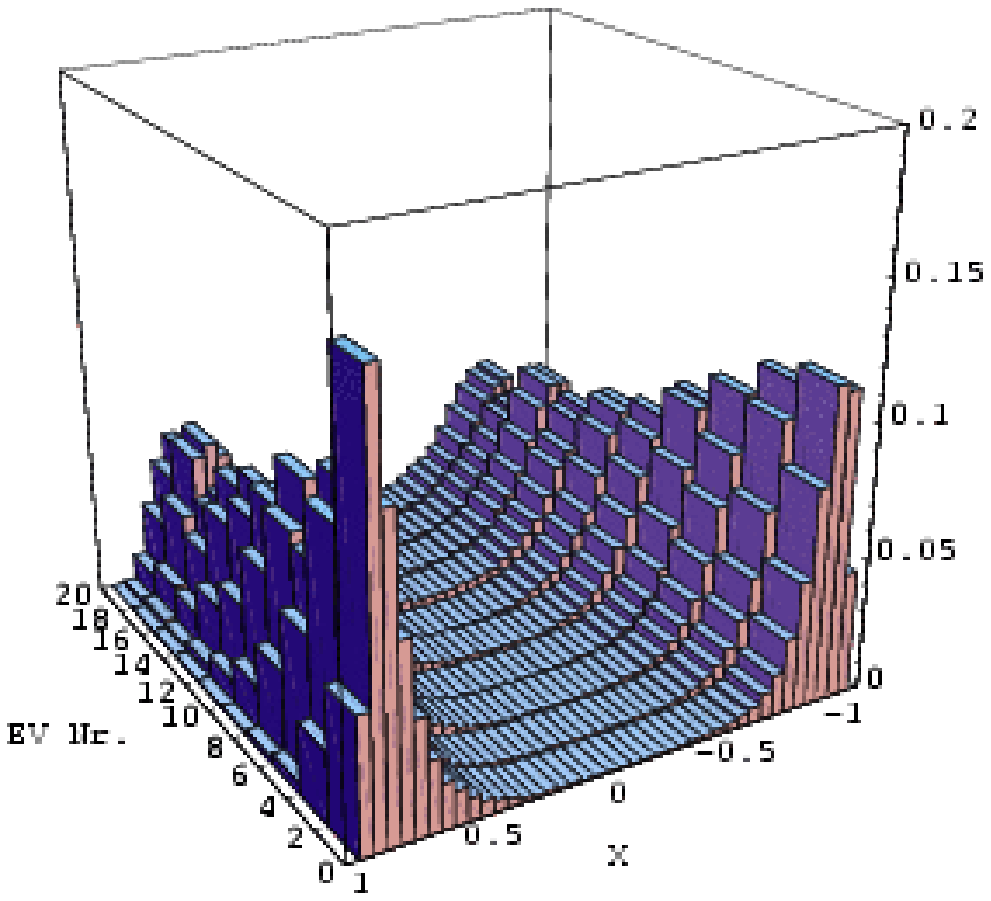,angle=0,width=7cm}&
\epsfig{file=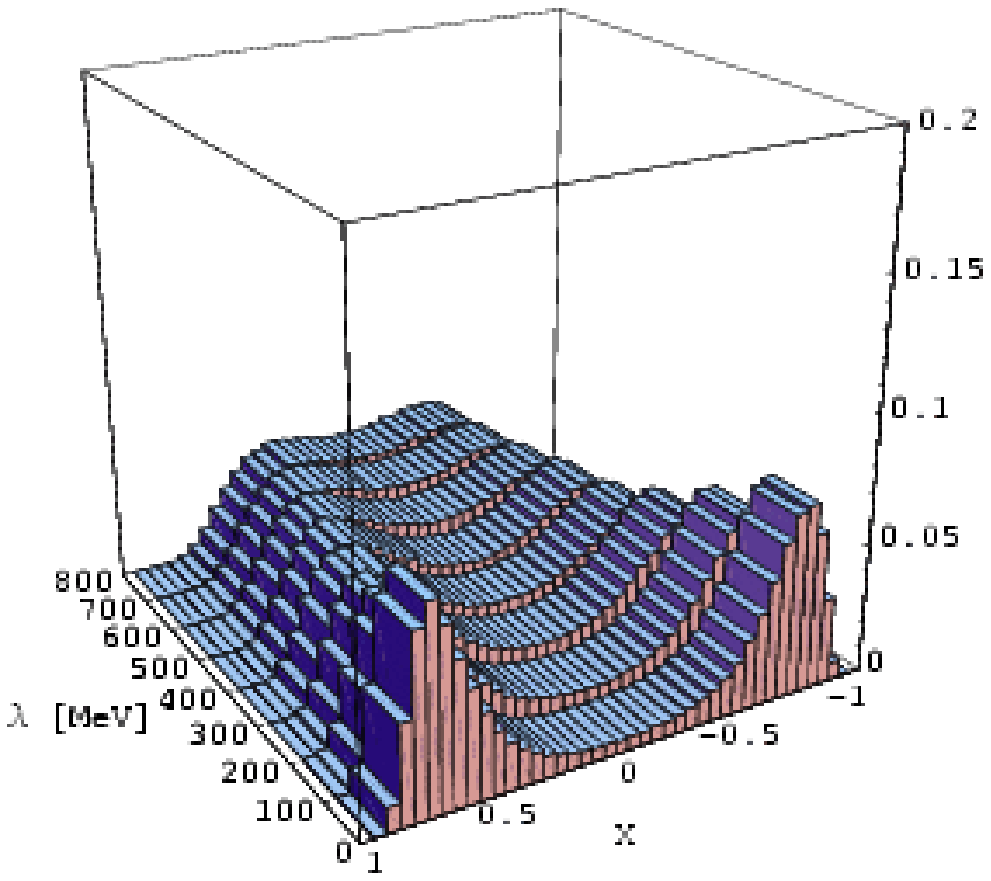,angle=0,width=7cm}\\
(a) & (b)\\
\epsfig{file=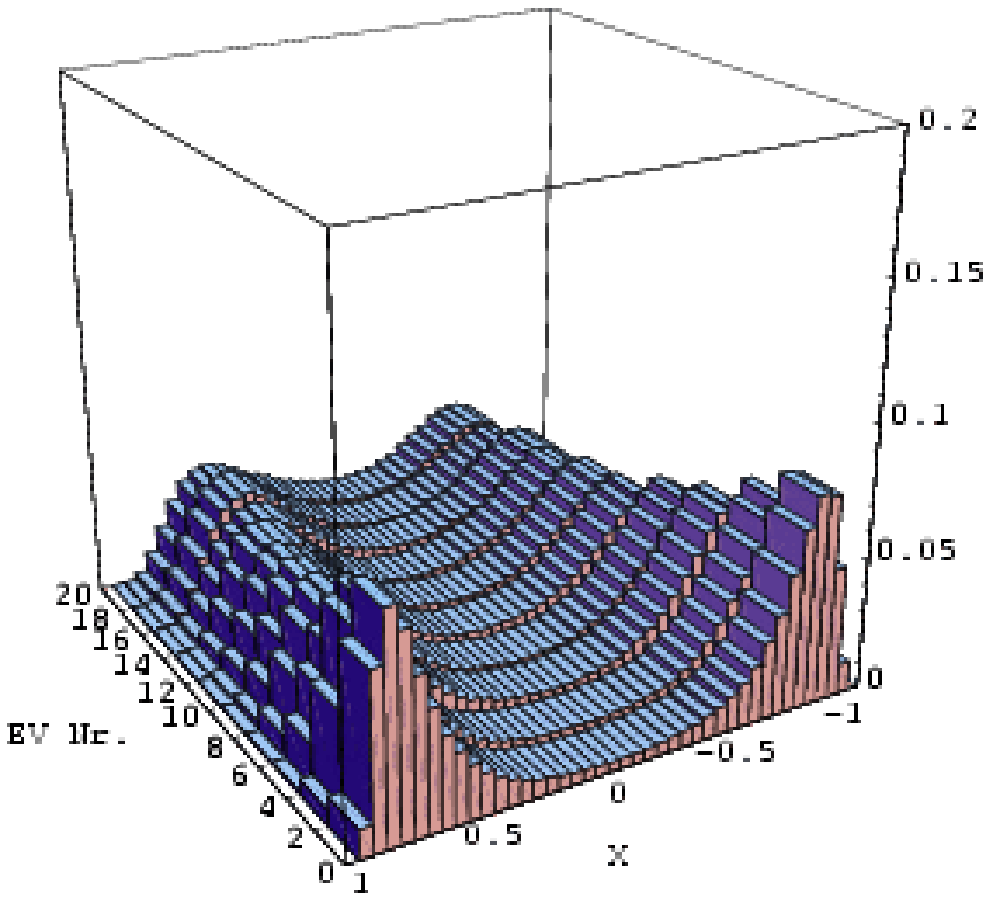,angle=0,width=7cm}&
\epsfig{file=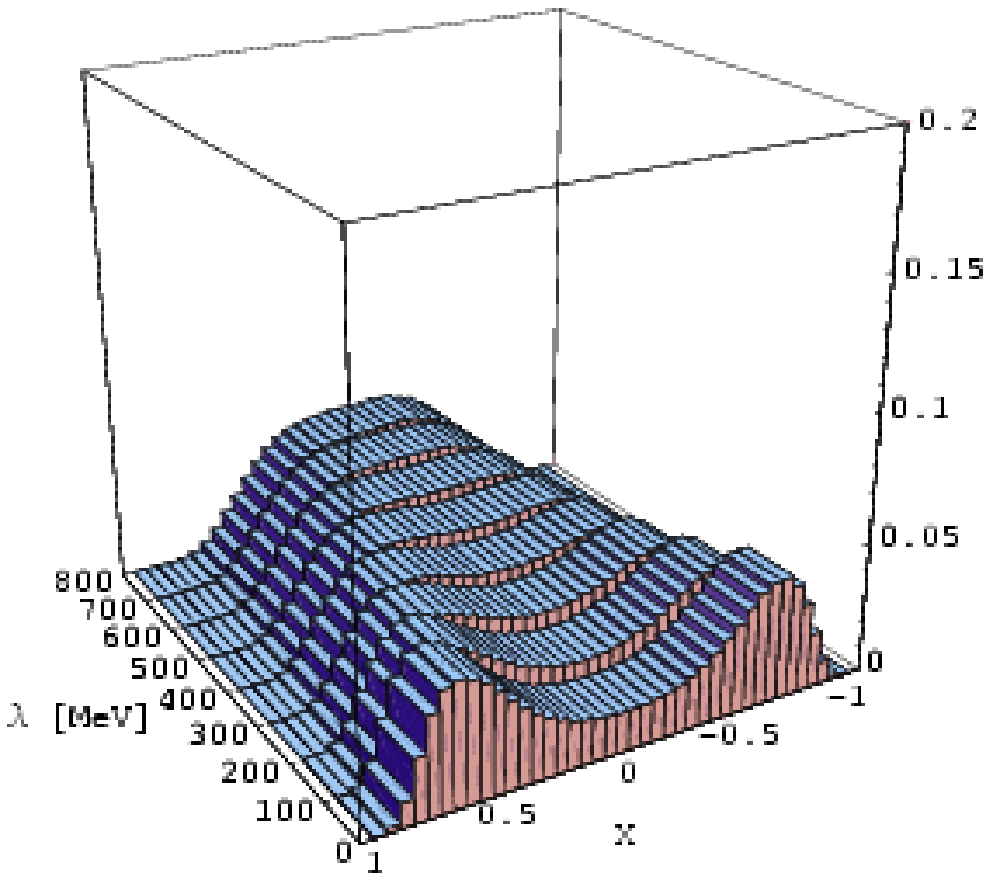,angle=0,width=7cm}\\
(c) & (d)
\end{tabular}
\end{center}
\caption{Normalized histograms of the local chirality in the $Q=0$ subsample 
(consisting of 37 configurations) of the $16^3\times32$ lattices generated 
at $\beta=8.45$. 
Left: for the lowest ten pairs (with positive and negative $\lambda$)  
averaged over the subsample.
Right: for all modes from the subsample, averaged over $\lambda$ 
bins of width $100$ MeV.
Upper row: with a cut for 1 \% of the sites with biggest scalar density $p(x)$,
lower row: with a cut for 6.25 \%  of the sites with biggest $p(x)$.}
\label{fig:all-mode-loc-chirality-high-dens}
\end{figure}
\begin{figure}[t]
\begin{center}
\begin{tabular}{cc}
\epsfig{file=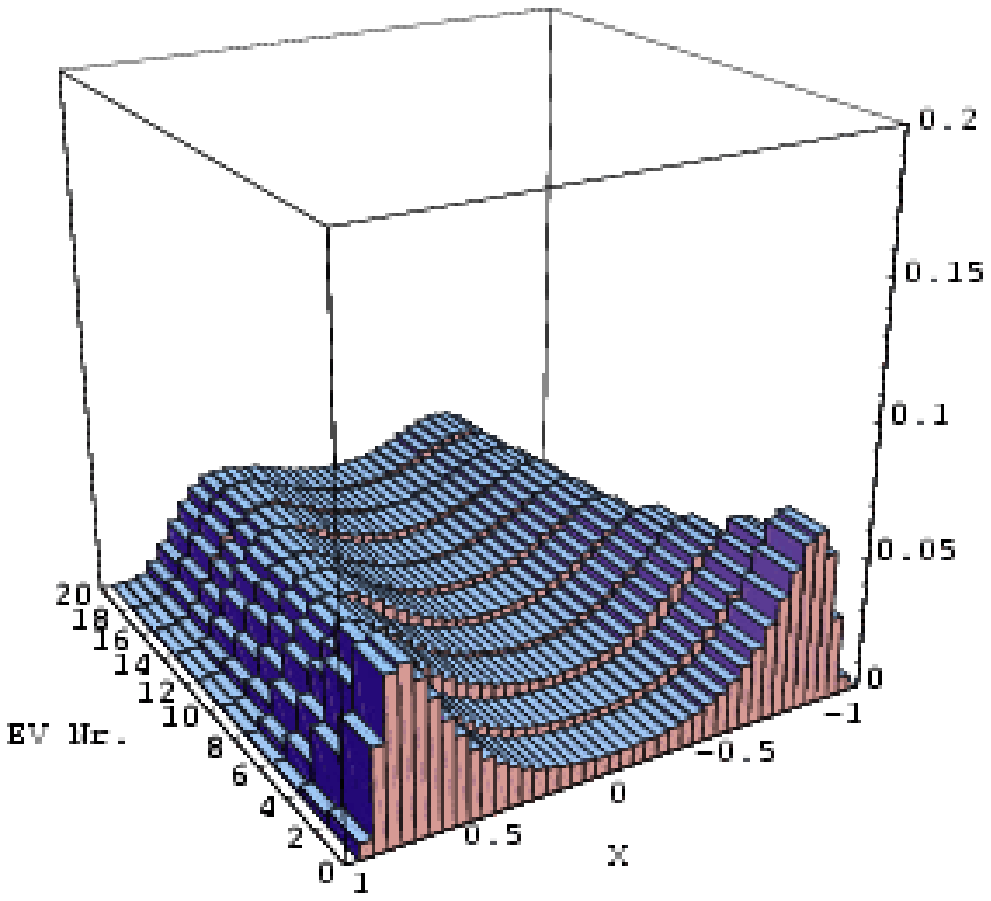,angle=0,width=7cm}&
\epsfig{file=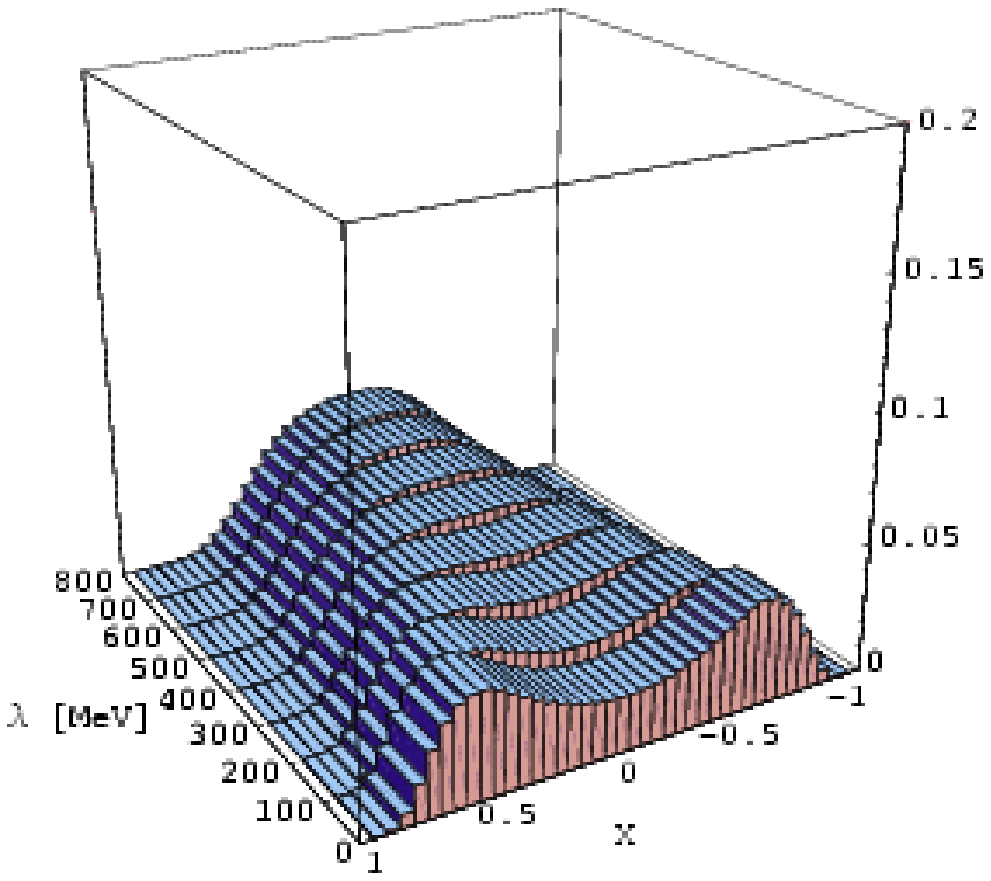,angle=0,width=7cm}\\
(a) & (b)\\
\epsfig{file=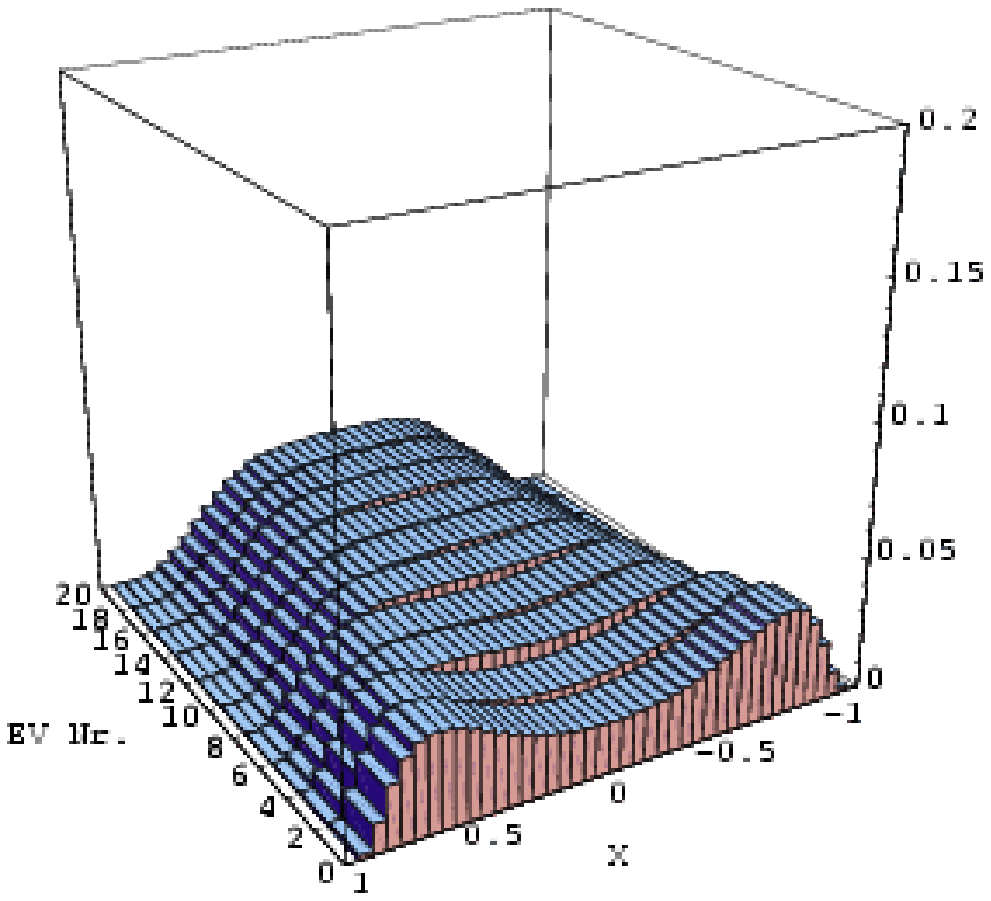,angle=0,width=7cm}&
\epsfig{file=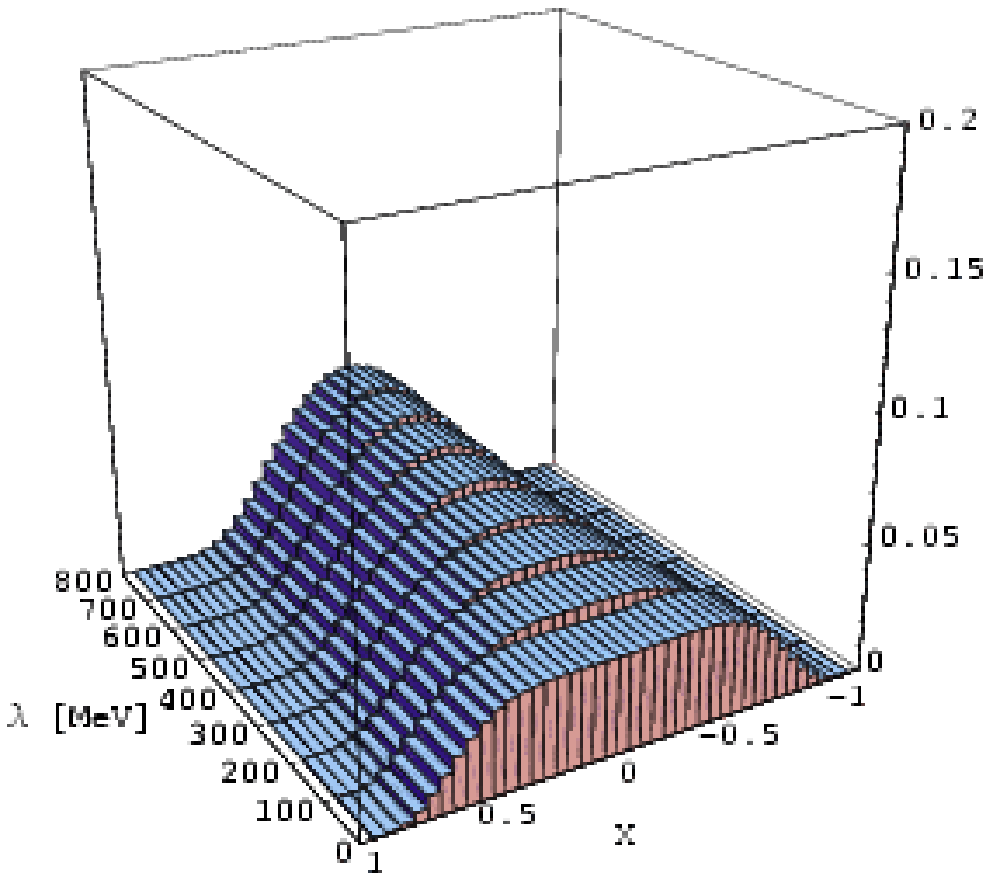,angle=0,width=7cm}\\
(c) & (d)
\end{tabular}
\end{center}
\caption{The same as Fig.~\ref{fig:all-mode-loc-chirality-high-dens}
but with less exclusive cuts with respect to the scalar density: 
Upper row: with a cut for 12.5 /
lower row: with a cut for 50 \% of the sites with biggest $p(x)$.}
\vspace*{3cm}
\label{fig:all-mode-loc-chirality-low-dens}
\end{figure}

In Fig.~\ref{fig:all-mode-loc-chirality-high-dens} 
and Fig.~\ref{fig:all-mode-loc-chirality-low-dens}
normalized histograms with respect to the local chirality $X(x)$ are shown
for the $Q=0$ subsample (containing 37 configurations) of the $16^3\times32$ 
lattices generated at $\beta=8.45$
with the lattice sites selected according to the scalar density $p(x)$.
The left figures show the 10 different $X$-histograms for the lowest 20 
individual non-zero modes (the pair with $\lambda$ and $-\lambda$ gives 
rise to the same histogram), whereas the right figures show the histograms 
where all non-zero modes belonging to a $\lambda$ bin (with a bin size 
100 MeV) are contributing to 8 histograms covering the spectral range 
up to 800 MeV. The different rows show different cuts applied
restricting the lattice sites $x$ to those with $p_{\lambda}(x) \geq p_{\rm cut}$.
The cuts are chosen such to represent 1 \%, 6.25 \%, 12.5 \% and 50 \% 
of all lattice sites.

For zero modes (not included here) the histograms would be $\delta$-functions
at $X=-1$ or $X=1$ independent of any cut with respect to the scalar density.  
Although being non-chiral when integrated over space-time, the first
non-zero modes locally may have a considerable degree of chirality 
(peaking in $X$ somewhere between 0.75 and 0.9) for the leading 1~\% or 6~\%
of the lattice sites with the highest scalar density. It turns out that with a higher, 
more restrictive cut the degree of local chirality can be enhanced. 
A similarly stronger effect can be seen towards the lower modes. That 
means that the 5 - 10 maxima of the scalar density of the 
lowest non-zero modes are 
localized in space-time region where the local chirality is strongly correlated 
to the topological charge density of the gluonic background. 
The highest modes, as we have found, have $O(20)$ local maxima. 
In contrast to the lower modes, the histogram 
of local chirality for the higher ones is dominated by a broad bump of low 
chirality $X(x)$ distributed around zero for all cuts in $p(x)$. 
This means that the maxima (localization) of the highest modes are not the 
result of pinning down at spots of topological charge density, rather they
are the result of waves scattering on a random gauge field background.

One can see that the eigenmodes with eigenvalues up to 200 MeV show an enhancement
of local chirality around $X=\pm~(0.5 ... 0.75)$ as long as one focusses on 
the 6 \% of the lattice points with the strongest scalar density.
If 50 \% of the lattice sites are included in the analysis, 
only the first two pairs of
non-zero eigenmodes still show some enhancement of local chirality around 
$X=\pm 0.75$.

We will later (at the end of Section~\ref{sec:4.4}) consider the correlation 
function between the topological charge in the gluonic background and the 
local chirality $X(x)$ of individual non-zero modes. The result establishes 
that they are interrelated over some distance in space-time.

\section{Topological charge density from the overlap operator}
\label{sec:topdensity}

\subsection{Two-point function of the all-scale and the mode-truncated density}
\label{sec:4.1}

The topological charge density of the gluonic field seems to
play a deciding role for
the localization of the lowest modes. In this Section we are studying the
properties of the overlap definition of the topological density according to
Eqs. (\ref{eq:eq-q}) and (\ref{eq:qlambdacut}).
The striking difference between the all-scale topological charge density 
and the mode-truncated one can be illuminated studying the point-to-point
correlator of the topological density. Formally, the definition of the
two-point function is the same for $q(x)$ and $q_{\lambda_{\rm cut}}(x)$:
\begin{equation}
C_{qq}(r) =  \frac{ \sum_{x,y} \langle~q(x)~q(y)~\rangle~\delta(r-|x-y|)}
               { \sum_{x,y} \delta(r-|x-y|) } \leq 0 \; 
	       \qquad \mathrm{for} \qquad r > 0 \; . 
\end{equation}
Reflection positivity, {\it i.~e.} the pseudoscalar nature of $q(x)$ together 
with positivity of the metric in Hilbert space, demands that the topological 
charge density correlator is negative for all distances, 
$C_{qq}(r) \leq 0 \; \; \rm{for} \; \;  r > 0 $. 
Of course, the vacuum expectation value $\langle~q(x)~q(y)~\rangle$ must be 
a function only of the distance 
$r = |x-y| \geq 0$, due to Euclidean rotational invariance.

While the topological charge correlator $C_{qq}(r)$ in the continuum should be 
negative for nonvanishing distance, the topological susceptibility, which is 
the integral over the correlator $\chi_t=\int dx~\langle~q(0)~q(x)~\rangle$,  
must be positive. To solve the 
apparent contradiction, formally divergent contact counterterms of the form  
$C_{qq}(x) \to C_{qq}(x) + c_1\delta(x)+c_2\Delta\delta(x)+c_3\Delta^2\delta(x)$ have to be introduced in 
the continuum theory \cite{Seiler:2001je}.
\begin{figure}[t]
\begin{center}
\begin{tabular}{cc}
\epsfig{file=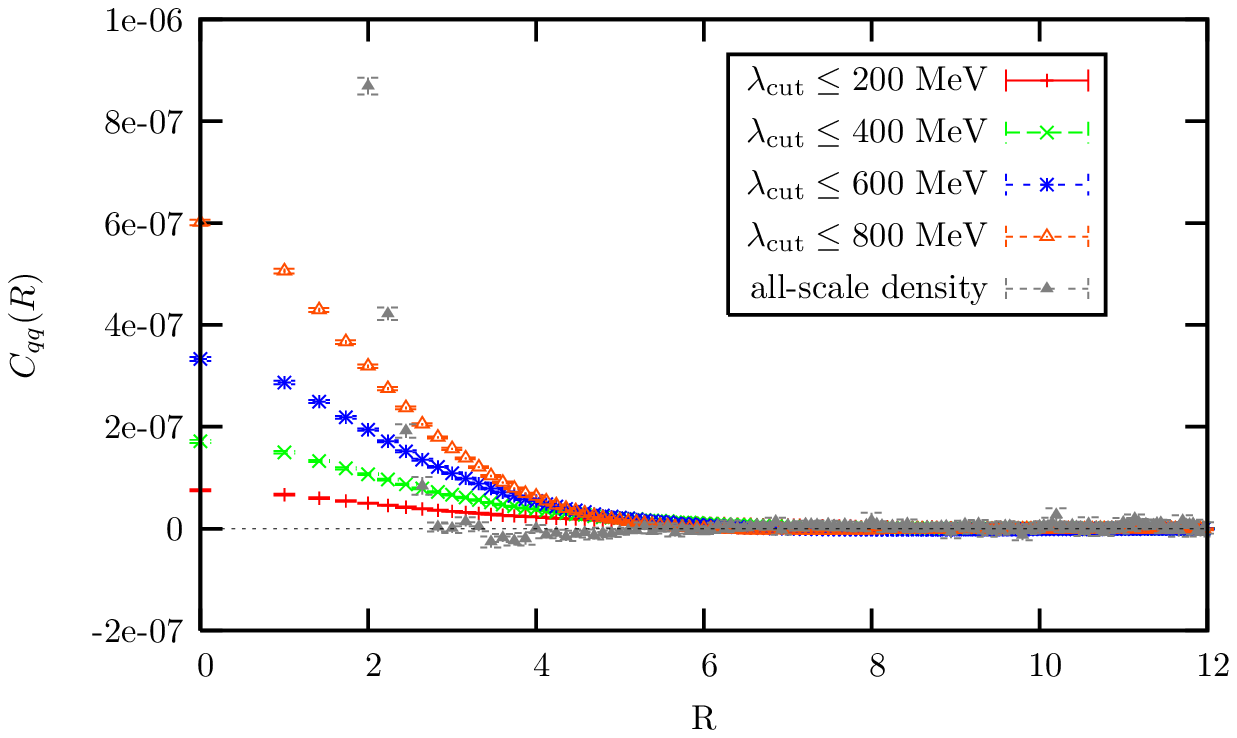,width=8cm}&
\epsfig{file=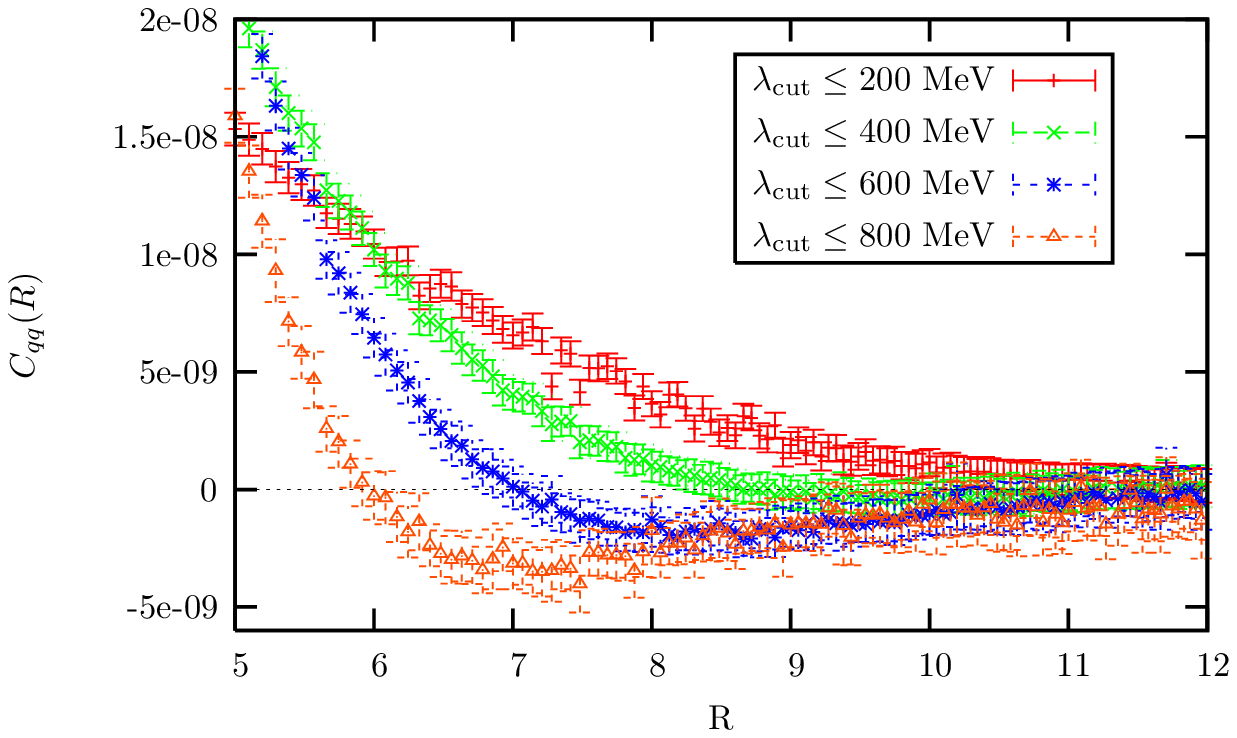,width=8cm} \\
(a) & (b)\\
\end{tabular}
\end{center}
\caption{The topological charge density correlator for the $12^3\times24$ 
lattice configurations at $\beta=8.10$.
(a) The correlator of the mode-truncated 
density $q_{\lambda_{\rm cut}}(x)$ for various $\lambda_{\rm cut}$ 
compared with the correlator of the all-scale density $q(x)$, measured 
only on 53 configurations.
The right figure (b) magnifies the region in $r$ where the mode-truncated 
correlators become negative for sufficiently large $\lambda_{\rm cut}$.}
\label{fig:qq-corr-trunc}
\end{figure}

Since overlap fermions are not ultralocal, the fermion action cannot be
strictly reflection positive. Because the overlap operator $D(0)$ has a
finite range in space-time (see Fig.~\ref{fig:locality}) practically 
independent of $\beta$, we expect that the correlator $C_{qq}(r)$ will 
have a positive core with a fixed width in lattice units. The negative 
tail beyond the core radius corresponds to the above mentioned theoretical 
requirement. This has been discussed recently in Ref.~\cite{Horvath:2005cv}.

For the correlator of the mode-truncated density $q_{\lambda_{\rm cut}}(x)$
one would expect that this structure is smoothed out and that below some 
$\lambda_{\rm cut}$ the corresponding two-point function will be positive 
at all distances.
In Fig.~\ref{fig:qq-corr-trunc} we show the correlator of the 
mode-truncated topological charge density for various $\lambda_{\rm cut}$ for the 
$12^3\times24$ lattice configurations generated at $\beta=8.10$. 
In Fig.~\ref{fig:qq-corr-trunc} (a) they are compared with the correlation
function of the all-scale topological density measured on 53 configurations. 
In Fig.~\ref{fig:qq-corr-trunc} (b)  
we show the correlator only for different mode-truncated densities, however 
magnified, in the region of $r$ where they start to develop a negative tail 
for sufficiently high cut-off. 
This happens at $R \ge 6~a$ for $\lambda_{\rm cut} = 800$ MeV,
at $R \ge 7~a$ for $\lambda_{\rm cut} = 600$ MeV, 
and at $R \ge 9~a$ for $\lambda_{\rm cut} = 400$ MeV. 
As long as it were positive, this correlator would be interpreted in
terms of the number density and the convoluted profile of effective 
topological charge
clusters. It is interesting, that higher resolution in the present case
(similarly to doing less smearing~\cite{Hasenfratz:1999ng} or doing restricted 
cooling~\cite{GarciaPerez:1998ru,EMIunpublished}) 
results in effectively attractive correlations between opposite sign 
topological charge clusters not only in full QCD but also in quenched 
gauge fields as emphasized in Ref.~\cite{Hasenfratz:1999ng}.
Since the total topological charge does not depend on $\lambda_{\rm cut}$,
the growing negative tail must be compensated in the susceptibility 
$\sum_x C_{qq}(|x|)$ by a growing positive core with increasing $\lambda_{\rm cut}$.

\begin{figure}[t]
\begin{center}
\begin{tabular}{cc}
\vspace*{-0.5cm}\epsfig{file=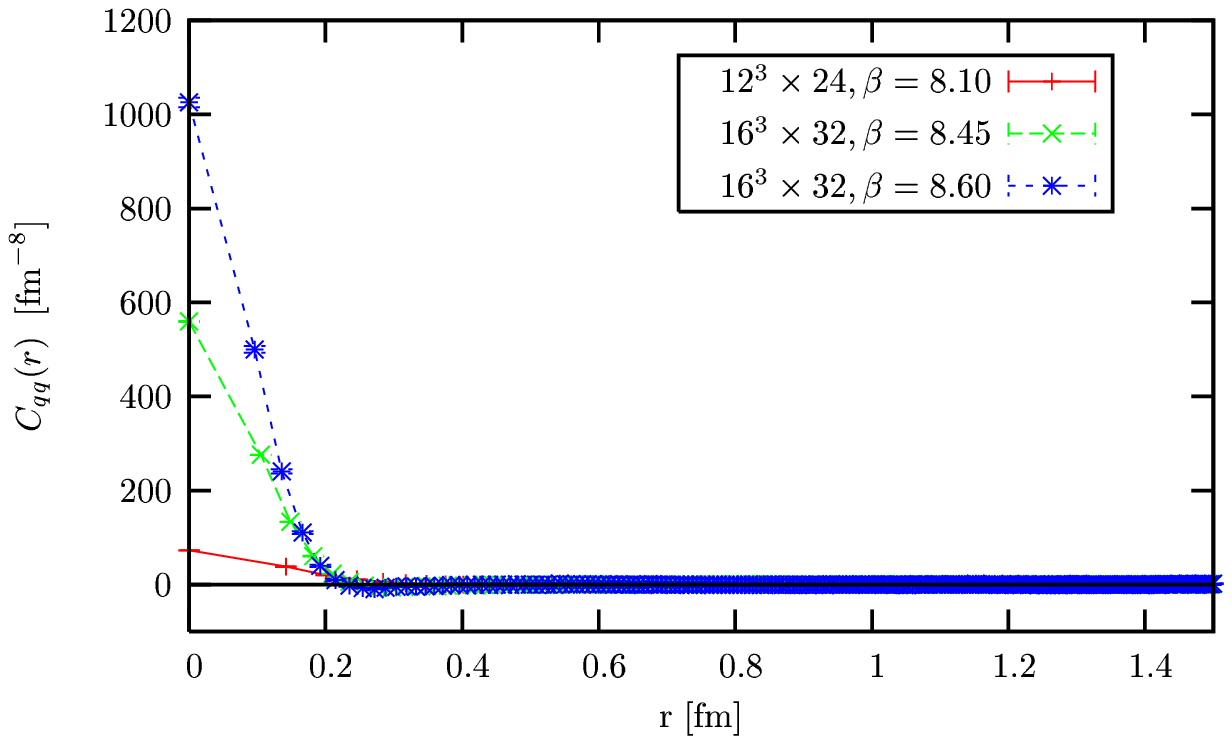,width=8cm}&
\epsfig{file=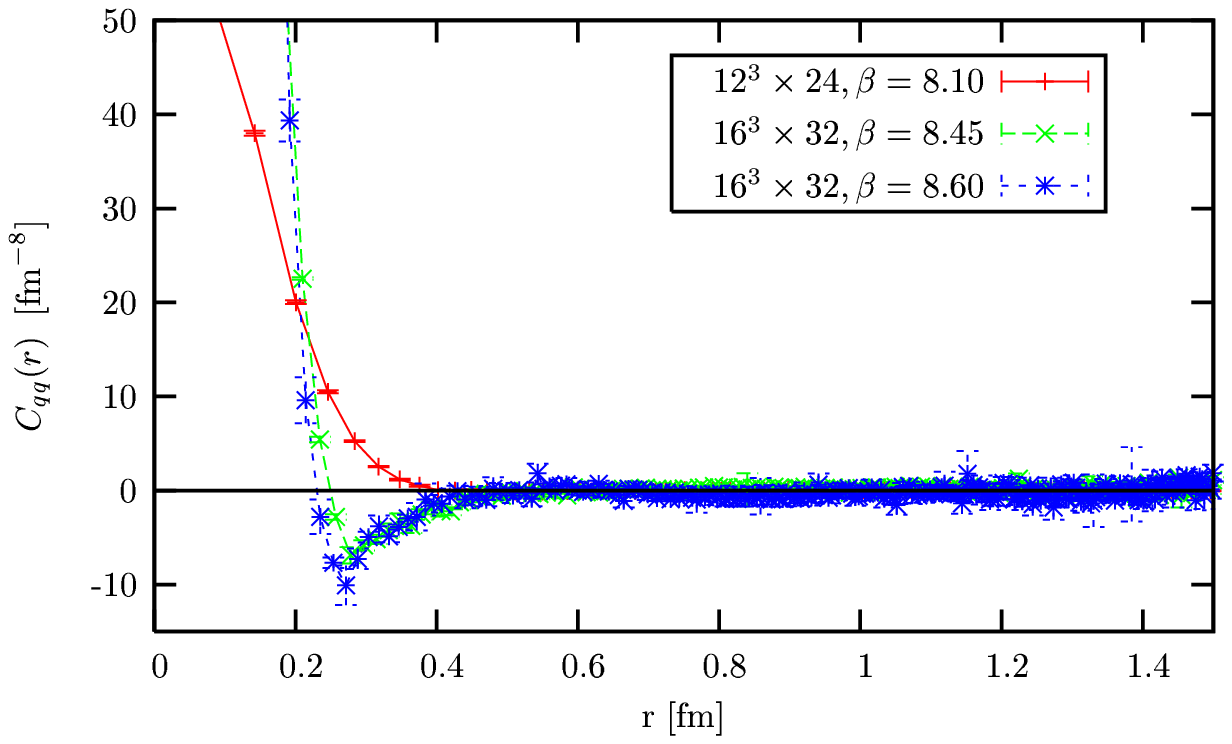,width=8cm} \\
(a) & (b)\\
\end{tabular}
\end{center}
\caption{Comparison of the topological charge density correlator based 
on the all-scale density $q(x)$ for the $12^3\times24$ lattice at $\beta=8.10$ 
(53 configurations) with the same correlator for the $16^3\times32$ 
lattice at $\beta=8.45$ (only 5 configurations) 
and $\beta=8.60$ (2 configurations). The right figure (b) magnifies the 
region in $r$ where the correlator becomes stronger negative when one gets closer 
to the continuum limit.}
\label{fig:qq-corr-full}
\end{figure}

In order to demonstrate how the correlation function of the all-scale 
topological charge density varies with the lattice spacing we compare in 
Fig.~\ref{fig:qq-corr-full}
the correlator for $12^3\times24$ lattices at $\beta=8.10$ (53 configurations)
with the correlator measured on $16^3\times32$ lattices at $\beta=8.45$ 
(only 5 configurations)~\cite{Koma:2005sw} and $16^3\times32$ lattices at
$\beta=8.6$ (2 configurations). An extrapolation of the function $a(\beta)$
to the highest $\beta$ gives $a(\beta=8.6)=0.096(1)$ .
The volumes of the first two lattices are approximately equal. 
The required negativity of the correlator develops only for a sufficiently 
fine lattice. The much bigger number of clusters at higher 
values $|q(x)| \approx 0.25~q_{\rm max}$ (see Section~\ref{sec:4.3})
seems to be necessary to achieve this.
Thus, the cluster multiplicity of the all-scale topological density
simply reflects the size of the lattice in lattice units.

\subsection{Two-dimensional profile seen with varying resolution} 
\label{sec:4.2}

As a vacuum expectation value,
the point-to-point correlation function obviously has to be rotational invariant. 
Individual lattice configurations, however, are  
richer in structure and necessarily have a locally anisotropic structure. 
This results from lower-dimensional structures present in the 
configurations that locally break rotational invariance.

\begin{figure}[t]
\begin{center}
\begin{tabular}{cc}
\epsfig{file=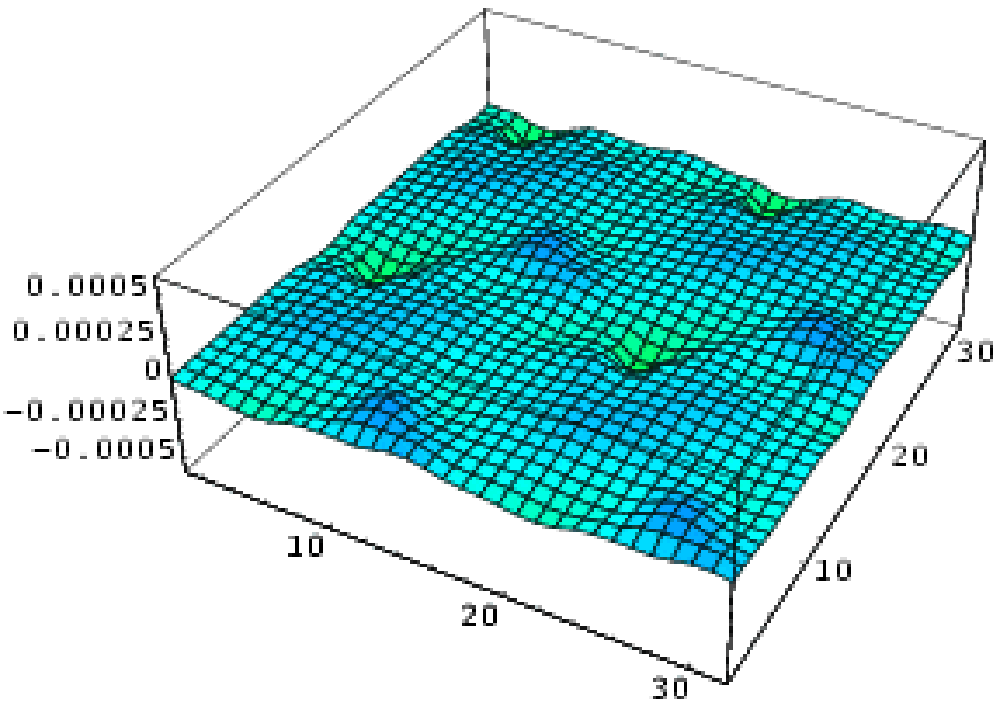,width=6.5cm}&
\epsfig{file=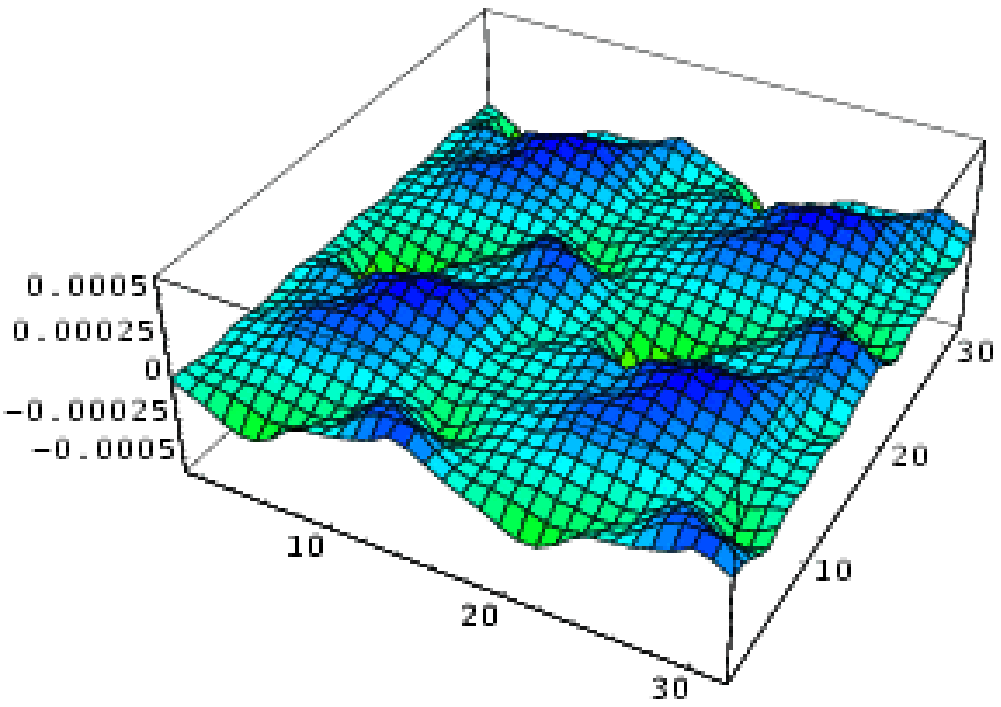,width=6.5cm}\\
(a) & (b)\\
\epsfig{file=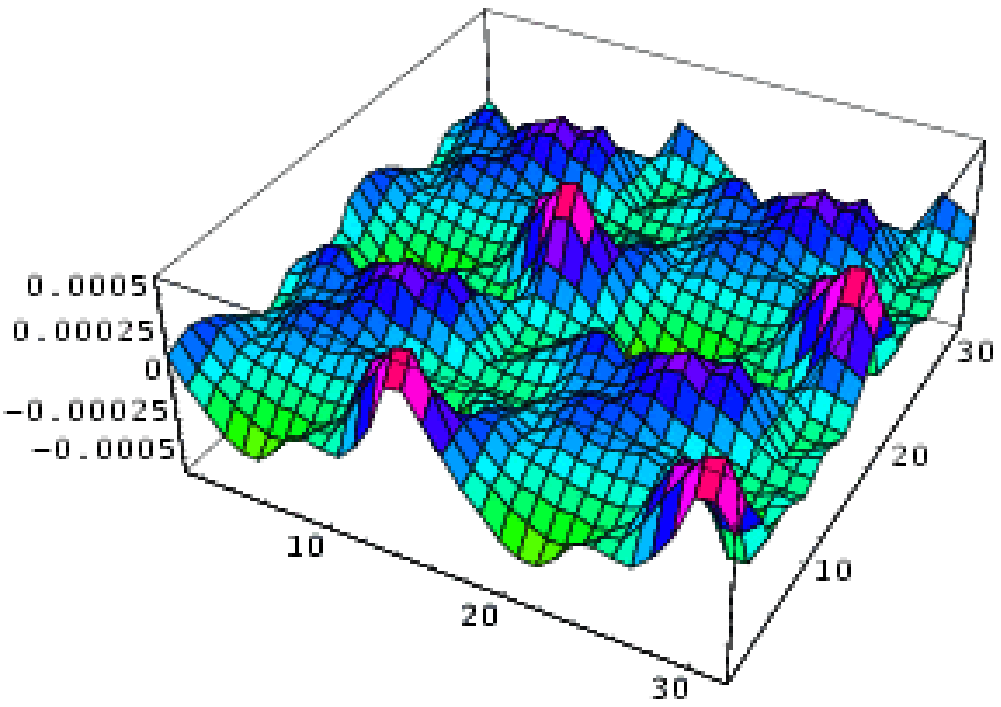,width=6.5cm}&
\epsfig{file=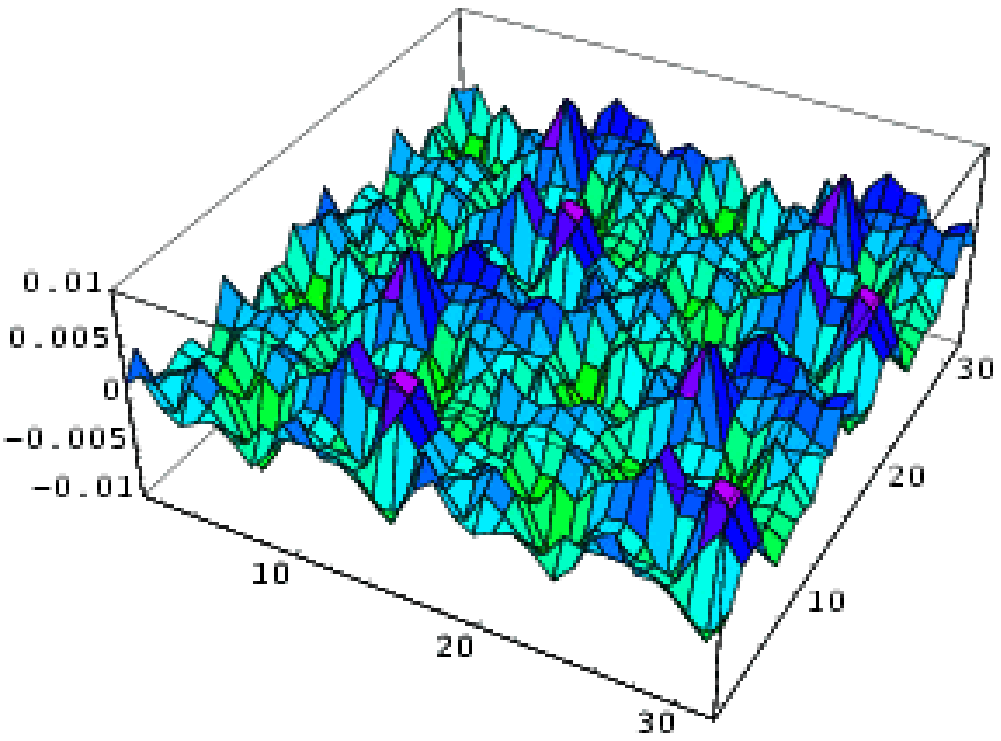,width=6.5cm}\\
(c) & (d)\\
\end{tabular}
\end{center}
\caption{Two-dimensional profile through the typical $Q=0$  configuration 
generated at $\beta=8.45$ on a $16^3 \times 32$ lattice. The profile is
periodically doubled in both directions for better visibility of extended
structures. The cuts show, from (a) to (c), the mode-truncated density 
for descending levels of truncation $\lambda_{\rm cut}  = 200$ MeV, 
$\lambda_{\rm cut} = 400$ MeV and $\lambda_{\rm cut} = 634$ MeV. 
The all-scale topological density 
is shown in (d). Note the 20-fold larger vertical scale in (d).}
\label{fig:2D-cut-all-trunc}
\end{figure}
As an example, a snapshot with four levels of resolution 
is presented in Fig.~\ref{fig:2D-cut-all-trunc} 
of the same two-dimensional section through the typical configuration, already 
considered in  Figs.~\ref{fig:mode-visualization} and \ref{fig:R-isosurface10},  
of the ensemble of $16^3\times32$ lattices at $\beta=8.45$. 
The respective resolution can be inferred from the corresponding 
$\lambda_{\rm cut}$. The two-dimensional section is periodically doubled
in the two directions such that the one-dimensional structure (apparently 
one ``ridge'' per elementary cell in this two-dimensional world) 
can be recognized. 
In particular, this ridge is 
clearly visible at 
$\lambda_{\rm cut} = 200$ and $400$ MeV, being already less clearly visible 
at $\lambda_{\rm cut}=634$ MeV. 
At a resolution of $400$ MeV and higher, also single peaks (a 0-dimensional 
structure in this two-dimensional world) on top of the ridges become 
visible. 
Describing the landscape from the perspective of the peaks of density in 
Fig.~\ref{fig:2D-cut-all-trunc} (c), the next-higher 
dimensional structures (``ridges'') are features becoming discernible
only at lower height of the density. 
In a cluster analysis, with $q_{\rm cut}$ chosen 
too large, one might miss the ridge catching only the peaks. 
Varying $q_{\rm cut}$ the dimension revealed by the cluster analysis
will be very reminiscent of the multifractal dimensions of the individual 
fermionic modes. Compared to these visible features of the three different
mode-truncated densities, the cut through the all-scale topological 
density is seemingly completely random. Nonetheless, structure can be 
discovered in the ``full'' density, too, simultaneously 
taking all four dimensions into account through the cluster algorithm,
as we shall see next.

\subsection{Cluster analysis of the all-scale topological charge density}
\label{sec:4.3}

In order to make the structural analysis more quantitative, the cluster analysis 
as described in Section~\ref{sec:2.5} is performed. 
Let us begin with the all-scale topological charge density.
The structure is changing with the threshold value $q_{\rm cut}$.
As the reference value for $q_{\rm cut}$ serves the maximum 
of the topological density in each configuration, $q_{\rm max}$. Thus, we take 
$0 < q_{\rm cut}/q_{\rm max} < 1$ as the running parameter. We remark that 
$q_{\rm max}$ for the mode-truncated density is weakly changing 
with $\lambda_{\rm cut}$ (around 0.005 in lattice units). For the all-scale
topological density $q_{\rm max}$ is a few times larger (see Table IV).
\begin{table}[t]
\vspace*{1cm}
\begin{center}
\begin{tabular}{|c|c|c|c|c|c|c|}
\hline
$\lambda_{\rm cut}$ & 200 MeV & 400 MeV & 600 MeV & 634 MeV & 800 MeV & all-scales \\
\hline
$\beta=8.10$  & $5.04(14)~10^{-3}$ & $6.45(14)~10^{-3}$ & $7.66(15)~10^{-3}$ &       & $8.88(14)~10^{-3}$ & $2.82(07)~10^{-2}$ \\
$12^3 \times 24$ &     &       &       &       &        &        \\ 
\hline
$\beta=8.45$  & $3.51(17)~10^{-3}$ & $4.45(18)~10^{-3}$ &       & $5.39(19)~10^{-3}$ &       & $1.97(06)~10^{-2}$  \\
$16^3 \times 32$ &     &       &       &       &       &        \\ 
\hline
\end{tabular}
\vspace{0.3cm}
\caption{The ensemble averages of $q_{\rm max}~a^4$ for the mode-truncated topological
charge density (for various $\lambda_{\rm cut}$) and for the all-scale density
for two lattice ensembles.}
\end{center}
\label{tab:qmax}
\end{table}

\begin{figure}[t]
\begin{center}
\begin{tabular}{cc}
\epsfig{file=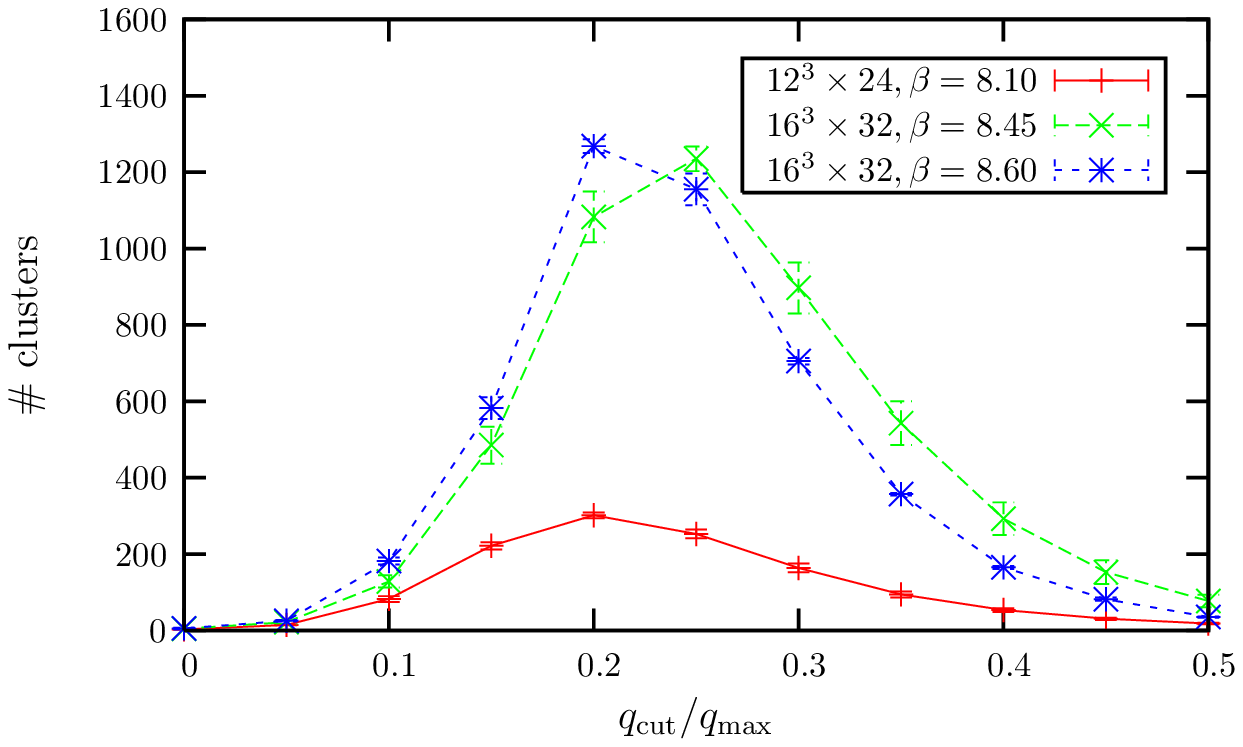,width=8cm}&
\epsfig{file=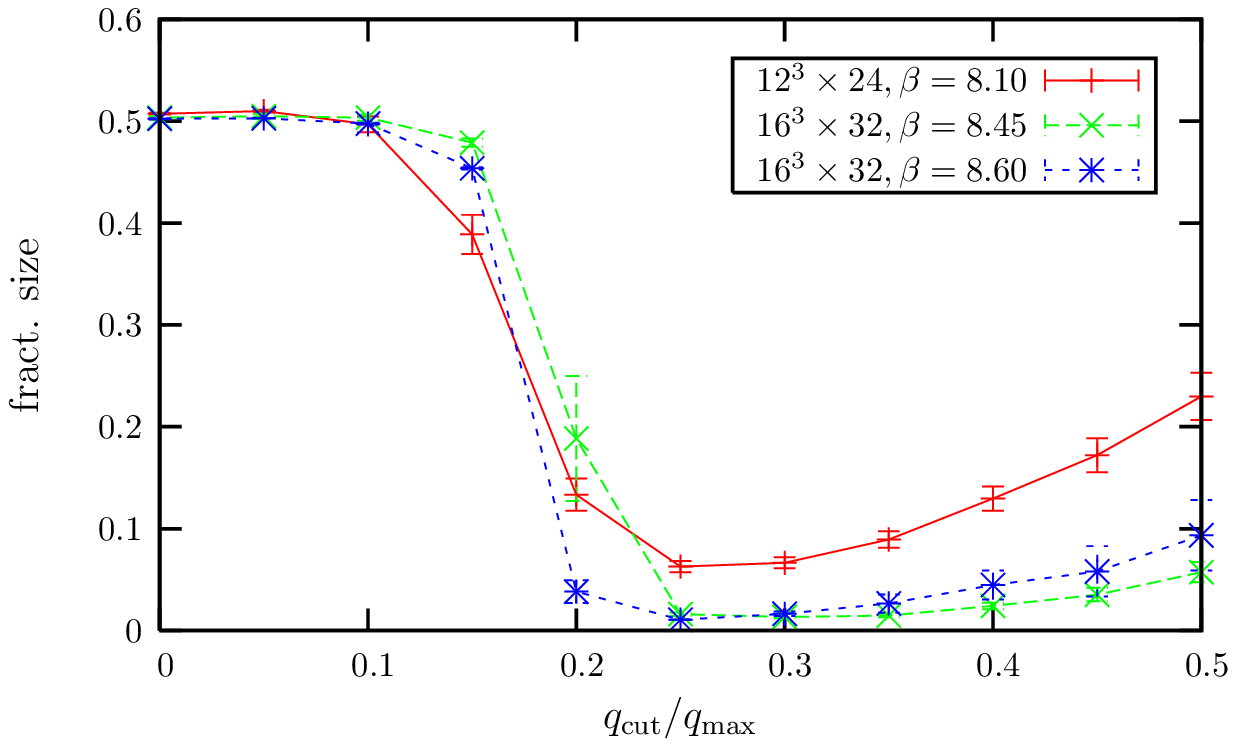,width=8cm}\\
(a) & (b)\\
\epsfig{file=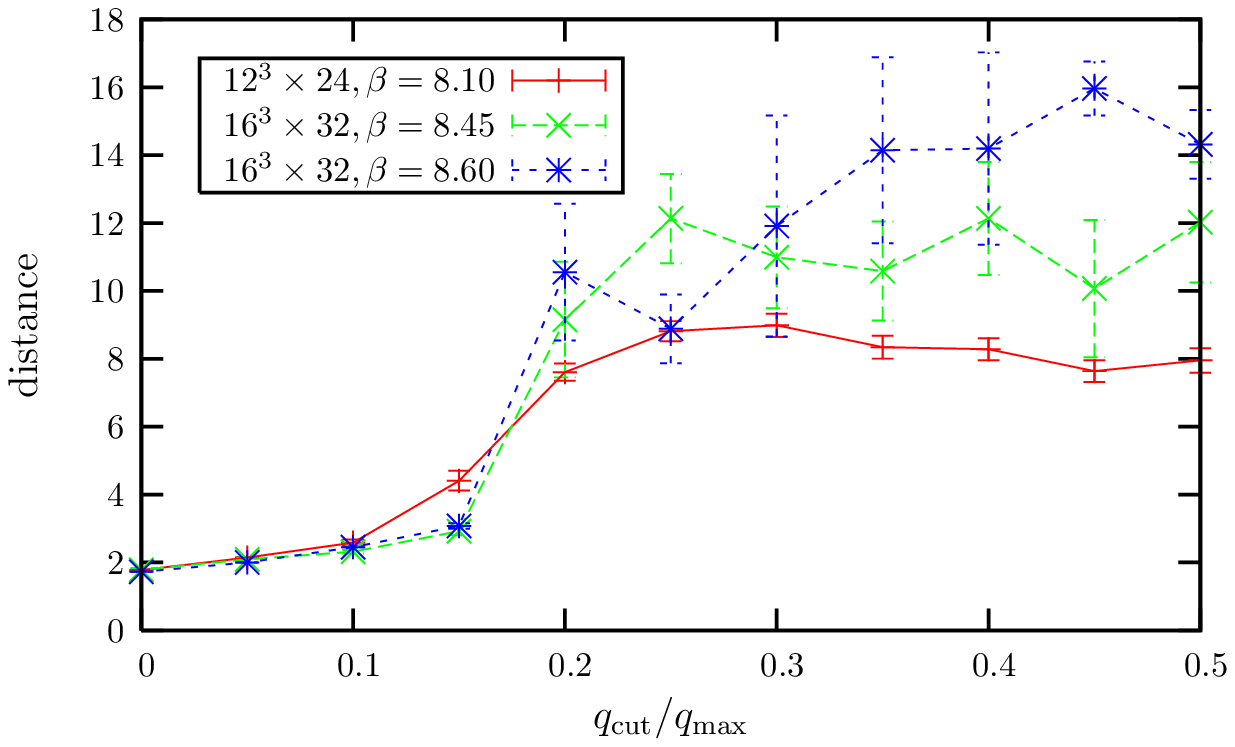,width=8cm}&
\epsfig{file=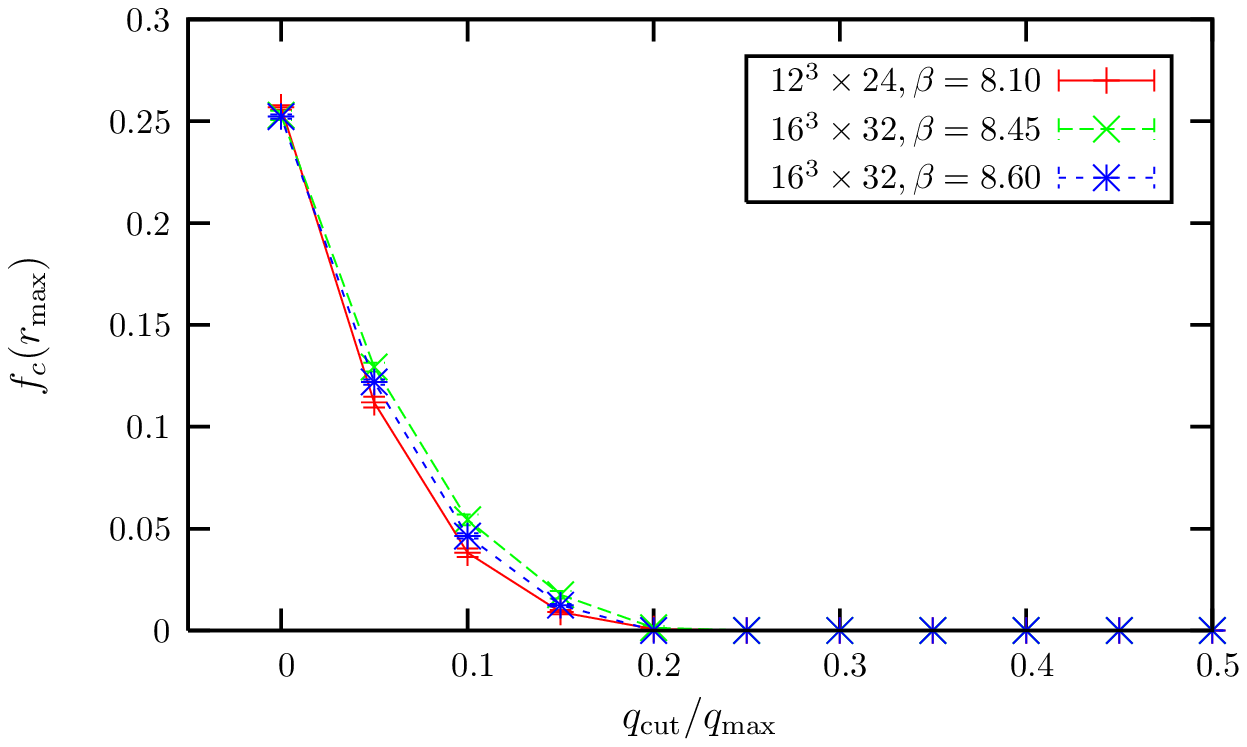,width=8cm}\\
(c) & (d)
\end{tabular}
\end{center}
\caption{Cluster analysis of the all-scale topological density: The $q_{\rm cut}$ 
dependence is shown of (a) the total number of separate clusters, (b) the size of 
the largest cluster relative to all clusters, (c) the distance between the two 
largest clusters in lattice units and (d) the connectivity 
(see section ~\ref{sec:2.3}). 
The data is plotted both for the $12^3\times 24$ lattice configurations at 
$\beta=8.10$ (53 configurations, in red), for the $16^3\times 32$ lattice 
configurations at $\beta=8.45$ (only 5 configurations, in green) and for the
$16^3\times 32$ lattice configurations at $\beta=8.60$ (only 2 configurations, 
in blue) being averaged over.} 
\label{fig:cluster-full-density}
\end{figure}
In Fig.~\ref{fig:cluster-full-density} the $q_{\rm cut}/q_{\rm max}$ dependence 
of the cluster multiplicity, of the relative size of the largest cluster with 
respect to all clusters, of the distance between the two biggest clusters and 
of the connectivity are presented. This is shown for two ensembles 
of practically the same volume, $12^3\times 24$ at $\beta=8.10$ and 
$16^3\times 32$ at $\beta=8.45$, and, in order to study the $\beta$ dependence
alone, for a few configurations of the same size $16^3\times 32$ in lattice 
units, but with a finer lattice spacing at $\beta=8.60$. 

The most striking difference between the first two cases with almost equal 
physical volume is the roughly four times larger multiplicity of clusters 
that is reached on the finer lattice. In both lattices percolation 
(nonvanishing connectivity $f(r_{\rm max})$) sets in close to where the 
multiplicity reaches its maximum. The percolation threshold $q_{\rm perc}$
is practically independent of the lattice spacing. The percolation finally 
ends with $q_{\rm cut} \to 0$ with only two, globally extended clusters of 
opposite sign of topological charge. With decreasing $q_{\rm cut}/q_{\rm max}$, 
but still above the percolation threshold at $q_{\rm cut}/q_{\rm max} = 0.2$, 
the number of clusters increases strongly with decreasing cut, in particular 
on the fine lattice. Since number and size of all clusters increase, the 
relative size of the biggest cluster decreases. This trend is reversed beyond 
the percolation threshold where the relative size of the largest cluster 
rapidly reaches 50 \%, practically independent of the lattice spacing. 
The distance between the two biggest clusters shown in 
Fig.~\ref{fig:cluster-full-density} (c) is given 
in lattice spacings. Before percolation the distance is therefore {\it only 
seemingly} bigger on the finer lattice. Taking the ratio 
$a(\beta=8.10)/a(\beta=8.45) \approx 1.4$ of the lattice spacings (see Table I)
into account, the average distance between the two leading clusters 
is approximately the same and independent of $q_{\rm cut}/q_{\rm max}$ all over 
the non-percolating regime. The finer the lattice, the faster the minimal 
distance between the two leading clusters is approaching the minimal 
distance of two lattice spacings as soon as $q_{\rm cut} < q_{\rm perc}$.
In the final state the two clusters are close to each other everywhere, kept 
as separate clusters only because of the different sign of $q(x)$. In other 
words, the remaining two global clusters of opposite charge are everywhere 
thin of $O(a)$ and entangling each other. The finite distance in lattice 
spacings is independent of $\beta$ reflecting nothing else than the 
non-ultralocality of the overlap operator.

For the two lattices $16^3\times32$ with different lattice spacing we find
a surprising similarity in all four plots. Even the distance between the 
biggest clusters coincides within large errors due to the low statistics 
(5 and 2 configurations for $\beta=8.45$ and $\beta=8.60$, respectively). 
In particular, the cluster multiplicity seems to depend mostly on the lattice 
size in lattice units.

\begin{figure}[t]
\begin{center}
\begin{tabular}{cc}
\epsfig{file=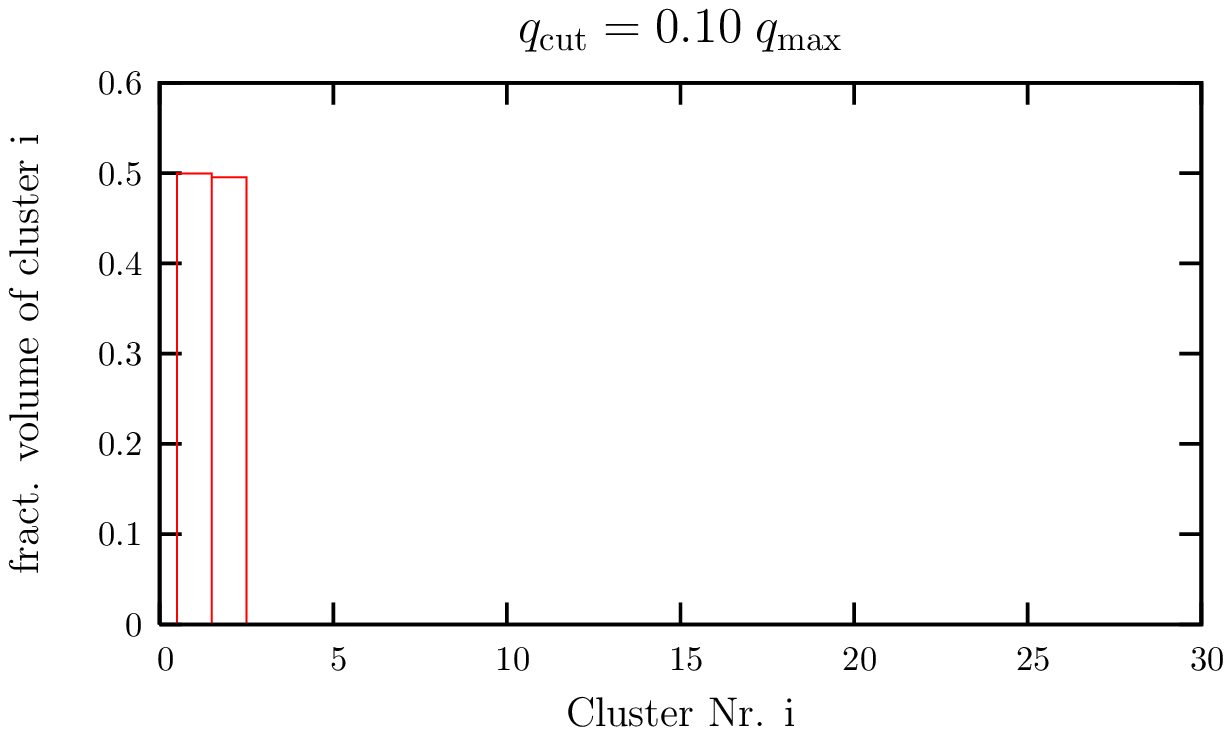,width=8cm}&
\epsfig{file=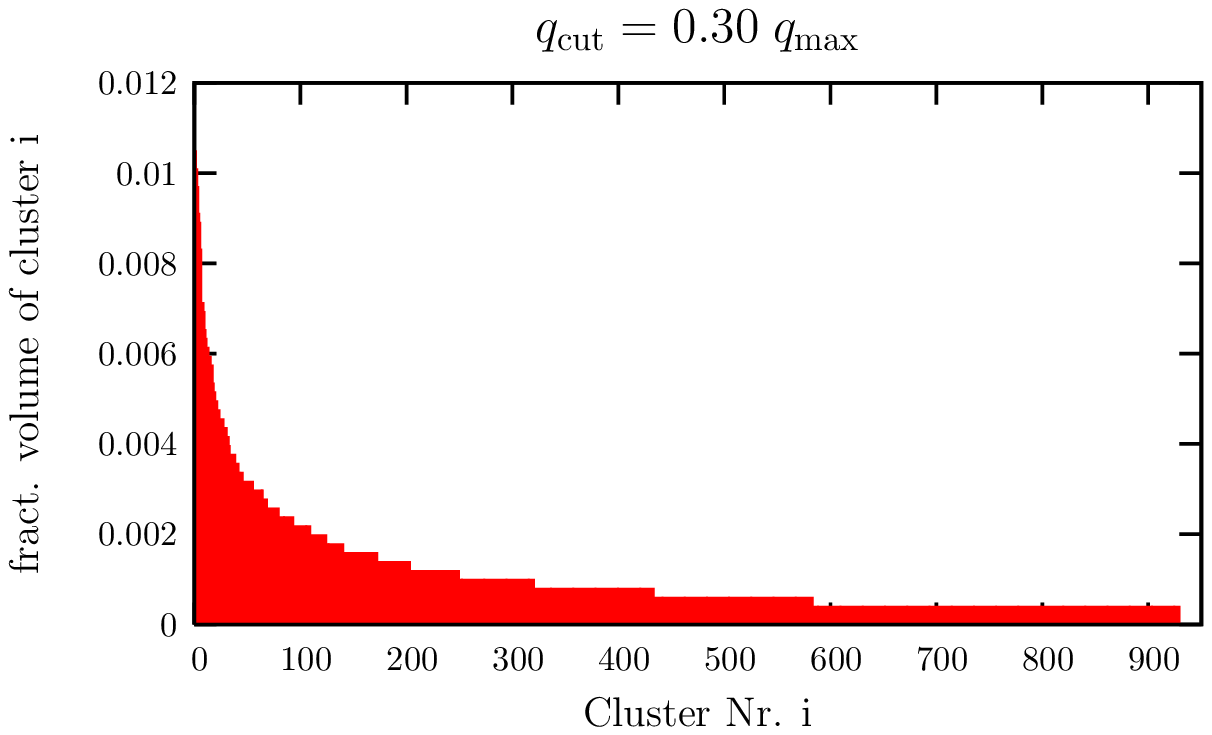,width=8cm}\\
(a) & (b)\\
\epsfig{file=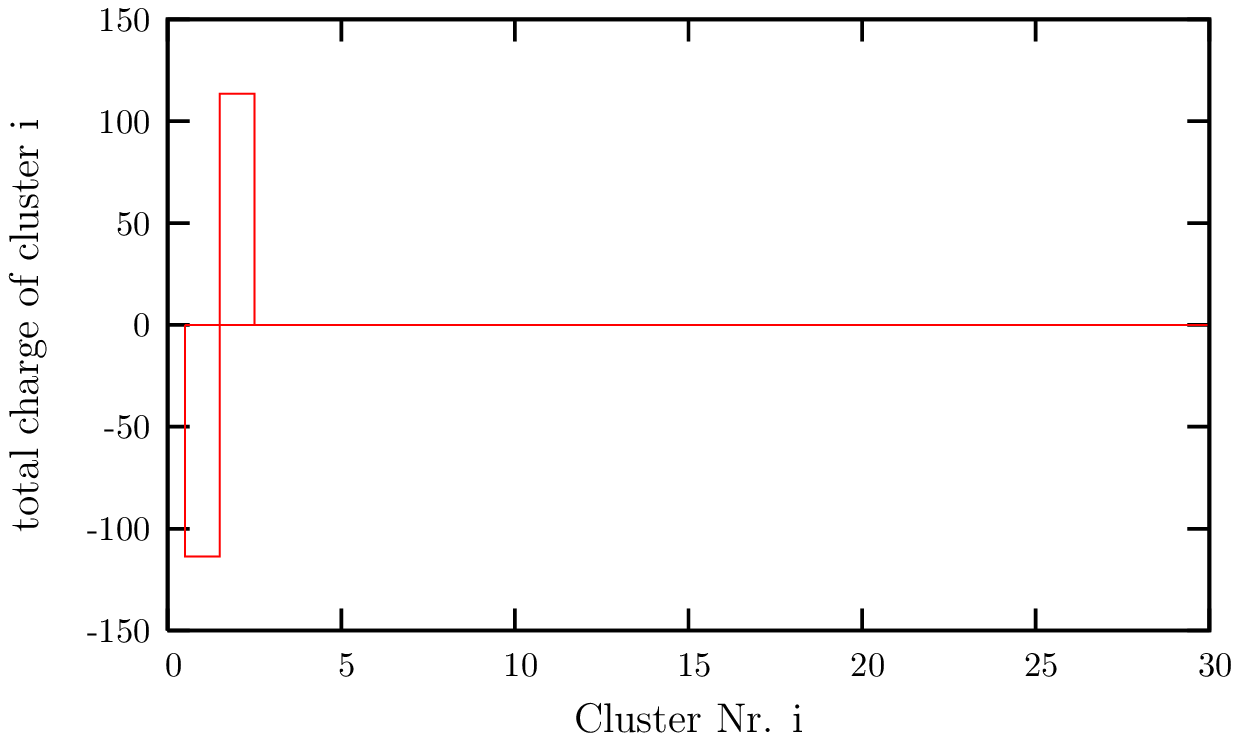,width=8cm}&
\epsfig{file=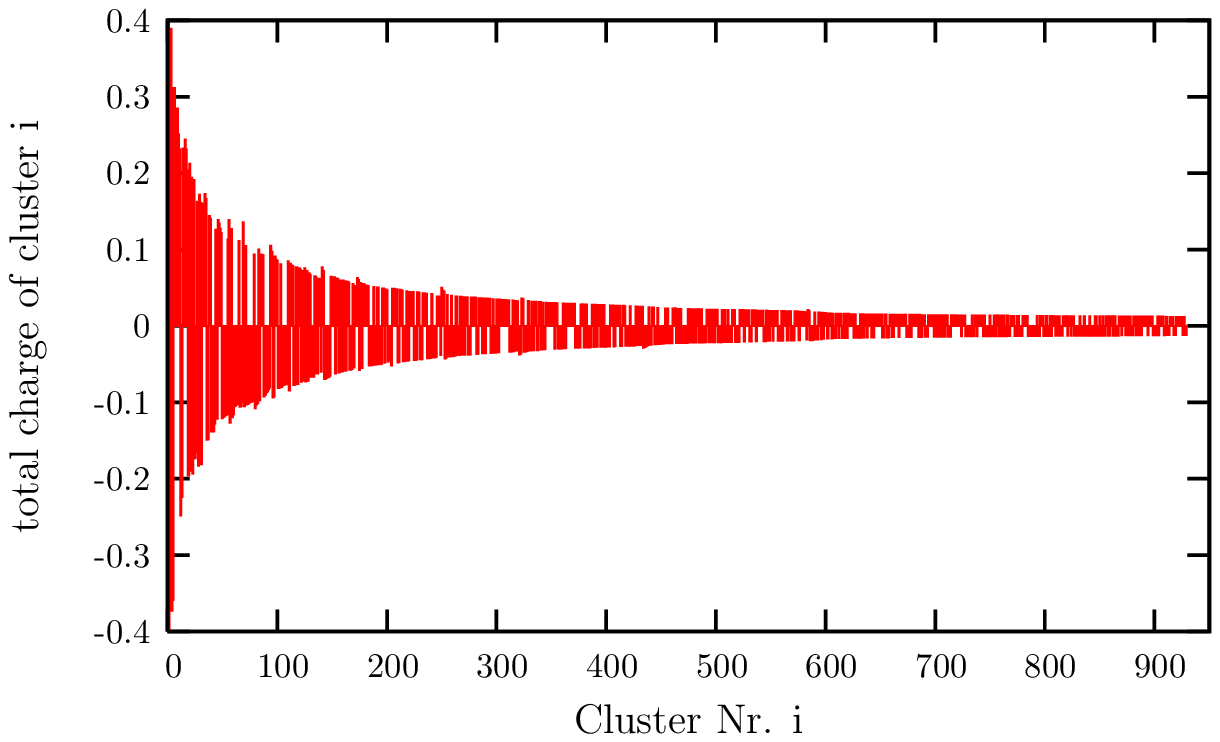,width=8cm}\\
(c) & (d)
\end{tabular}
\end{center}
\caption{Cluster analysis of the all-scale topological density: 
The fractional volume ((a)+(b)) and the total charge
((c)+(d)) of the corresponding clusters, ranked according to their 
occupied volume, is shown for two values of $q_{\rm cut}/q_{\rm max}$:  
in the percolated regime (left) for 
$q_{\rm cut}/q_{\rm max}=0.10$ 
and in the cluster-separated regime (right) for 
$q_{\rm cut}/q_{\rm max}=0.30$.
The plot represents one single configuration 
of the $16^3 \times 32$ lattice ensemble generated at $\beta=8.45$.}
\label{fig:clustercharges-full-density}
\end{figure}
How the whole $q$-cluster composition with respect to charge {\it and} 
volume varies between the percolating regime ($q_{\rm cut}=0.1~q_{\rm max}$) and the 
non-percolating regime ($q_{\rm cut}=0.3~q_{\rm max}$), 
is illustrated in Fig.~\ref{fig:clustercharges-full-density}.
As long as percolation has not set in, there is a broad spectrum of
cluster volumes.  
The spectrum of cluster topological charges is surprisingly similar,
up to the random sign. The ranking of clusters is according to the 
cluster volume, and the modulus of the cluster charges almost follows this ranking.
Thus, the average charge density in each cluster is approximately the same.
When percolation is completed, there are only two, oppositely charged
clusters left over with approximately equal volume. 
This peculiar situation has been first
discovered and discussed by Horvath {\it et al.}~\cite{Horvath:2003yj}.

\subsection{Dimensionality and multifractality of the all-scale topological charge}
\label{sec:4.5}

In this subsection a more explicit description of the fractal dimension and 
the multifractal properties of this topological density will be given. 
These properties are difficult to visualize in three dimensions 
because of the close
packing and the huge multiplicity (at the percolation threshold) 
of the corresponding $q$-clusters. 

\begin{figure}[t]
\begin{center}
\hspace*{-1.0cm}\epsfig{file=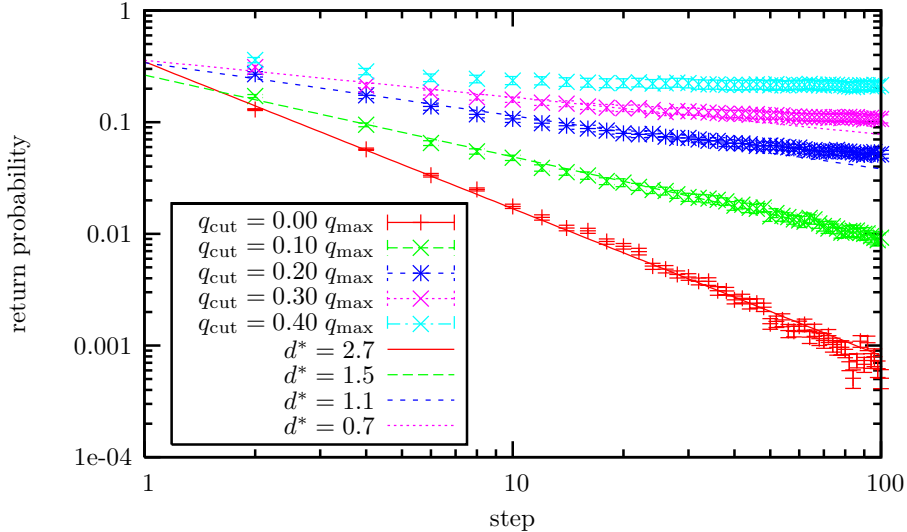,angle=0,width=13.0cm, clip=}
\end{center}
\caption{The probability of return to the cluster center $x_c$ for random walks, 
restricted to topological clusters of the all-scale topological charge 
density, shown as a function of the step number. The clusters are distinguished
by different cuts with respect to the density that they enclose. The cases 
$q_{\rm cut}/q_{\rm max}=0.4$, $0.3$, $0.2$, $0.1$, and $0.0$ are considered on 
the coarser lattice $12^2\times24$ at $\beta=8.10$. The results of the power fits 
(see the curves) are labeled by the effective dimensions $d^{*}$.}
\label{fig:returnprobability}
\end{figure}

For our purpose we use the random walker method described in section~\ref{sec:2.6}.
In Fig.~\ref{fig:returnprobability} we show in a double-logarithmic 
plot the decay of the return probability on the  $12^3\times 24$ lattice.  
In an open $d^{*}$-dimensional space the return probability after $\tau$ steps would 
follow the power law $P({\vec 0},\tau)=1/(2 \pi \tau)^{d^{*}/2}$.
It turns out, that by extending the clusters ({\it i.e.} by  
chosing $q_{\rm cut}/q_{\rm max}$ lower and lower) the power of decay
becomes larger. The corresponding effective fractal dimensions 
are ranging from
$d^{*}=2.7$ for low $q_{\rm cut}/q_{\rm max} \approx 0$,
over $d^{*}=1.5$ at $q_{\rm cut}/q_{\rm max}=0.1$
and $d^{*}=1.1$ at $q_{\rm cut}/q_{\rm max}=0.2$
where the $q$-cluster multiplicity reaches its maximum,  
to $d^{*}=0.7$ at $q_{\rm cut}/q_{\rm max}=0.3$ where the clusters are really 
restricted to the immediate neighborhood of the maxima. 
For a survey of dimensionalities of clusters of the topological charge
see Tables V and VI in the Appendix.
Even if the values of the fractal dimensions $d^{*}$ provided by the
random walker algorithm do not exactly coincide with the dimensions obtained 
from other methods, 
the tendency of the results of the
random walk method indicates that the landscape of the all-scale topological 
charge is multifractal and becomes nearly three-dimensional only far below 
the percolation threshold. Percolation sets in at 
$q_{\rm cut} \approx 0.25 q_{\rm max}$, 
when the effective dimension is still low as $d^{*} = 1 - 1.5$ .

\begin{figure}[t]
\begin{center}
\begin{tabular}{cc}
\epsfig{file=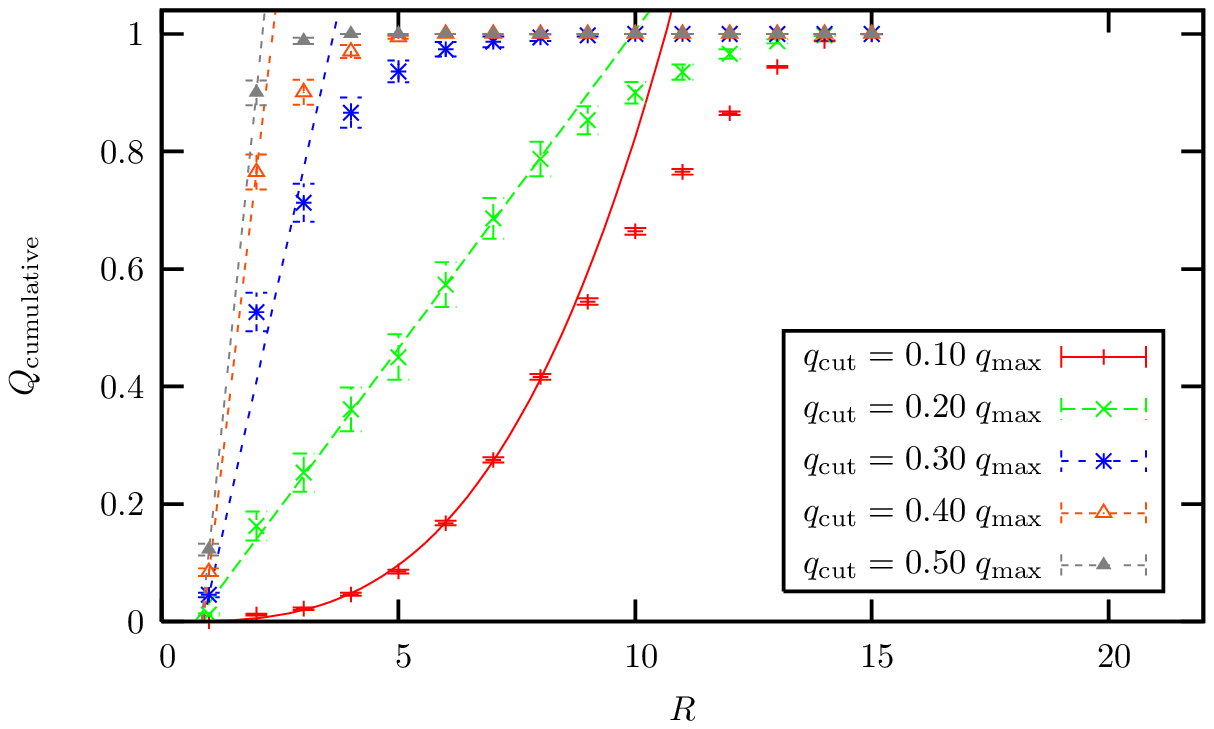,width=8cm}&
\epsfig{file=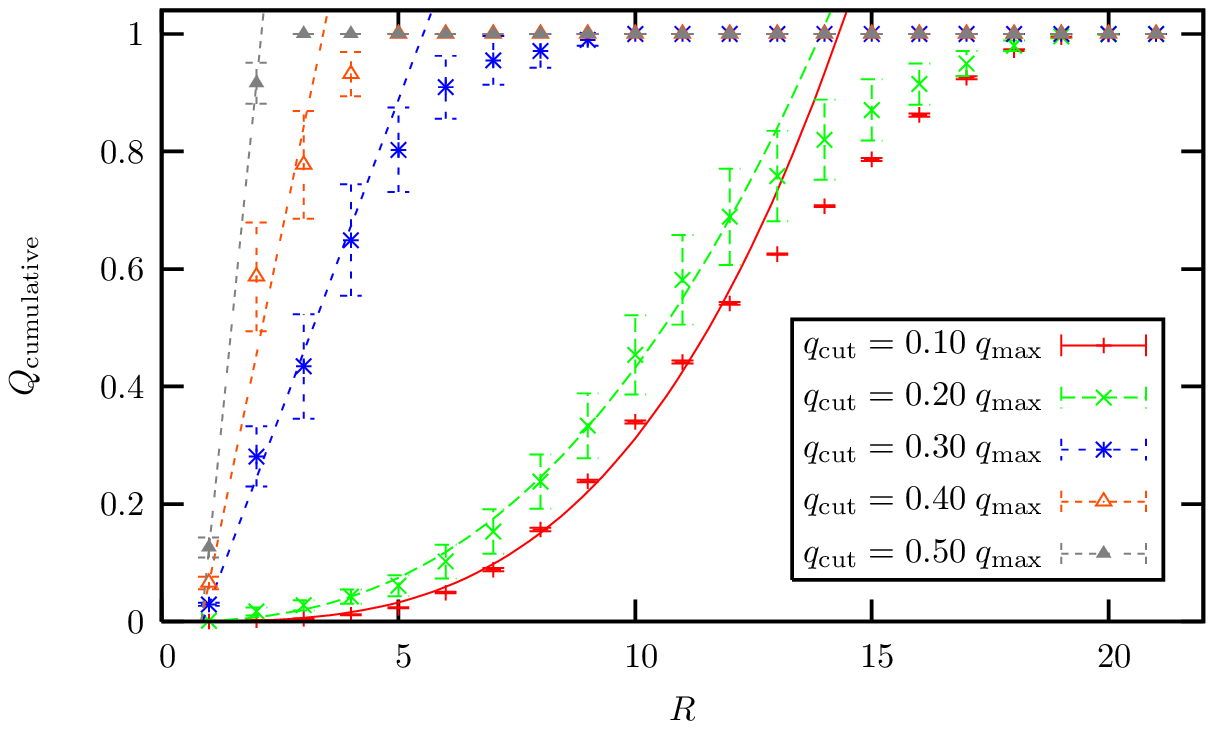,width=8cm}\\
(a) & (b)
\end{tabular}
\end{center}
\caption{The covering sphere method (see text): (a) for the coarse lattice 
$12^3\times24$ at $\beta=8.10$, (b) for the fine lattice $16^3\times32$ at 
$\beta=8.45$. The fitted power behavior of the accumulated charge jumps from 
one-dimensional to three-dimensional at some percolation threshold 
$q_{\rm cut}/q_{\rm max}$ (see text).} 
\label{fig:cluster-accumulated-charge}
\end{figure}

A second argument comes from a consideration of the cumulative cluster 
charge. One considers a collection of $q$-clusters primarily defined  
for a given cut-off $q_{\rm cut}/q_{\rm max}$. 
One covers the set of clusters with a four-dimensional 
sphere of radius $R$ centered at the center $x_c$ of one particular cluster
({\it i.e.} the maximum of $|q(x)|$ inside the cluster $c$). One sums that 
part of the cluster charge that is located in the intersection of 
the cluster and the covering sphere. 
The relative cumulated cluster charge is the ratio between the cumulated charge
to the full charge of the cluster and it depends on the radius $R$. 
The dependence of the relative cumulated cluster charge on the radius $R$ 
will start with some power and finally reach unity at saturation. The
observed power behavior can give information about the fractal dimension
$d^{*} < d$ of a $q$-cluster. 
The initial growth with $R$ may be parametrized 
as $R^{d^{*}}$, where $d^{*}$ describes the local dimension near the 
cluster center and depends on the cut-off $q_{\rm cut}/q_{\rm max}$.
One can see in Fig.~\ref{fig:cluster-accumulated-charge} that a linear fit 
works very well as long as $q_{\rm cut}/q_{\rm max}$ is sufficiently high such 
that the clusters are isolated and do not percolate. The beginning percolation 
is characterized by an initial growth fitted like $Q_{\rm cumul} \sim R^3$. 

One might expect problems when $R$ is so large that the sphere covers
many different three-dimensional branches of a $q$-cluster which might 
appear disconnected inside 
the sphere but are actually globally connected (and therefore counted
in the cumulative charge according to the cluster membership). 
This would happen at very low cut-off $q_{\rm cut}$ when only the two 
largest clusters have survived. 
Only then the cumulated charge of each of the clusters would grow like 
$Q_{\rm cumulative} \sim R^4$ although the local 
dimension is $d^{*}=3$~\footnote{This consideration was the motivation 
to propose the random walker method described in Section~\ref{sec:2.6}. 
The walkers are not allowed to penetrate to or return from another branch
of the same cluster.}. 

One can see in Fig.~\ref{fig:cluster-accumulated-charge} that the
transition from linear to cubic behavior happens
for the coarser lattice (left panel)  
somewhere between $q_{\rm cut}/q_{\rm max}=0.1$ and $0.2$ . 
The actual power behavior for the lowest shown cut-off  is like 
$Q_{\rm cumulative} \sim R^{3.1}$. On the finer lattice,
for the higher cut-off $q_{\rm cut}/q_{\rm max}=0.2$ the situation is already 
close to the $d^{*}=3$ regime. The fitted power behavior is like 
$Q_{\rm cumulative} \sim R^{2.5}$. 
For $q_{\rm cut}/q_{\rm max}=0.1$ it is like $R^{3.3}$.

For a survey of dimensionalities of the topological density (as well as
of overlap eigenmodes) inferred from the covering sphere method 
see Table VI (and Table VIII) in the Appendix.

\subsection{Cluster analysis of the mode-truncated topological charge density}
\label{sec:4.4}

\begin{figure}[t]
\begin{center}
\begin{tabular}{cc}
\epsfig{file=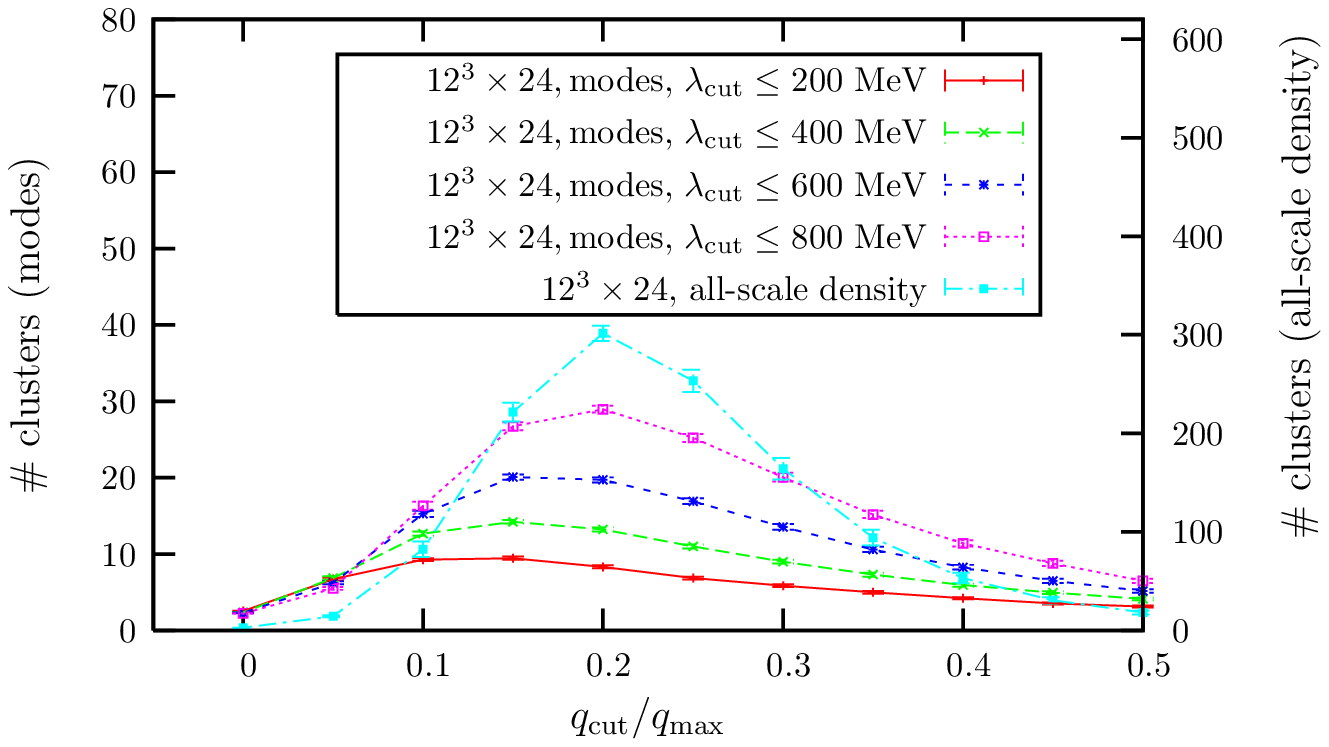,width=8.5cm}&
\epsfig{file=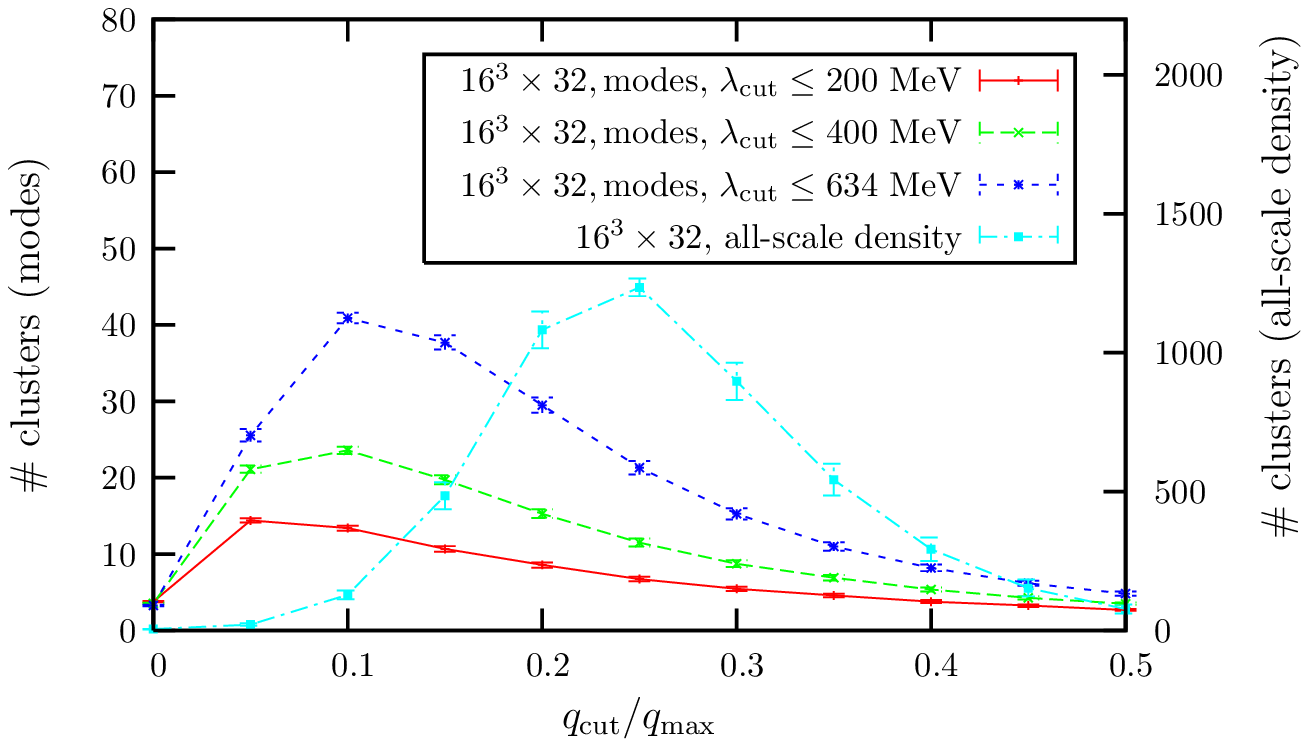,width=8.5cm}\\
(a) & (b)\\
\hspace*{-0.8cm}\epsfig{file=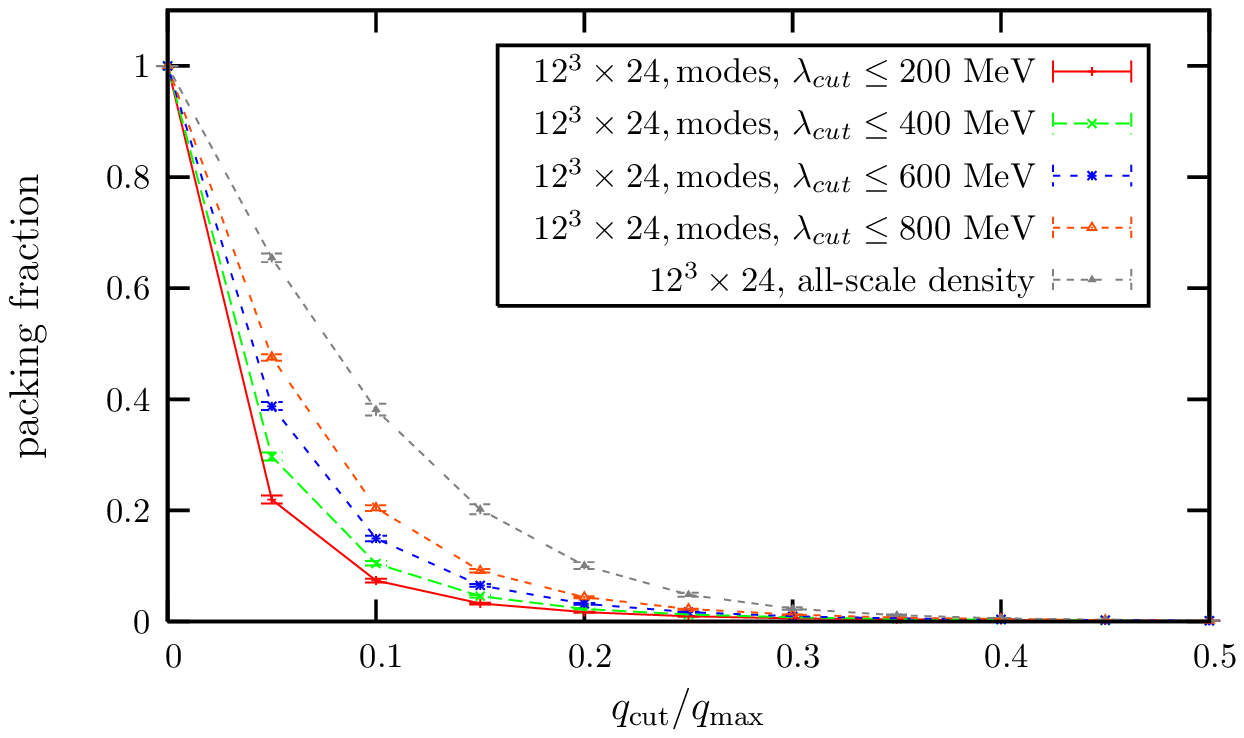,width=8cm}&
\hspace*{-0.9cm}\epsfig{file=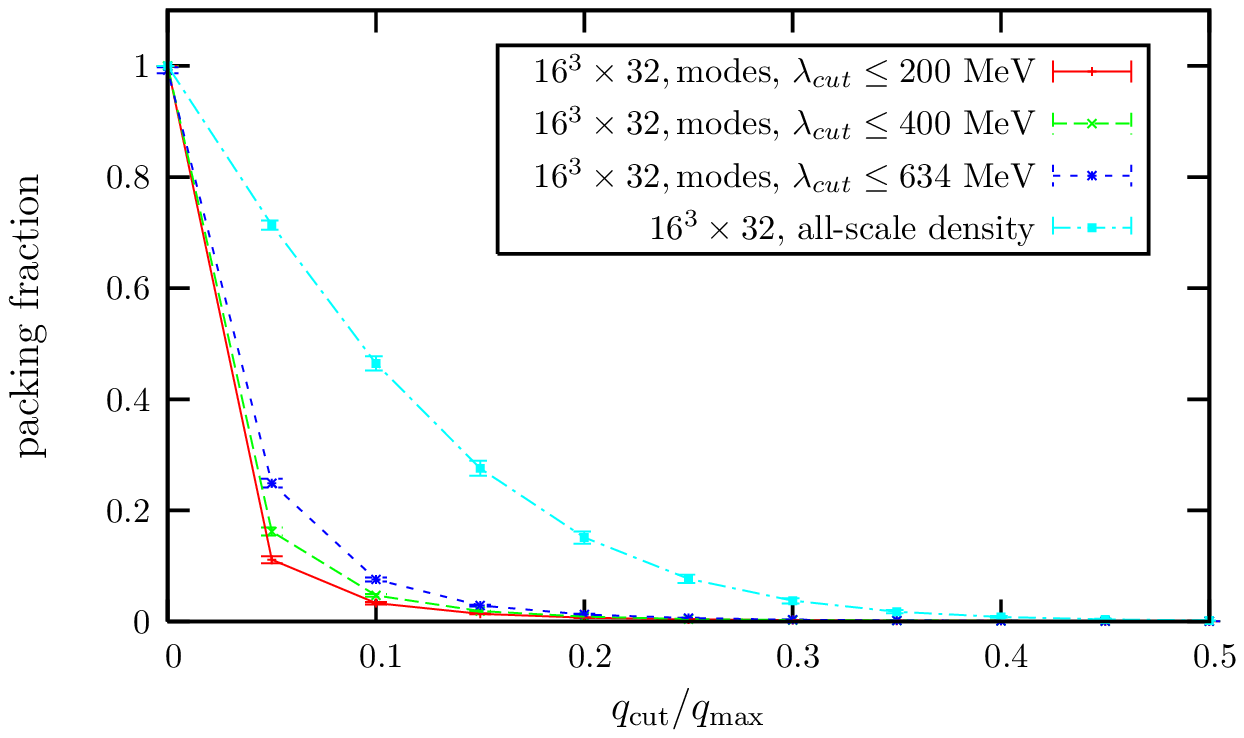,width=8cm}\\
(c) & (d)\\
\vspace*{0.4cm}
\hspace*{-0.8cm}\epsfig{file=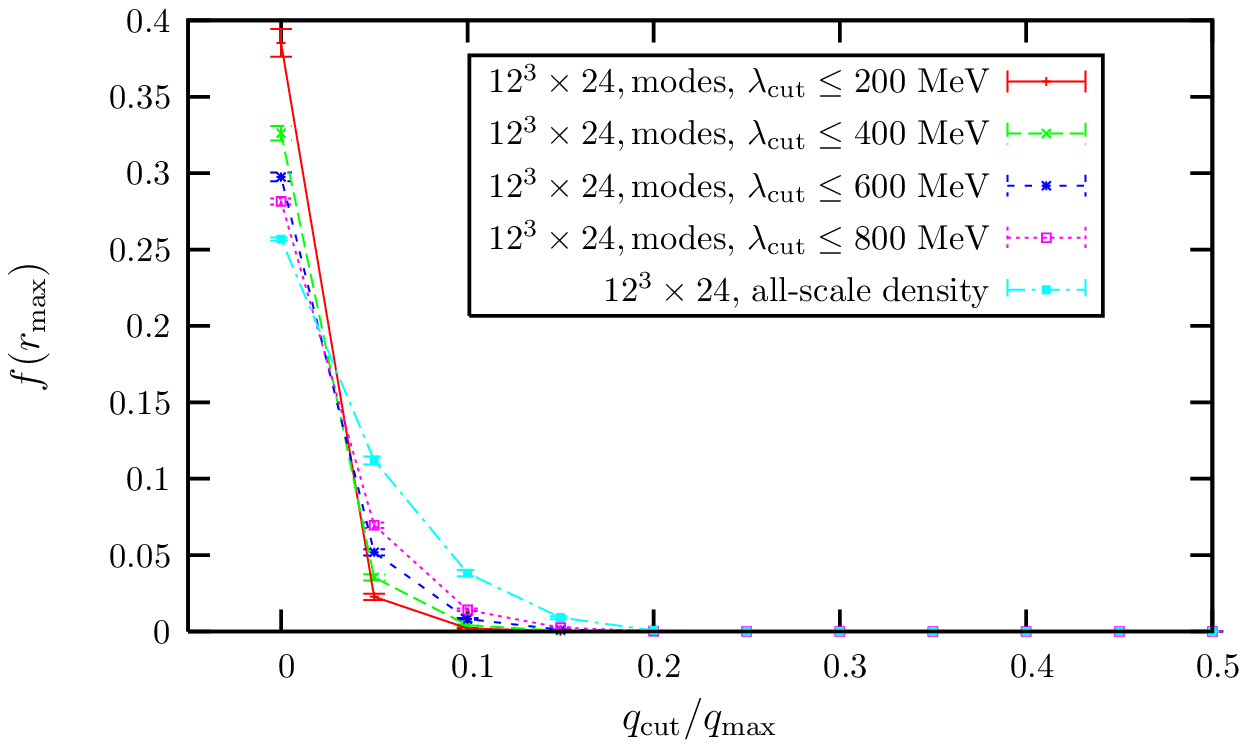,width=8cm}&
\hspace*{-0.9cm}\epsfig{file=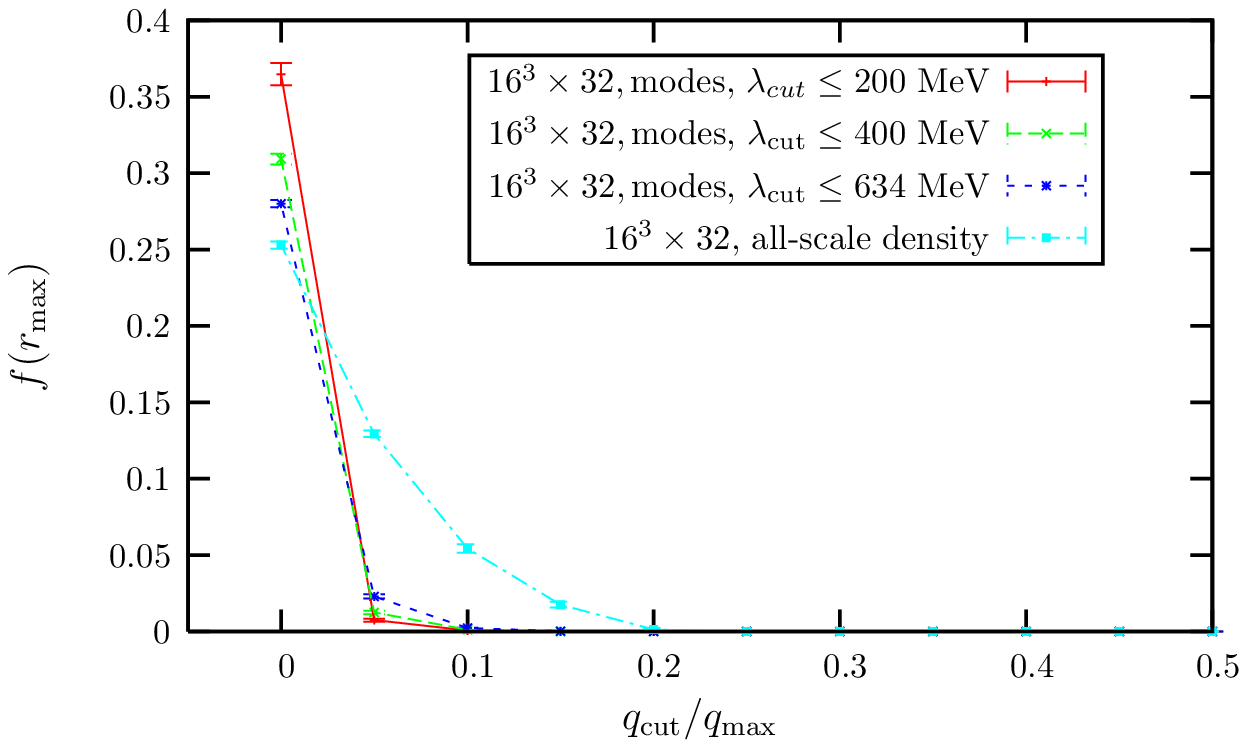,width=8cm}\\
(e) & (f)
\end{tabular}
\end{center}
\caption{Comparison of the cluster structure of the mode-truncated density
with three resp. four cut-offs $\lambda_{\rm cut}$ with the
all-scale topological charge density. The effect of $\beta$ at the same physical volume 
can be seen by comparing the left side ($12^3\times24$ lattices at $\beta=8.10$) 
with the right side ($16^3\times 32$ lattices at $\beta=8.45$).
Top row: the $q_{\rm cut}$ dependence of the total number of clusters 
((a) and (b)). Note the different multiplicity scale on the right for clusters 
of the all-scale topological charge density.
Middle row: the packing fraction of all clusters together ((c) and (d)).
Bottom row: the connectivity ((e) and (f)).} 
\label{fig:cluster-full-vs-trunc-density_1-3}
\end{figure}
Let us now discuss the effect of mode-truncation on the cluster structure of the 
topological charge density.  We compare in
Fig.~\ref{fig:cluster-full-vs-trunc-density_1-3} for the
two ensembles, $12^3 \times 24$ at $\beta=8.10$ (left)
           and $16^3 \times 32$ at $\beta=8.45$ (right) with 
	   practically equal physical volume,
the ensemble averages of the cluster multiplicity, of the fractional volume
filled by {\it all} clusters (packing fraction) and of the connectivity
$f(r_{\rm max})$. 
In each panel we show this for 3 or 4 cut-offs $\lambda_{\rm cut}$ 
together with the same
quantity defined for the all-scale density. 
Notice the different scale for the cluster multiplicity for the all-scale 
density on the right of the upper panels.
For the coarser lattice presented on the left and the finer lattice presented 
on the right, the tendency is the same.
The maximal number of clusters is much smaller for the mode-truncated 
density than for the all-scale density. It rises with the cut-off 
$\lambda_{\rm cut}$. For a fixed cut-off $\lambda_{\rm cut}$ in MeV, it is 
approximately 40 to 100 \% larger on the finer lattice than on the coarser lattice.
Thus, the cluster multiplicity defined with a fixed ultraviolet cut-off 
$\lambda_{\rm cut}$, 
is strongly discretization dependent.
The maximum of cluster multiplicity is reached at ever smaller 
$q_{\rm cut}/q_{\rm max}$ for the finer lattice.
The packing fraction, {\it i.e.} the fraction of volume filled by the clusters,
for the all-scale density rises faster with decreasing $q_{\rm cut}$ on the finer 
lattice than on the coarser. On the other side, for the mode-truncated densities,
the final rise of the packing fraction happens almost simultaneously for 
$q_{\rm cut} < 0.1~q_{\rm max}$ for all cut-offs $\lambda_{\rm cut}$ under 
consideration. This means that, as long as $q_{\rm cut}/q_{\rm max} > 0.1$, 
the clusters remain well separated.
These differences are washed out on the coarser lattice. 
On the finer lattice the maximal cluster multiplicity (or the percolation 
threshold) for the all-scale density is found at much higher 
$q_{\rm cut}/q_{\rm max}$ 
value than for the mode-truncated density including the highest investigated 
$\lambda_{\rm cut}=634$ MeV. On the coarser lattice this
happens to the all-scale and the mode-truncated density
at the same $q_{\rm cut}/q_{\rm max}$. 

\begin{figure}[t]
\begin{center}
\begin{tabular}{cc}
\epsfig{file=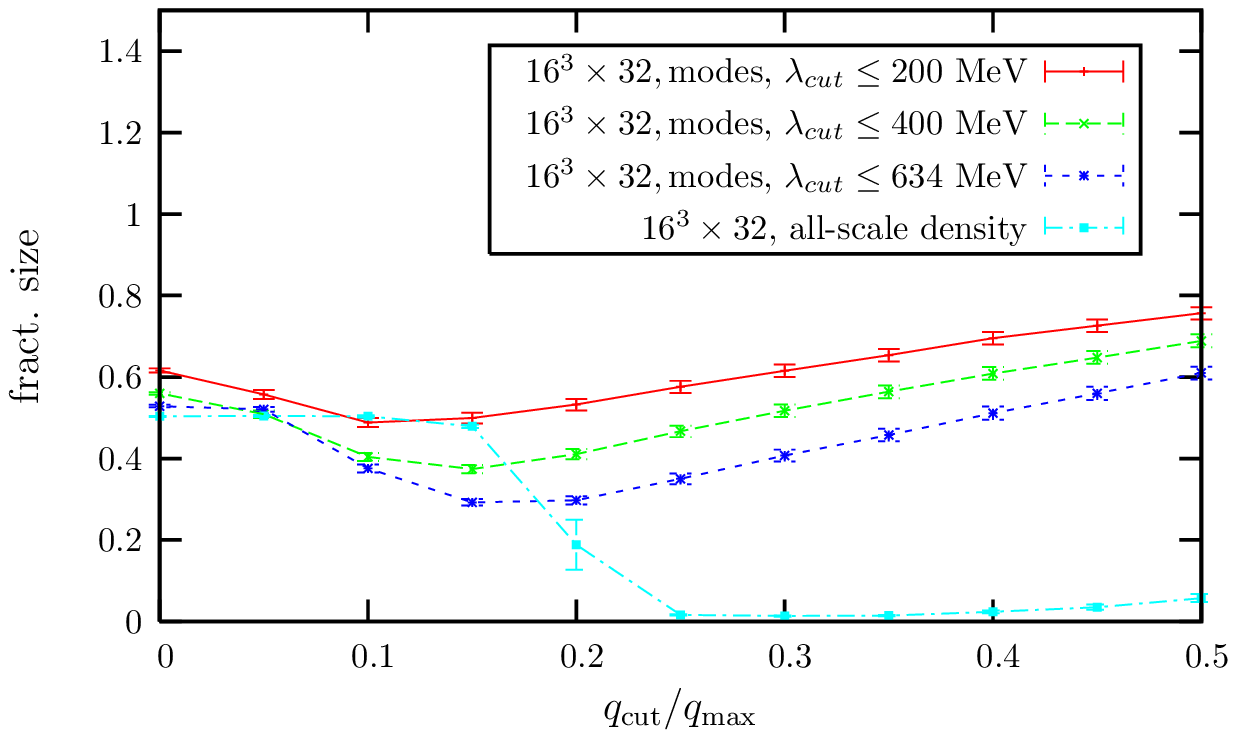,width=8cm}&
\epsfig{file=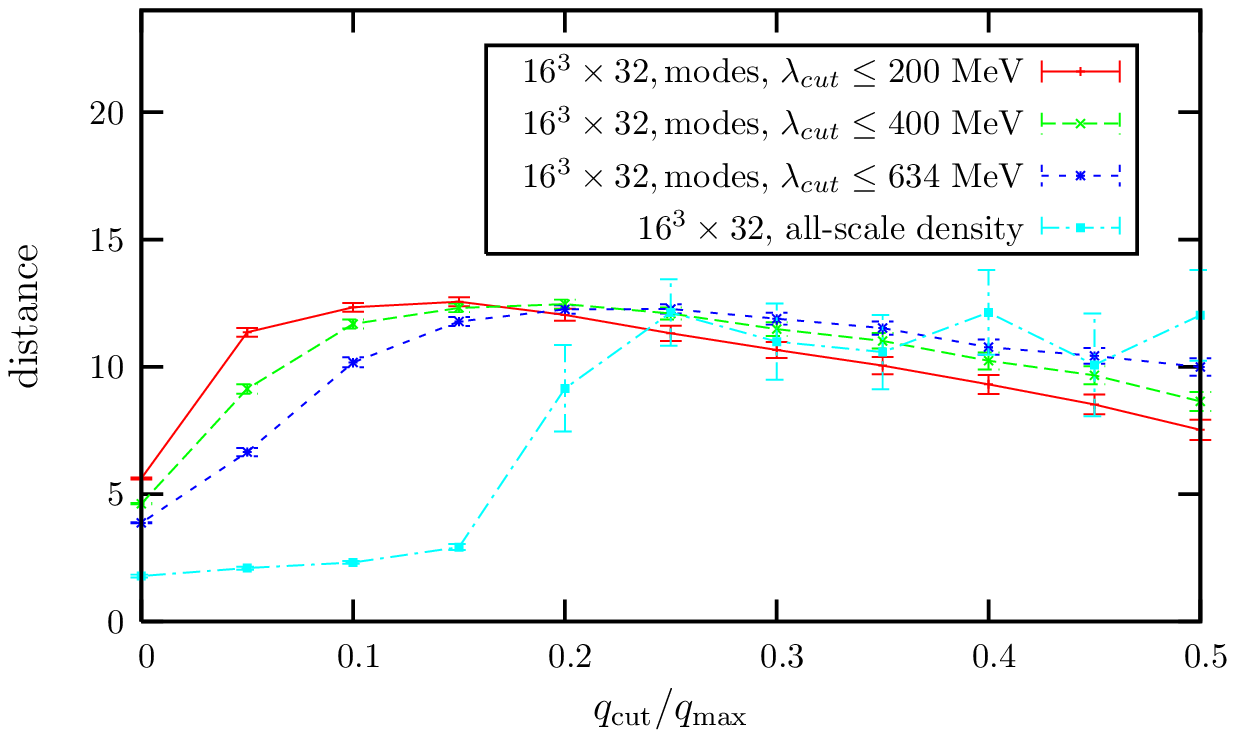,width=8cm}\\
(a) & (b)\\
\end{tabular}
\end{center}
\caption{Comparison of the cluster structure of the all-scale topological charge 
density with the mode-truncated density for three values of the cut-offs 
$\lambda_{\rm cut}$: (a) the fractional size of the largest cluster and 
(b) the distance between the two largest clusters in lattice units, both as 
function of $q_{\rm cut}/q_{\rm max}$. The data refers to the $16^3\times32$ 
lattice at $\beta=8.45$ only.}
\label{fig:cluster-full-vs-trunc-density_4}
\end{figure}
For the finer lattice only, in Fig.~\ref{fig:cluster-full-vs-trunc-density_4} 
the relative size of the largest cluster and the distance between the two largest 
clusters is plotted in dependence on $q_{\rm cut}/q_{\rm max}$. Each panel shows
this for various $\lambda_{\rm cut}$ and for the all-scale density.  
The behavior for the different cut-offs $\lambda_{\rm cut}$ of mode-truncated 
densities is similar. 
The main difference of the mode-truncated density compared to the non-truncated 
(all-scale) density
is that the relative size of the largest cluster remains relatively constant 
before and across the percolation threshold although the number and size of all 
clusters increases with decreasing $q_{\rm cut}$ before percolation sets in.
The largest cluster is always much bigger than the other clusters, and
this effect is the stronger the lower the cut-off $\lambda_{\rm cut}$ is.
It can be clearly seen that the distance $d(a,b)$ between the two largest 
clusters $a$ and $b$ of the mode-truncated density 
does not drop rapidly to 2 lattice 
spacings immediately after the onset of percolation. This is because these clusters 
remain well-separated even at $q_{\rm cut}/q_{\rm max} < 0.1$.

All these observations point towards the conclusion that the mode-truncated density 
allows to define $q$-clusters of topological charge that, although they also begin 
to percolate at some (relatively low) height (cut-off $q_{\rm cut}$), remain 
well-separated even at this level of the density. One of the clusters is much larger than the
remaining ones. These properties they have in common with the 
$R$-clusters that will be defined on the basis of the local (anti-)selfduality 
$R(x)$ in Section~\ref{sec:5.3}.  
Thus, these clusters allow an interpretation close to the traditional 
``lumpy'' instanton or caloron gas picture, 
however with the tendency to coalesce into larger objects of 
same-sign topological charge density (``instanton clumping'').
At higher $\lambda_{\rm cut}$ an attractive correlation between opposite
charge clusters appears.

Given this cluster interpretation of the mode-truncated topological density,
we return to the question whether the chirality $X_n(x)$ of the $n$-th 
non-zero mode is locally correlated with the topological charge density.
To answer this question we consider the correlation function between the density 
$q_{\lambda_{\rm cut}}(y)$ and the local chirality $X(y)$ over some distance 
$r$, $C_{qX}(r;\lambda_{\rm cut},n)$ defined as
\begin{equation}
C_{qX}(r;\lambda_{\rm cut},n) =  \frac{ \sum_{x,y} 
\langle~q_{\lambda_{\rm cut}}(x)~X_n(y)~\rangle~\delta(r-|x-y|)}
               { \sum_{x,y} \delta(r-|x-y|) } \; . 
\label{eq:qX-correlator}
\end{equation}

\begin{figure}[t]
\begin{center}
\hspace*{-1.0cm}\epsfig{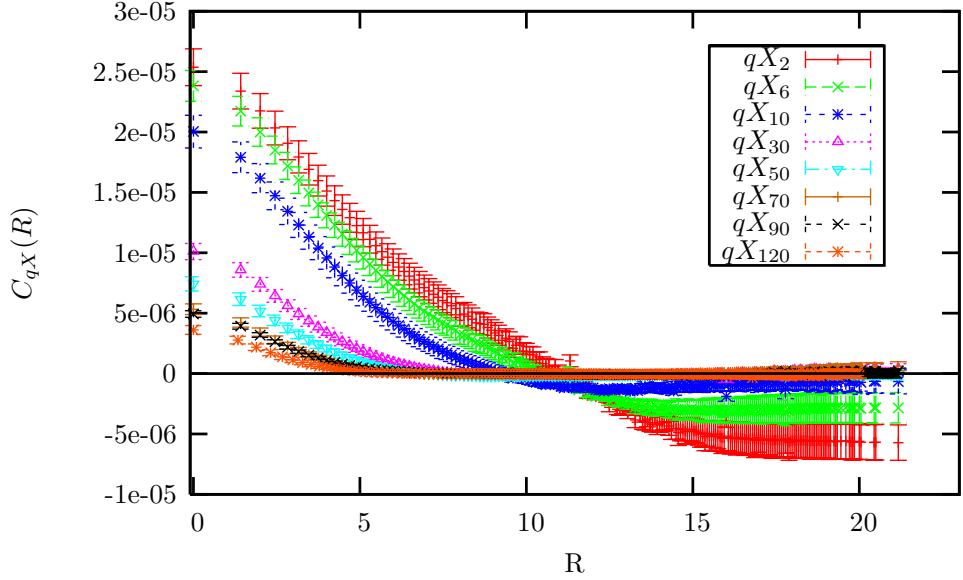}
\end{center}
\caption{The correlation function between the mode-truncated topological charge
density for $\lambda_{\rm cut}=100$ MeV and the local chirality of selected
non-zero modes from the lowest one to the 120th mode. 
The correlation function is an average over the 37 configurations
with $Q=0$ on the $16^3\times32$ lattice generated at $\beta=8.45$.}  
\label{fig:qXcorrelator}
\end{figure}

The result is shown in Fig.~\ref{fig:qXcorrelator} for $\lambda_{\rm cut}=100$ MeV.
The local chirality that enters the correlator in the figure is evaluated
for the 2nd, the 6th, the 10th, the 30th, the 50th etc. up to the 120th
non-zero mode. We keep in mind that this mode-restricted topological charge
density is influenced (on average) by the lowest 9 or 10 pairs of non-zero
modes. We see that the positive correlation of the (collective) topological
charge density on one hand with the local chirality of individual overlap 
modes up to the 10th mode on the other hand 
is strong and ranges over up to ten 
lattice spacings (similar to the distance where the correlation function $C_{qq}$
of $q_{\lambda_{\rm cut}}$ changes the sign. At larger distance the local 
chirality gets anticorrelated to the mode-truncated topological charge density. 
Although weaker and over shorter distance, the 30th and higher modes are still 
positively correlated with the sign of the truncated topological charge density. 
For these higher modes the anticorrelation is absent at larger distance.

\section{The infrared field strength tensor}
\label{sec:UVfilterF}

\subsection{Basic idea}
\label{sec:5.1}

We have considered in Section~\ref{sec:topdensity} the topological 
charge density as represented by the overlap operator according to 
Eqs.~(\ref{eq:eq-q}) or (\ref{eq:qlambdacut}). We have seen that 
the structure of the mode-truncated topological charge density
strongly depends on the number of included lowest eigenmodes. 
Restricting to fewer modes is tantamount to an ultraviolet filtering 
applied to the topological density. Our point of view as explained in 
Section~\ref{sec:introduction} is that the low-lying modes represent 
the physically relevant degrees of freedom for hadronic physics. 
Thus, filtering should be suitable to highlight the {\it important} structures.
Here we briefly discuss another filtering method similar to what  
was proposed by Gattringer~\cite{Gattringer:2002gn} and is suggested 
by the newly discussed representation of the field strength tensor through
the overlap Dirac operator~\cite{Liu:2006wa,Liu:2007hq}.
The main purpose is to get an ultraviolet filtered, infrared field strength 
tensor. This would allow to assess the local degree of (anti-)selfduality.
The local selfduality or antiselfduality of a gauge field is an important 
feature in 
all semiclassically motivated vacuum models, and the search for it in 
generic lattice configurations may play a similar (or superior) 
role compared to the search for lumps of topological charge, which was
the main strategy of access before.

The starting point is the observation that in the continuum the square of 
the Dirac operator $D$, projected with the help of a combination of 
$\gamma_{\mu}$ matrices, represents the field strength tensor
\begin{equation}    
F_{\mu\nu}(x) = -\frac{1}{4}~{\rm tr}_{\rm Dirac} \left(\sigma_{\mu\nu}~D^2(x) \right) \; .
\end{equation}
Here ${\rm tr}_{\rm Dirac}$ is the trace taken over spinor indices.
In Ref.~\cite{Gattringer:2002gn}
the chirally improved Dirac operator~\cite{Gattringer:2000qu} was used 
to evaluate this in the simplified local form
\begin{eqnarray}
F^a_{\mu\nu}(x) & \propto & \sum_{\lambda} \lambda^2 f^a_{\mu\nu}(x,x|\lambda) \; ,
\nonumber \\
f^a_{\mu\nu}(x,x|\lambda) & = & -\frac{i}{2} \psi^{\sigma~c~*}_{\lambda}(x)
         \sigma_{\mu~\nu}^{\sigma~\sigma^{\prime}}
      T^a_{c~c^{\prime}}~\psi^{\sigma^{\prime}~c^{\prime}}_{\lambda}(x)  \;
\end{eqnarray}
with the color generator $T^a$ in the fundamental representation.
Zero modes do not contribute to this representation of the field 
strength. The ultraviolet filtering consists now in including only a 
certain number of low-lying modes in this spectral representation.  

Here we explore the application of this formula to the eigenmodes of the 
overlap Dirac operator $D(0)$ and compare the revealed structure with that
shown by the ultraviolet filtered topological density.
Obviously, a normalization factor remains undetermined, in particular because 
only a subset of low-lying modes will be included in the filter. 

In two recent 
papers~\cite{Liu:2006wa,Liu:2007hq}, a representation of field strength and 
action density has been discussed, in a way analogous to 
Eqs. (\ref{eq:eq-q}) and (\ref{eq:qlambdacut}), which give
the representation of the topological charge density {\it directly} 
in terms of the overlap operator.

\subsection{Local (anti-)selfduality of the infrared field strength tensor}
\label{sec:5.2}

\begin{figure}[t!]
\begin{center}
\begin{tabular}{cc}
\hspace*{-1cm}\epsfig{file=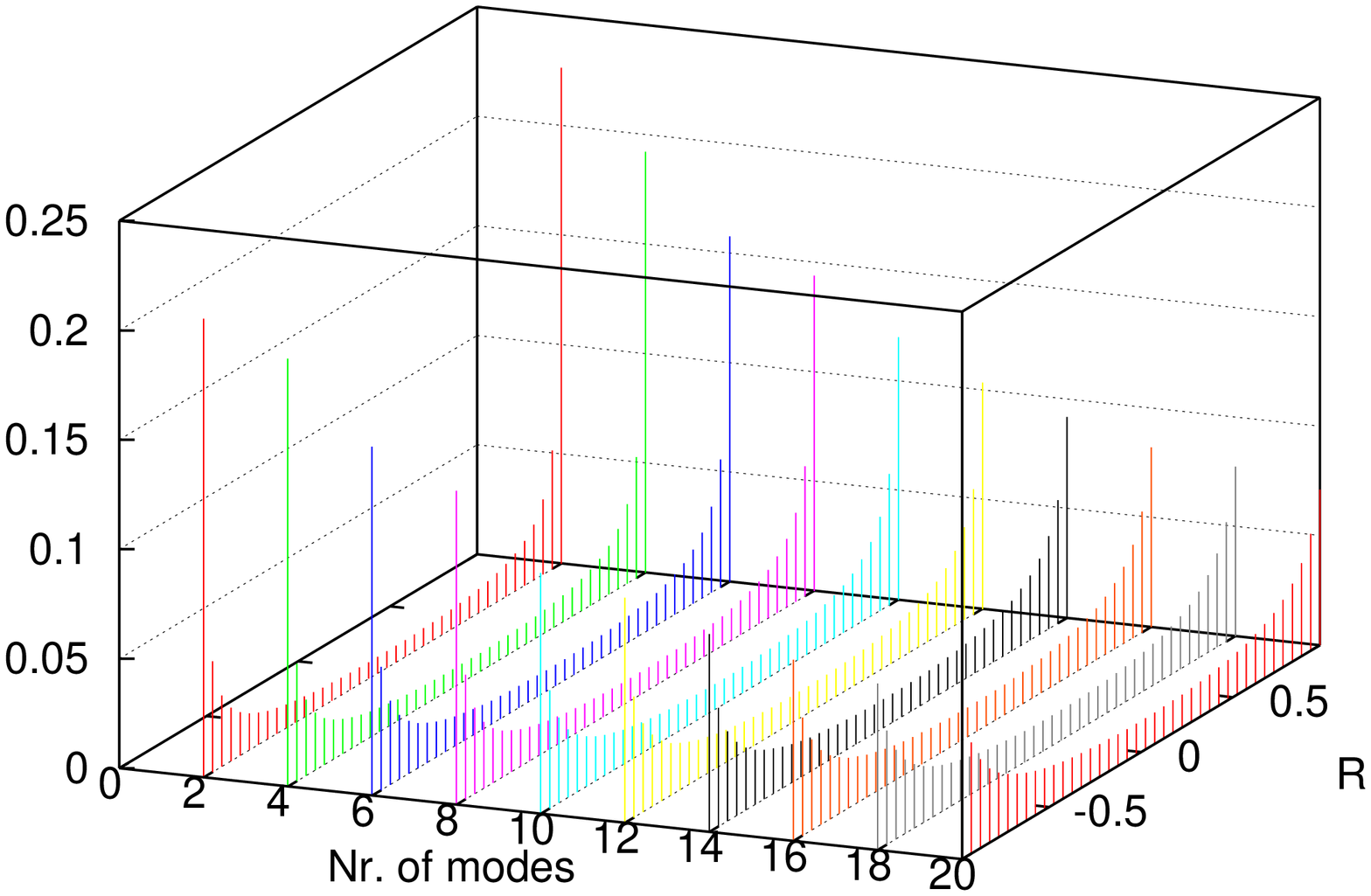,width=7cm,angle=-90}&
\hspace*{-1.5cm}\epsfig{file=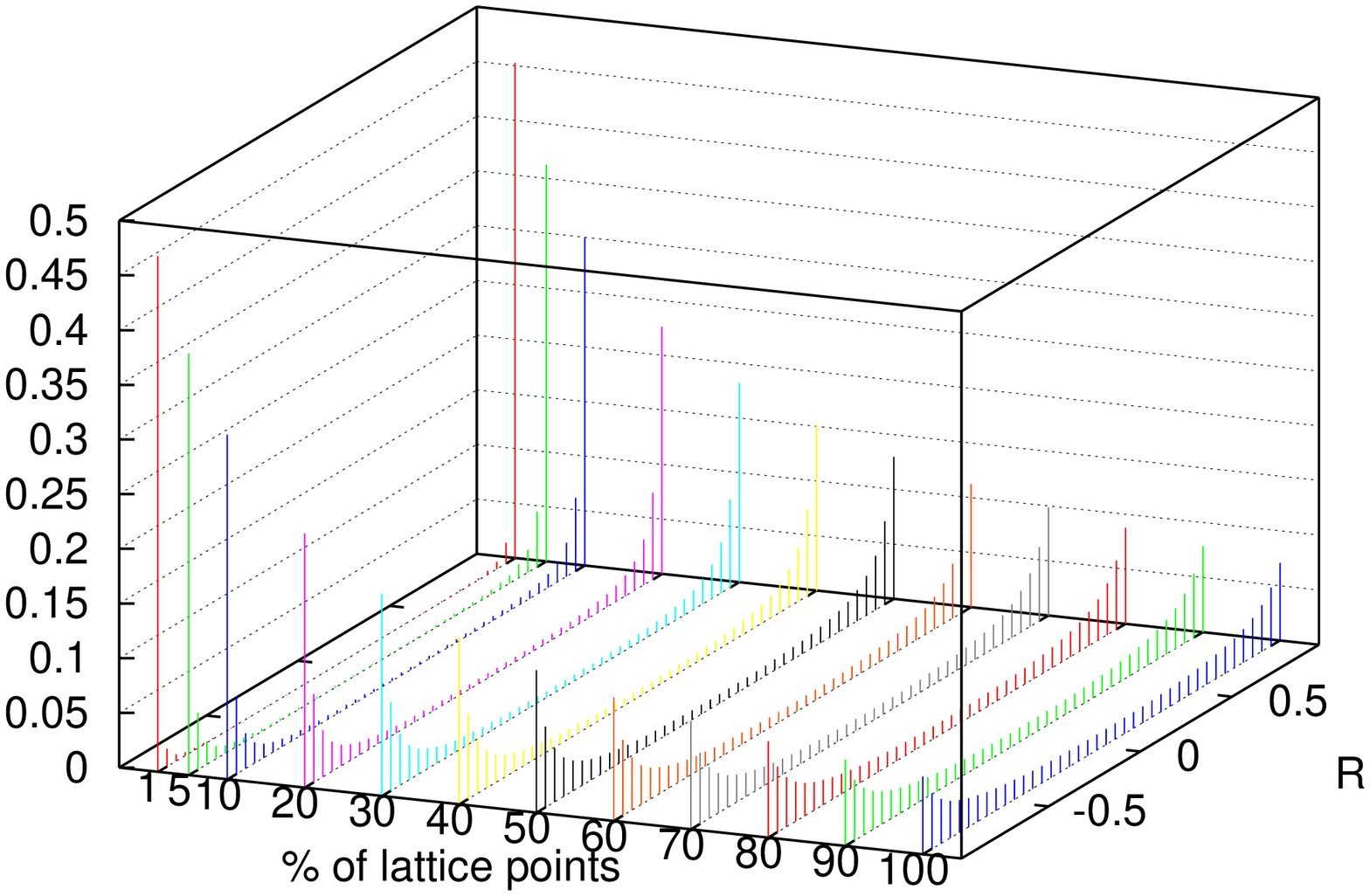,width=7cm,angle=-90}\\
(a) & (b)\\
\end{tabular}
\end{center}
\caption{Normalized histograms with respect to the local (anti-)selfduality 
of the field strength tensor in the $Q=0$ subsample consisting of 37 
configurations generated on the $16^3\times32$ lattice at $\beta=8.45$, 
(a) taken over 100 \% of the lattice sites, but depending on the number of non-zero 
modes (2 - 20) included in the filter, 
(b) applying various cuts specifying the infrared action density for the case of 
20 eigenmodes included in the filter.}
\label{fig:Gattringer-histograms}
\end{figure}

We now study the contribution of individual non-zero modes to the filtered
field strength tensor. 
The normalization of the infrared field strength $F^a_{\mu\nu}(x)$ 
does not come automatically. It could be obtained requiring that the 
infrared topological charge density 
\begin{equation}
\tilde{q}_{\rm IR}(x) \propto \Tr \left( F_{\mu\nu}(x)~\Tilde{F}_{\mu\nu}(x) \right) 
\end{equation}
with the dual field strength 
$\Tilde{F}_{\mu\nu} = \frac{1}{2} \epsilon_{\mu\nu\rho\sigma}~F_{\rho\sigma}$
fits the ultraviolet filtered topological density $q(x)_{\lambda_{\rm cut}}$ 
(for a suitable cut-off $\lambda_{\rm cut}$). The latter can be directly 
derived from the overlap operator following Eq. (\ref{eq:qlambdacut}). 
A less elaborate normalization for $Q \ne 0$ configurations
could be that the total topological charge obtained 
from $\Tr \left( F_{\mu\nu}(x)~\Tilde{F}_{\mu\nu}(x) \right)$ 
should reproduce the index $Q$ of $D(0)$ for the given configuration. 

One can define analogously the infrared action density 
\begin{equation}
\tilde{s}_{\rm IR}(x) \propto \Tr \left( F_{\mu\nu}(x)~F_{\mu\nu}(x) \right) 
\end{equation}
in order to compare it with $\tilde{q}_{\rm IR}(x)$. With both infrared 
densities at hand, one can ignore the quest for normalization. 
It is now possible to turn the interest to the local (anti-)selfduality based on  
a ratio similar to Eq. (\ref{eq:r_for_X}),
\begin{equation}
r(x) = \frac{ \tilde{s}_{\rm IR}(x) - \tilde{q}_{\rm IR}(x) }
            { \tilde{s}_{\rm IR}(x) + \tilde{q}_{\rm IR}(x) } \; .
\label{eq:localselfdual}
\end{equation}
This is converted to $R(x)$, 
\begin{equation}
R(x) = \frac{4}{\pi} \arctan \left(\sqrt{r(x)}\right) - 1 \in [-1,+1] \; ,
\label{eq:arctanR}
\end{equation}
analogously to Eq. (\ref{eq:arctan}).

In Fig.~\ref{fig:Gattringer-histograms} histograms are presented with respect 
to $R(x)$ applying various cuts 
specifying the local action density $\tilde{s}_{\rm IR}(x)$. 
Firstly, sharp peaks at $R= \pm 1$ are observed that become weaker with the 
inclusion of more low-lying pairs of non-zero modes. We notice that even with 
$10$ pairs included these peaks are still visible. They become more pronounced 
again when one focusses on part of the lattice sites applying a cut 
with respect to the infrared action density $\tilde{s}_{\rm IR}(x)$.

\subsection{Cluster analysis of selfdual and antiselfdual domains}
\label{sec:5.3}

\begin{figure}[t]
\begin{center}
\begin{tabular}{cc}
\epsfig{file=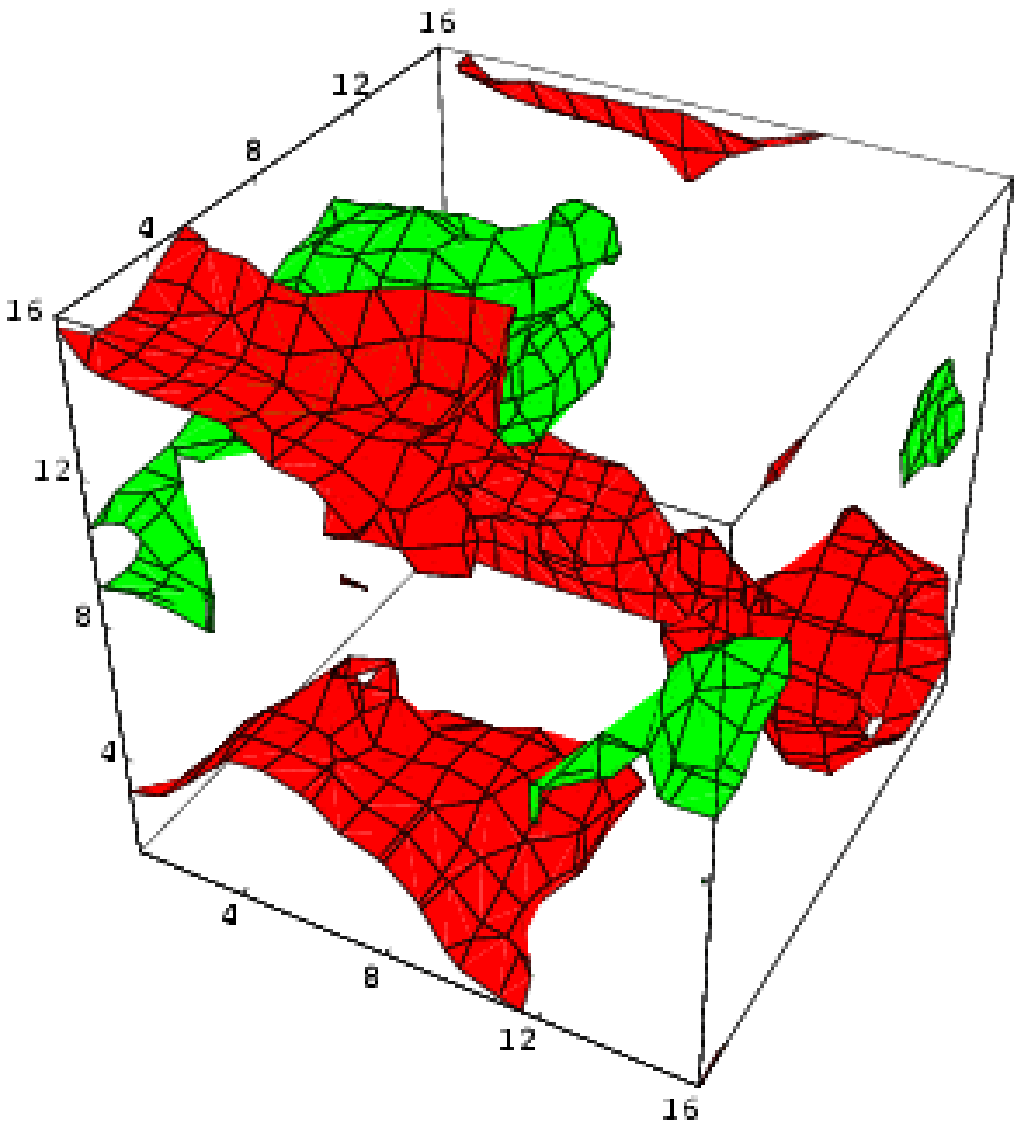,angle=0,width=7cm}&
\epsfig{file=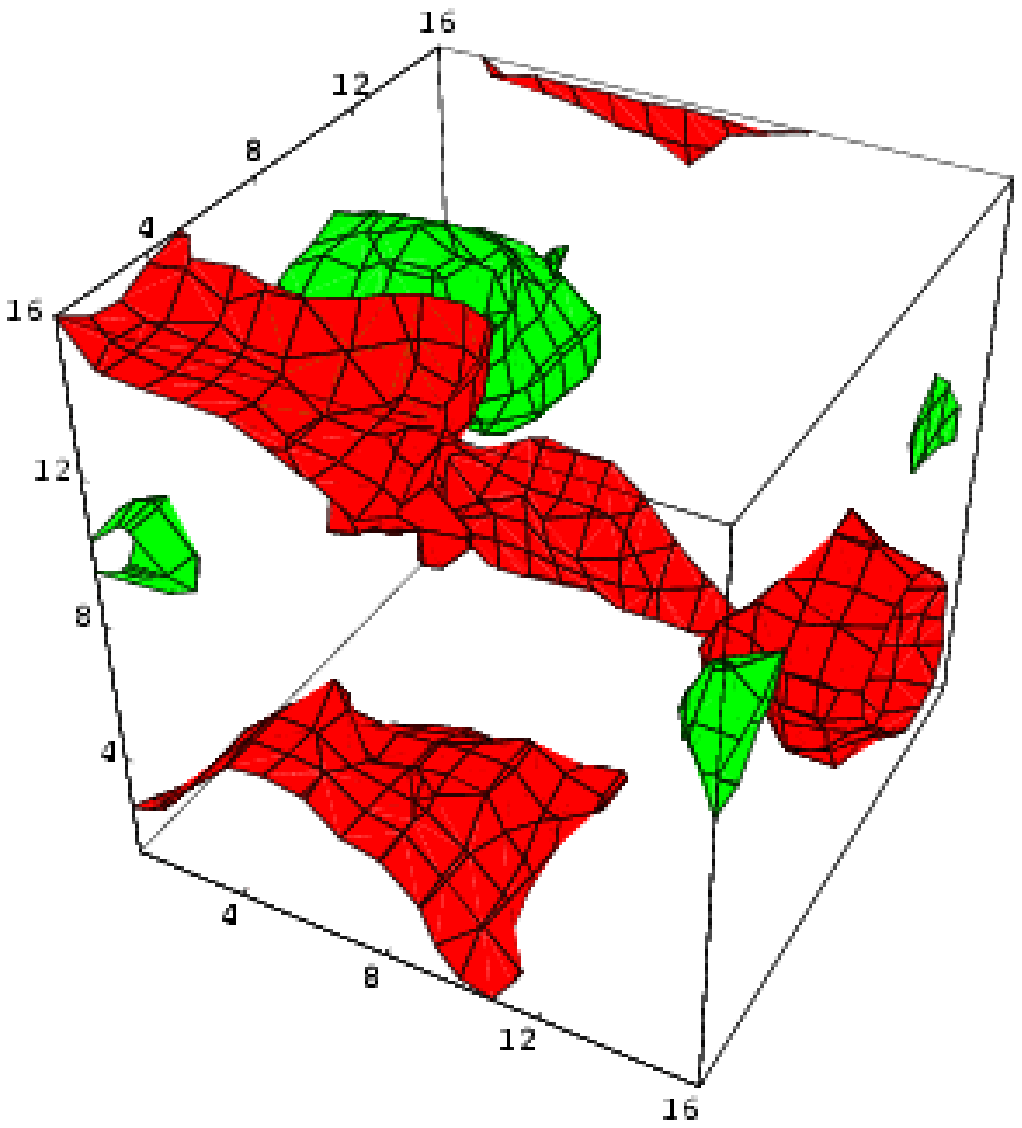,angle=0,width=7cm}\\
(a) & (b)\\
\epsfig{file=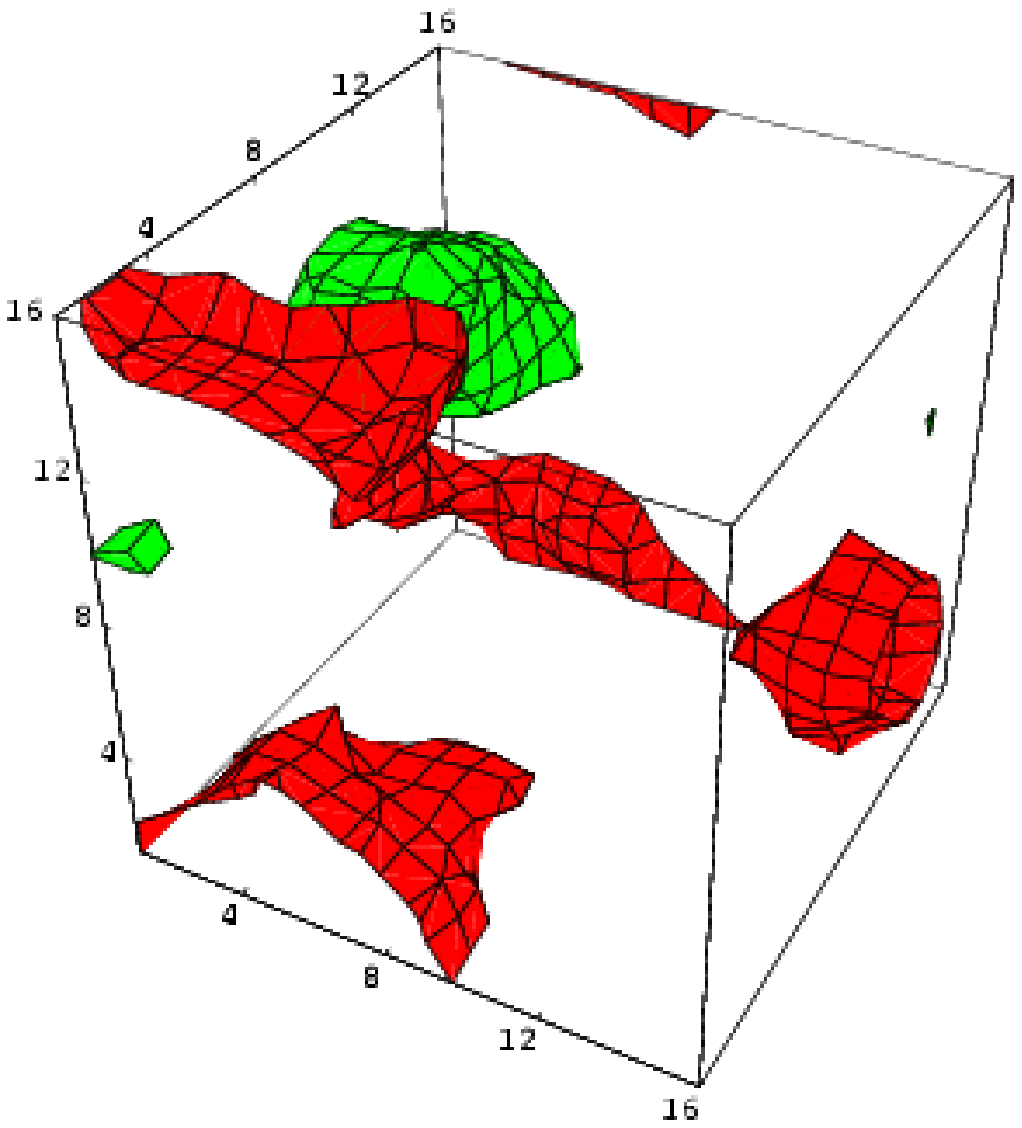,angle=0,width=7cm}&
\epsfig{file=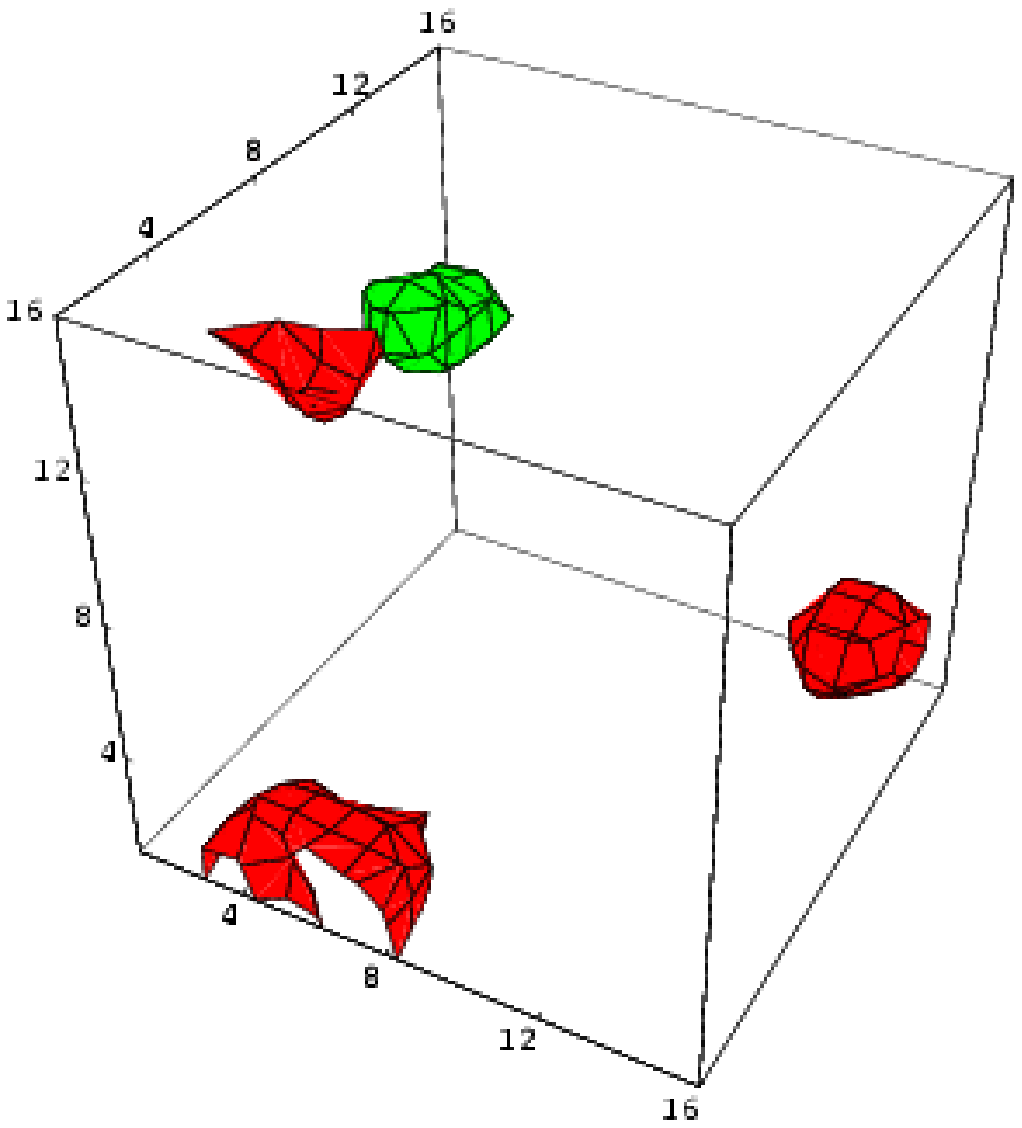,angle=0,width=7cm}\\
(c) & (d)
\end{tabular}
\end{center}
\caption{Isosurface plots of $R(x)$ in the order of decreasing tolerance 
with respect to deviations from perfect (anti-)selfduality, 
(a) $R_{\rm cut}=0.97$, (b) $R_{\rm cut}=0.98$, (c) $R_{\rm cut}=0.99$ 
and (d) $R_{\rm cut}=0.999$.
The figure shows the timeslice $t=6$ of the typical $Q=0$ configuration 
on the $16^3\times32$ lattice generated at $\beta=8.45$. Red and green 
surfaces enclose regions where antiselfduality or selfduality is better 
fulfilled than in the rest of the volume. The filter is using the 5 lowest 
pairs of non-zero eigenmodes.}    
\label{fig:R-isosurface10}
\end{figure}

Since the close neighborhood of $R= \pm 1$ of these histograms corresponds 
to lattice points where the infrared field strength tensor is nearly 
(anti-)selfdual, it is interesting to see whether they 
are completely disconnected or form connected regions in space-time.
We construct isosurface plots 
of $|R(x)| = R_{\rm cut} = 1 - \varepsilon$  
demonstrated in Fig.~\ref{fig:R-isosurface10} for a filter 
including the lowest 5 pairs of non-zero modes. The bubbles denote regions 
where the infrared field strength tensor is more (anti-)selfdual inside 
than outside. Red refers to antiselfdual regions, green to selfdual ones.
Similar clusters have been seen when we have plotted clusters of the 
ultraviolet filtered topological charge density.

\begin{figure}[t]
\begin{center}
\begin{tabular}{cc}
\epsfig{file=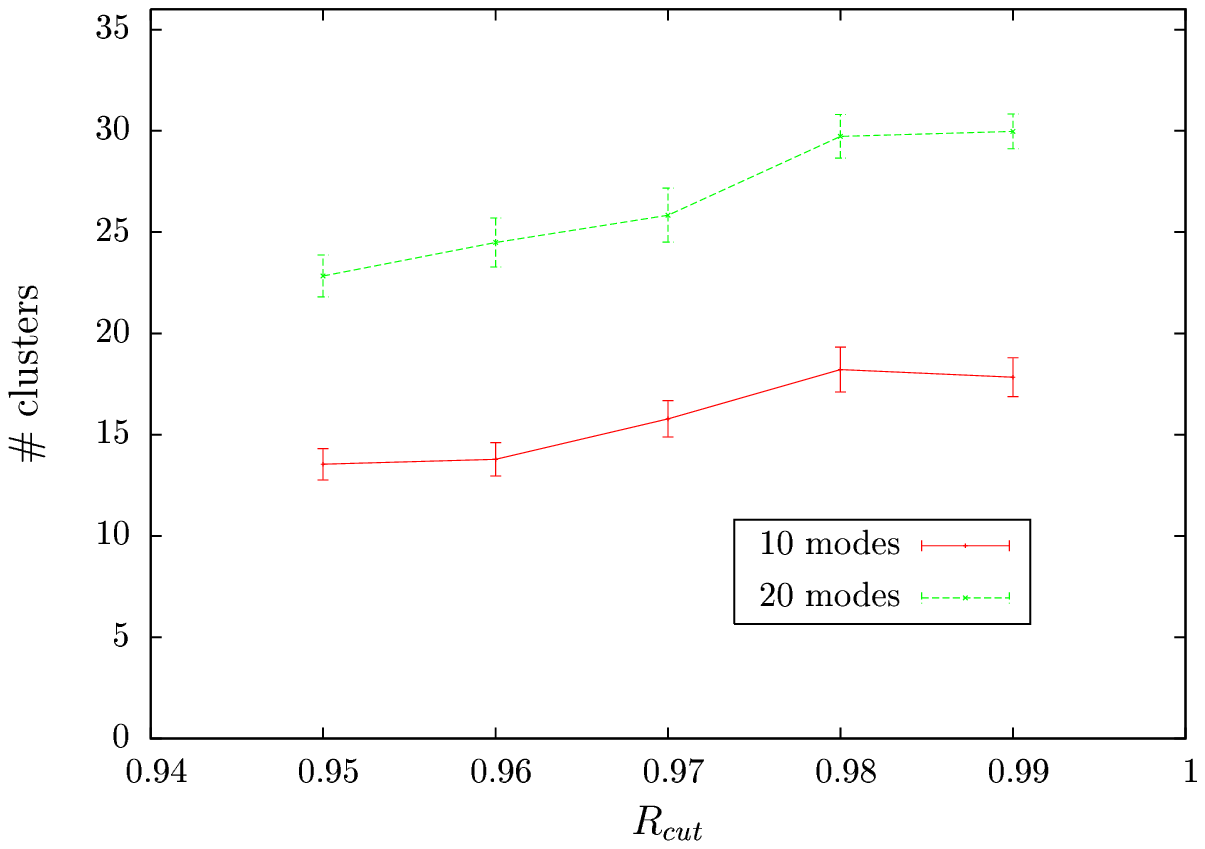,width=8cm}&
\epsfig{file=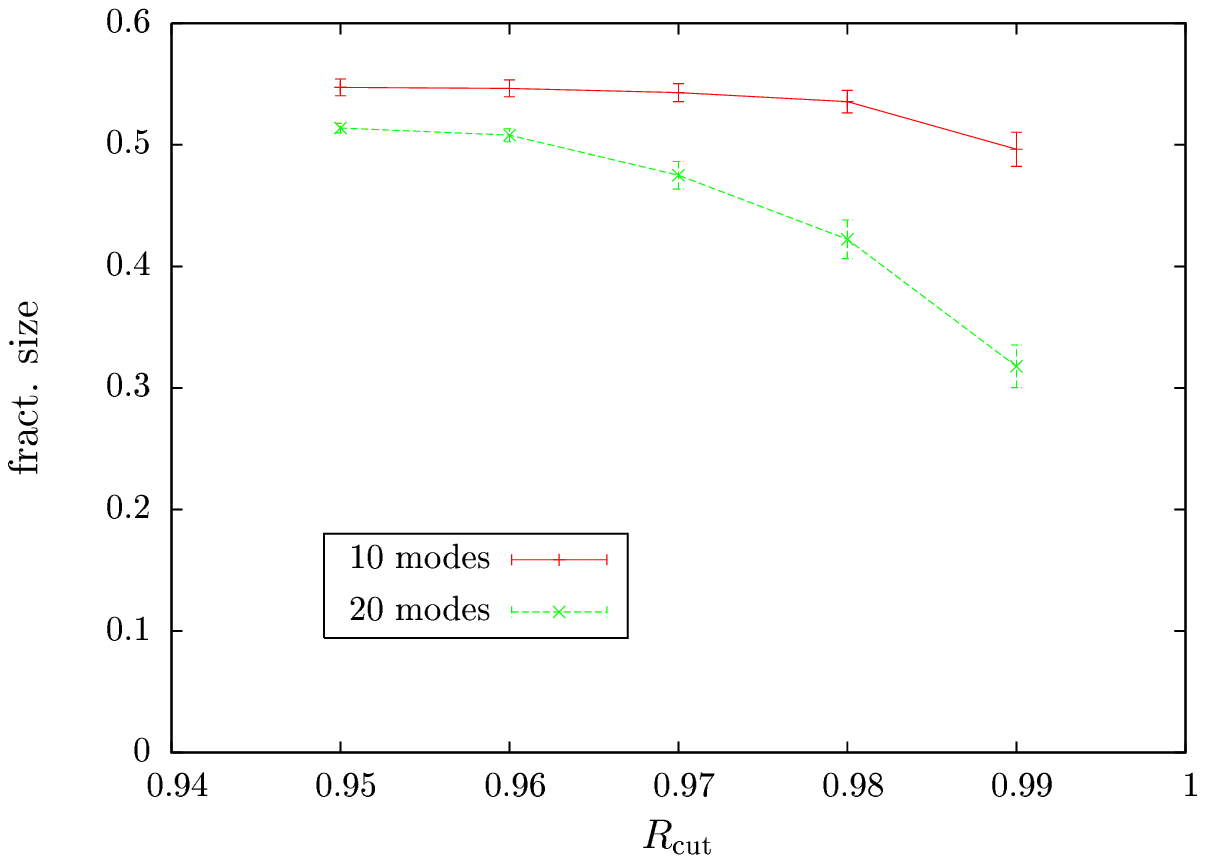,width=8cm}\\
(a) & (b)\\
\epsfig{file=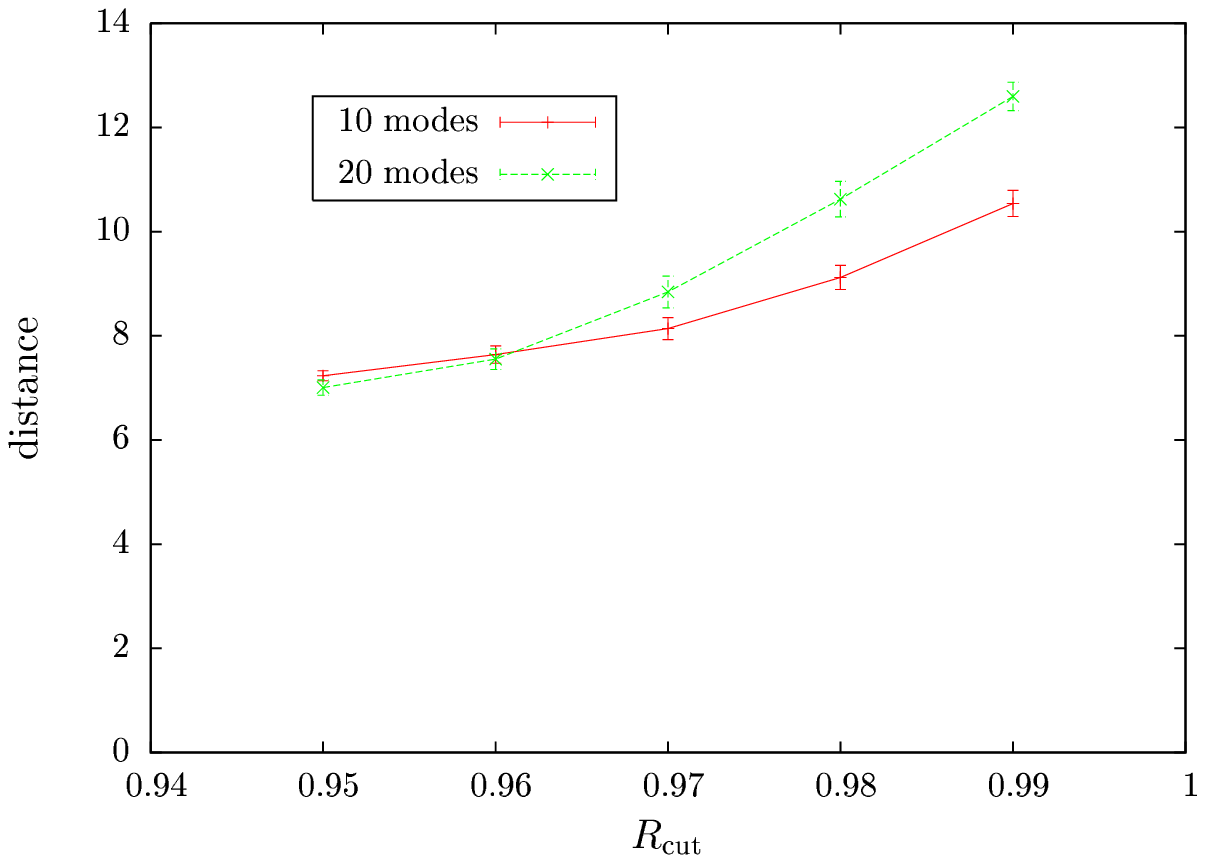,width=8cm}&
\epsfig{file=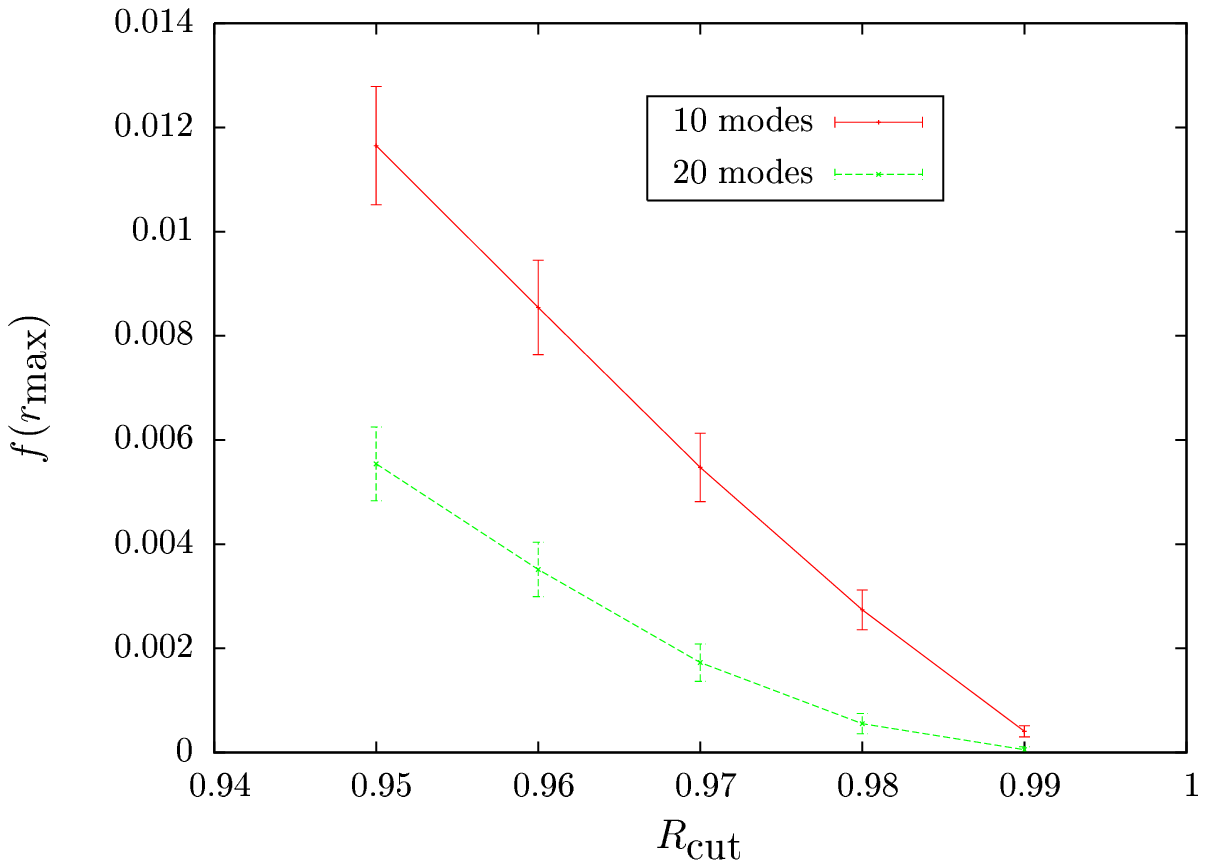,width=8cm}\\
(c) & (d)
\end{tabular}
\end{center}
\caption{$R$-cluster analysis using a running cut-off $R_{\rm cut}$
on the basis of the filtered action density $s_{\rm IR}$ and 
topological charge density $q_{\rm IR}$ used to define $R(x)$.
The $R_{\rm cut}$ dependence is shown of (a) the total number of 
separate clusters, (b) the size of the largest cluster relative 
to all clusters, (c) the distance between the two largest clusters 
and (d) the connectivity (see text). The data is averaged over the 
$16^3\times32$ lattice ensemble generated at $\beta=8.45$.}
\label{fig:R-cluster}
\end{figure}

In order to get a more quantitative characterization of the vacuum, it is useful
to perform a cluster analysis on the basis of the quantity $R(x)$. 
In Fig.~\ref{fig:R-cluster} we show how the cluster composition changes with 
varying $R_{\rm cut}$. 
When percolation begins close to $R_{\rm cut}=0.98$ the number of separate 
$R$-clusters starts to slightly decrease 
with decreasing $R_{\rm cut}$. This means that allowing more tolerance with 
respect to violation of (anti-)selfduality inside the clusters 
leads to the growth and 
coalescence of already identified clusters, rather than to the detection of
further, isolated clusters. 
One can see that the distance between the two largest clusters remains relatively 
large (decreasing only from 12 to 7), meaning that a lot of space-time remains 
unoccupied by the almost (anti-)selfdual $R$-clusters, even in the percolating 
regime signaled by the connectivity becoming $f(r_{\rm max}) \ne 0$. 
The unoccupied regions are regions where the infrared field strength 
tensor is violating selfduality or antiselfduality more
than the given tolerance $R_{\rm cut}$ permits.

It turns out again that one largest cluster exists, which is rather large already 
before the onset of percolation, occupying $40$ \% of the total volume 
of all clusters. 
This $R$-cluster rapidly becomes the absolutely dominating one with respect to 
the volume (more than 50 \%) after percolation has set in. 
This fact (topological ``clumping'') remains unexplained in all instanton/caloron 
models which are based on uncorrelated (anti-)selfdual objects of $Q=\pm 1$,
eventually taking the attractive correlation of {\it opposite} charge cluster 
into account (that is actually seen at higher resolution $\lambda_{\rm cut}$).

\begin{figure}[t]
\begin{center}
\begin{tabular}{cc}
\hspace*{-0.5cm}\epsfig{file=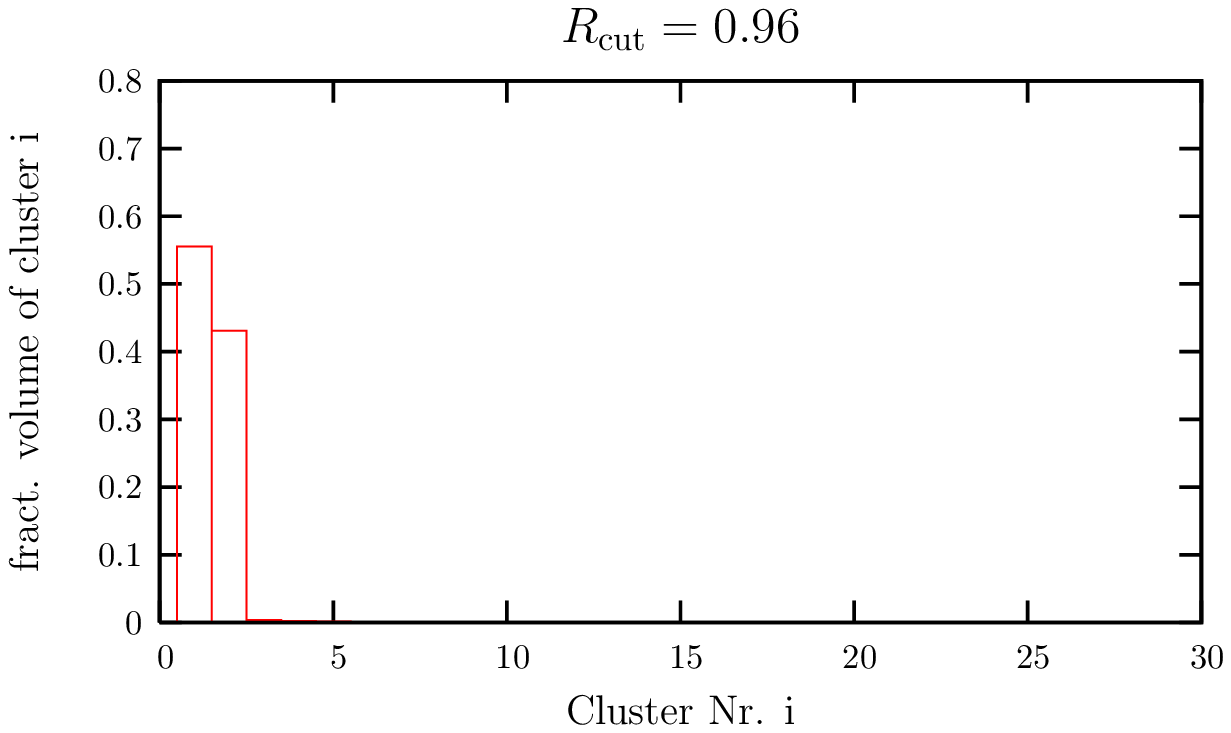,width=8.4cm}&
\hspace*{-0.3cm}\epsfig{file=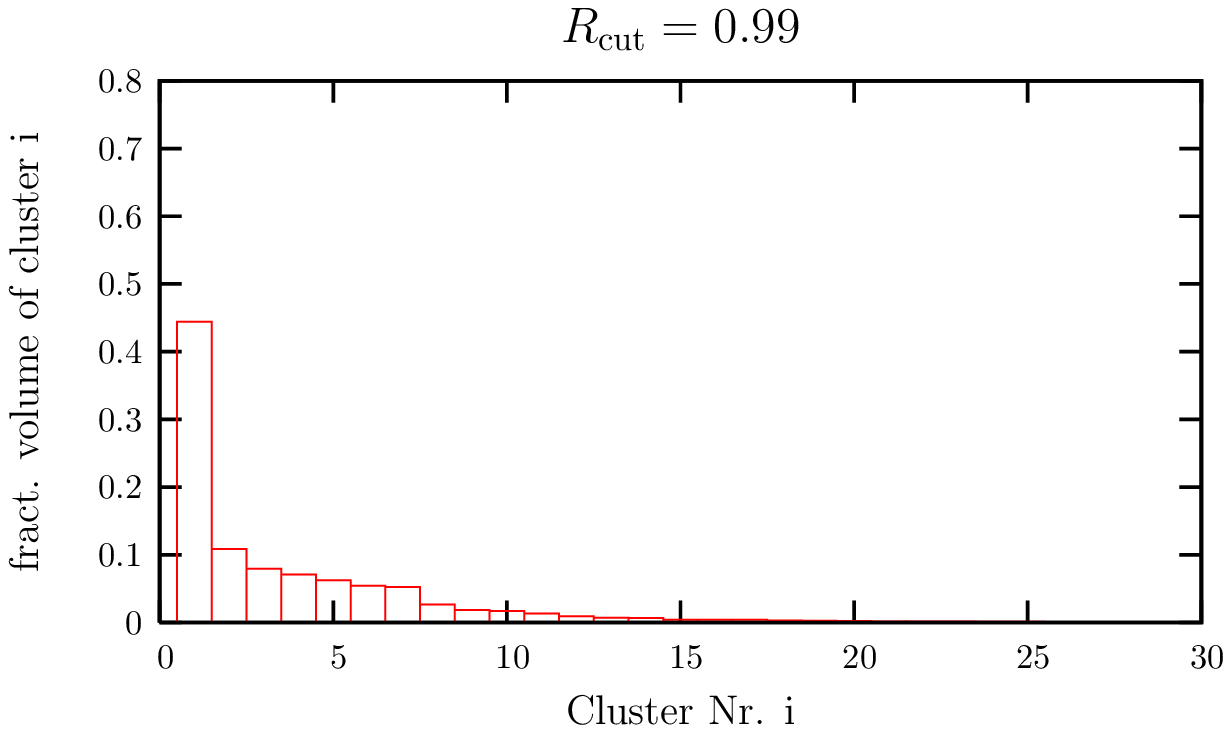,width=8.4cm}\\
(a) & (b)\\
\epsfig{file=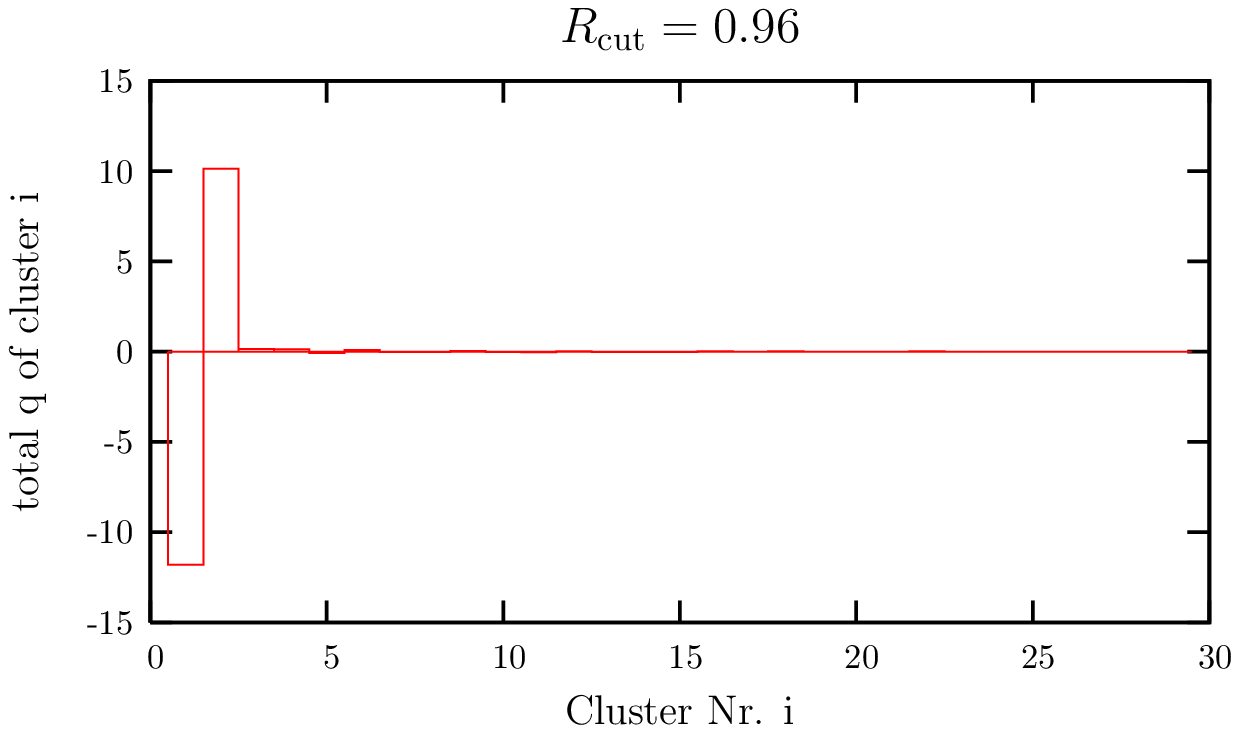,width=8cm}&
\epsfig{file=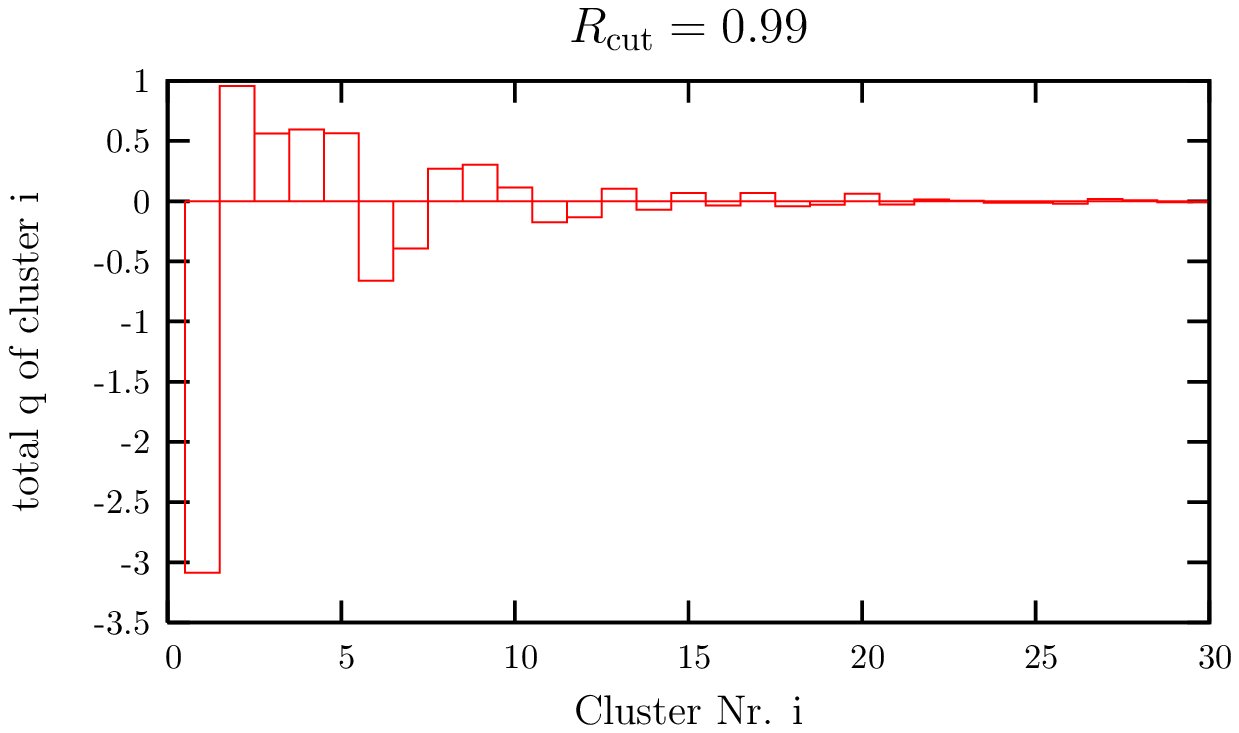,width=8cm}\\
(c) & (d)
\end{tabular}
\end{center}
\caption{Results of the $R$-cluster analysis at $R_{\rm cut}=0.96$ (left)
and $R_{\rm cut}=0.99$ (right) concerning the relative volume of all emerging 
clusters ((a) and (b)) and concerning the charge inside the clusters ((c) and (d)). 
The cluster charge is defined by the all-scale topological charge density 
(\ref{eq:eq-q}).
The distributions describe the same $Q=0$ configuration 
from the $16^3\times32$ lattice ensemble generated at $\beta=8.45$ already 
considered in Fig.~\ref{fig:mode-visualization} and \ref{fig:R-isosurface10}. }
\label{fig:R-composition}
\end{figure}

\begin{figure}[t!]   
\begin{center}
\begin{tabular}{cc}
\epsfig{file=pics/R-isosurface-nev-10-16x32-03_0126-0.99-6.ps,width=7.0cm}&
\hspace*{0.8cm}\epsfig{file=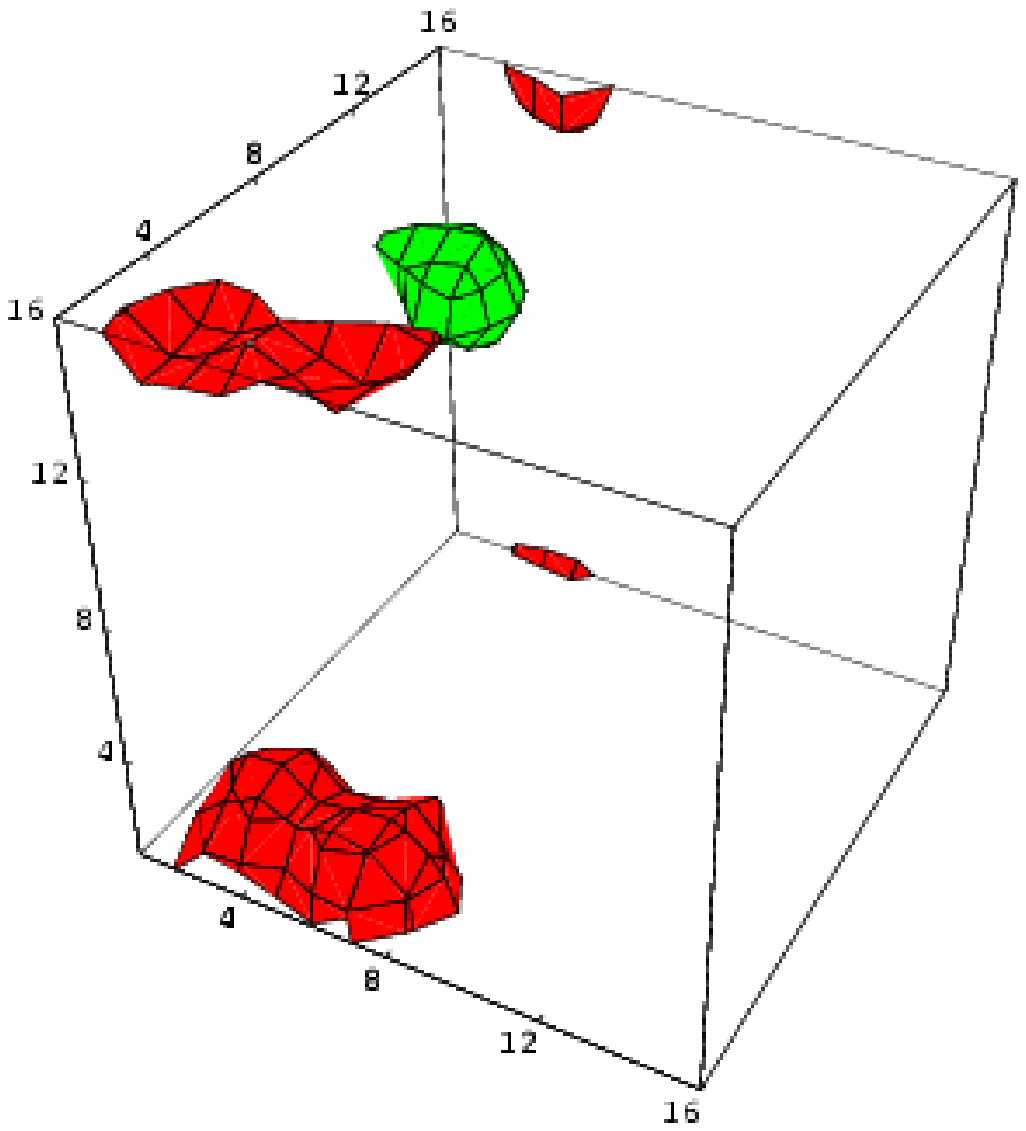,width=7.0cm}\\
(a) & (b)\\
\epsfig{file=pics/R-isosurface-nev-10-16x32-03_0126-0.98-6.ps,width=7.0cm}&
\hspace*{0.8cm}\epsfig{file=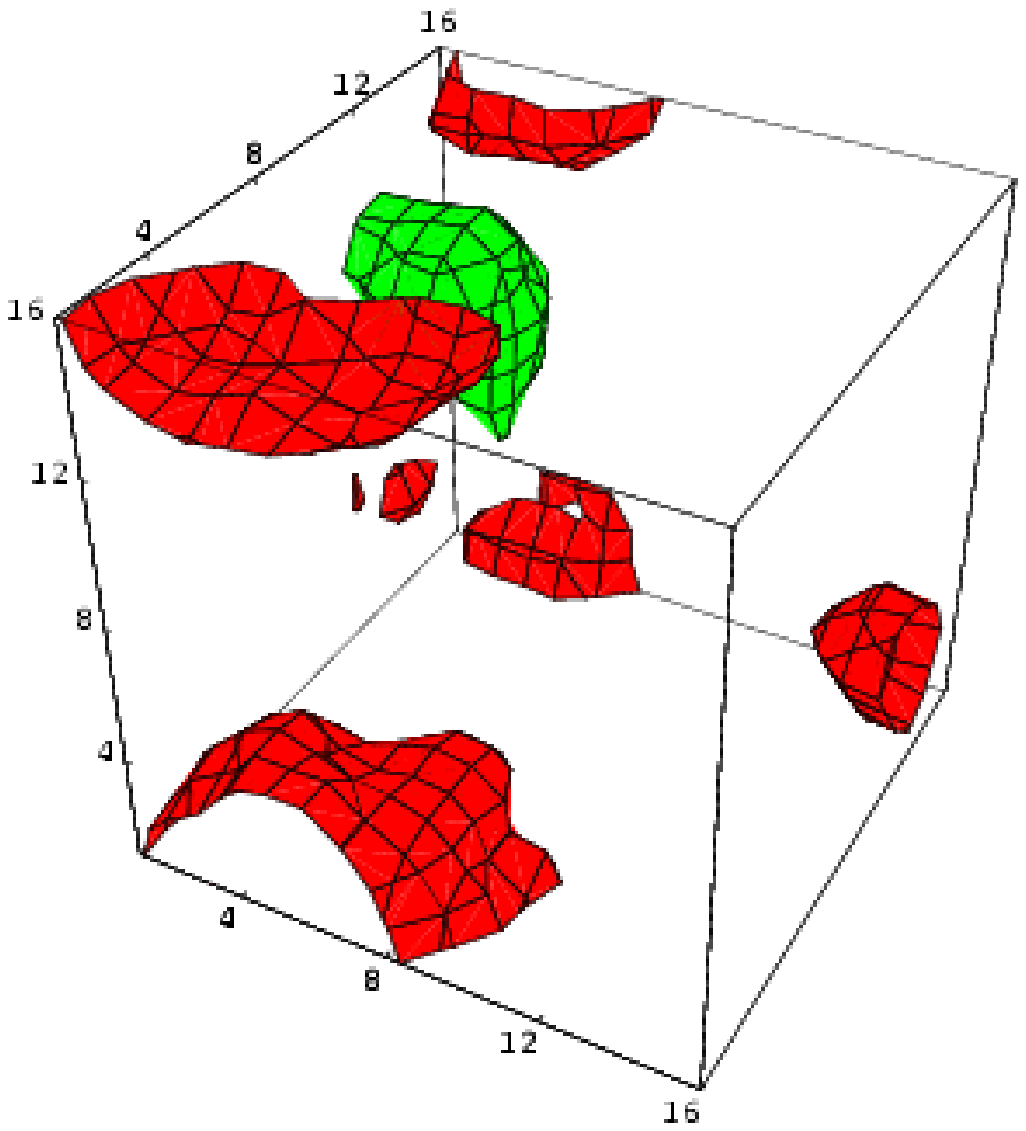,width=7.0cm}\\
(c) & (d)\\
\end{tabular}
\end{center}
\caption{Direct comparison between the isosurface plots with respect to $R$ (left:
with $R_{\rm cut}=0.99$ in (a) and $R_{\rm cut}=0.98$ in (c))
and the isosurface plots with respect to the truncated topological density 
$q_{\lambda_{\rm cut}}$ with $\lambda_{\rm cut}=200 $ MeV (9 pairs of non-zero
modes)
(right, with $q_{\rm cut}/q_{\rm max}=0.2$ in (b) 
and $q_{\rm cut}/q_{\rm max}=0.1$ in (d)).
All subfigures show the same time slice $t=6$ of the typical $Q=0$ configuration 
of the $16^3\times32$ ensemble at $\beta=8.45$. The similarity is  
found in all timeslices.\vspace*{1cm}}
\label{fig:compare-isosurfaces}
\end{figure}

\begin{figure}[t!]
\begin{center}
\begin{tabular}{ccc}
\hspace*{-0.8cm}
\includegraphics[width=5.5cm]{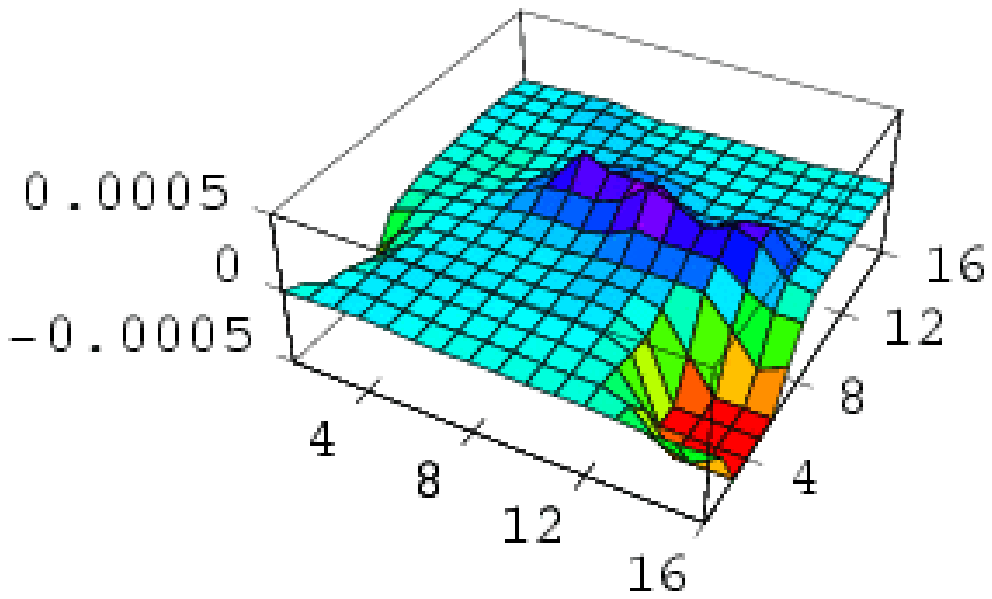}&
\vspace*{-0.4cm}\includegraphics[width=5.5cm]{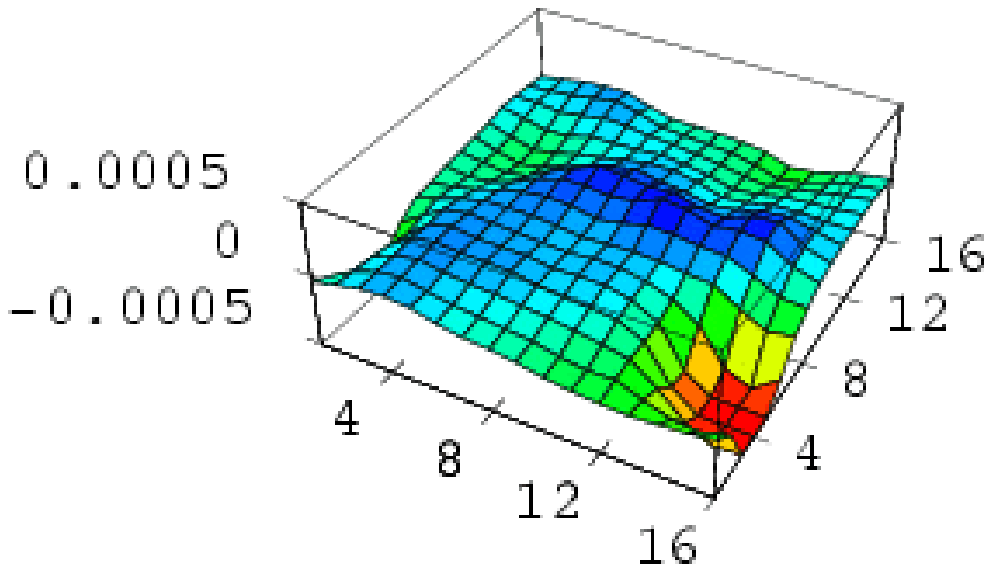}&
\includegraphics[width=5.5cm]{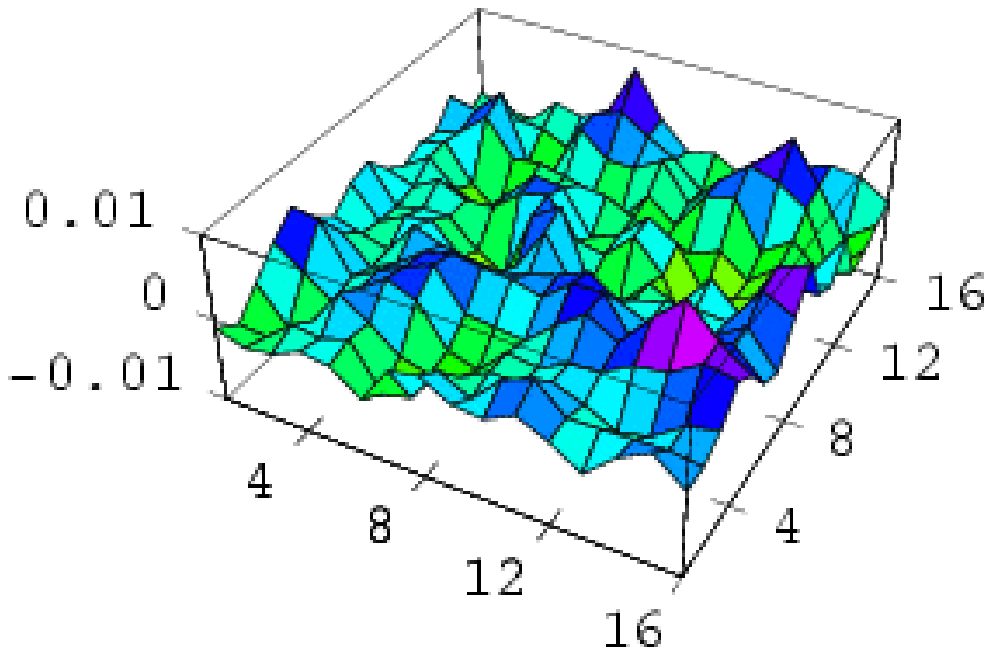}\\
(a) & (b) & (c)\\
\multicolumn{3}{c}{
\begin{tabular}{cc}
\includegraphics[width=5.5cm]{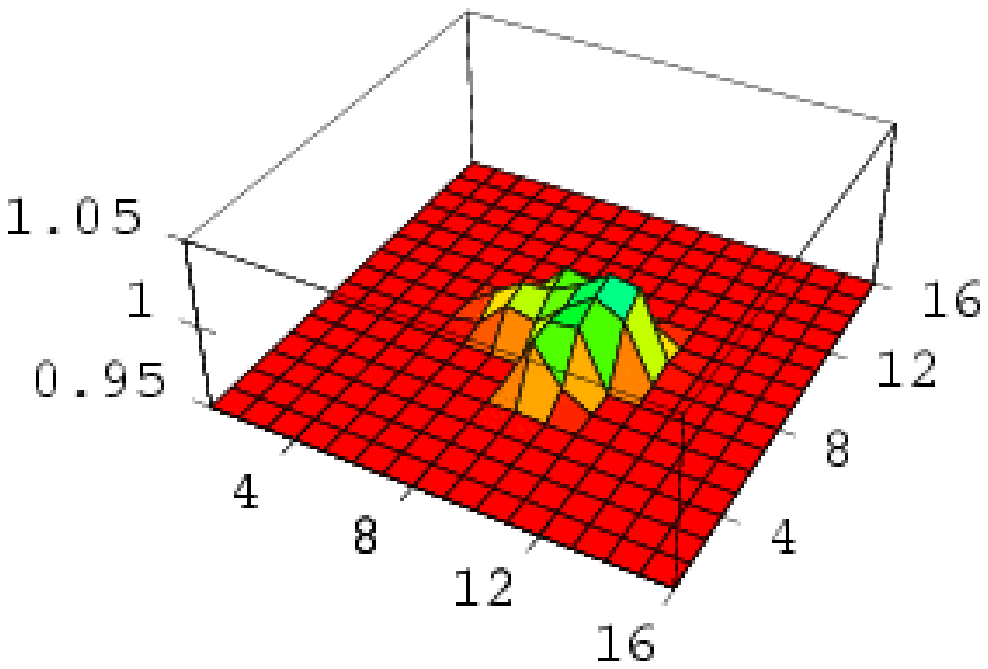}&\includegraphics[width=5.5cm]{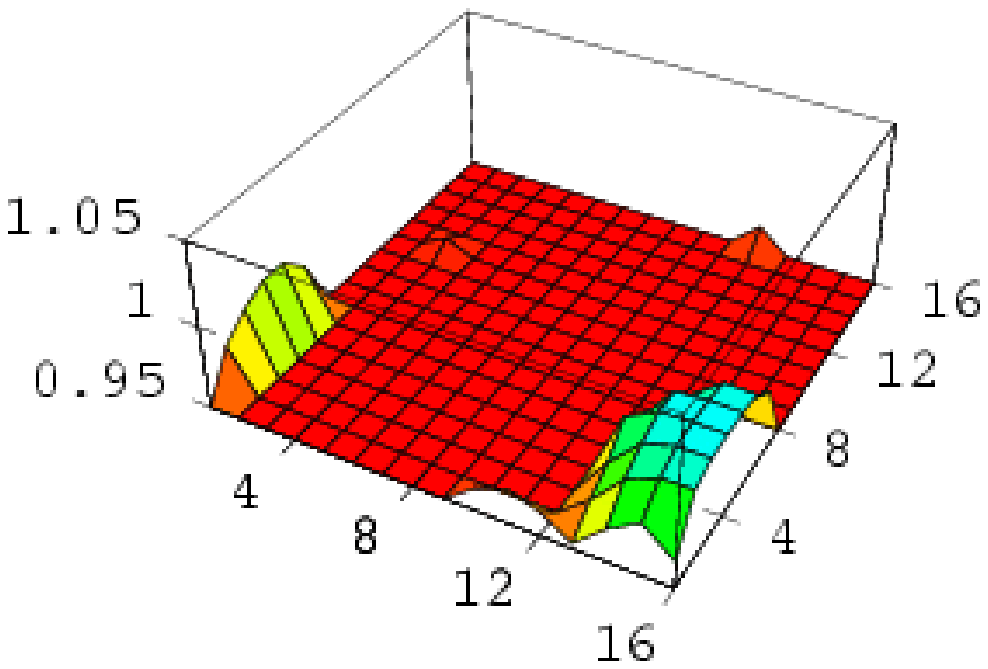}\\
(d) & (e)\\
\end{tabular}
}\\

\end{tabular}
\end{center}
\caption{Two-dimensional section through the typical $Q=0$ configuration
generated at $\beta=8.45$ on a $16^3 \times 32$ lattice.
Direct comparison of (a) $\tilde q_{\rm IR}(x)$ based on the infrared field strength
tensor based on 10 non-zero modes (5 pairs), (b) the mode-truncated density 
$q_{\lambda_{\rm cut}}(x)$ with $\lambda_{\rm cut}= 200$ MeV (for this 
configuration it amounts to 9 pairs of non-zero modes), 
(c) the all-scale charge density $q(x)$.
Subpanels (d) and (e) show the positive and negative part of the local
(anti-)selfduality variable $R(x)$. 
Note the 20-fold larger vertical scale in (c) compared to (b). 
The scale of (a) is freely adapted to that of (b).}
\label{fig:2dcuts_with_R}
\end{figure}

How the whole cluster collection changes between the percolating 
($R_{\rm cut}=0.96$) and the non-percolating regime ($R_{\rm cut}=0.99$), 
with respect to charge volume, is illustrated in 
Fig.~\ref{fig:R-composition} for the $Q=0$ configuration already considered 
in Figs.~\ref{fig:mode-visualization} and \ref{fig:R-isosurface10}.  
In order to define the cluster charge, the all-scale topological charge density 
$q(x)$ (\ref{eq:eq-q})  has been summed over the sites belonging to the 
respective connected $R$-clusters. 

Before percolation sets in (at $R_{\rm cut}=0.99$) one finds in this particular
configuration $O(20)$ significant clusters. Among them one cluster has an 
exceedingly large 
volume and two have a cluster charge $Q_{\rm cluster}=1$ or $-3$. 
Deep in the percolation regime  
(at $R_{\rm cut}=0.96$), there are essentially only two significant $R$-clusters
left over which share the total occupied volume in the ratio $55/45$. Their cluster 
charges are large and must compensate each other. Even in this state,
the distance between the two clusters is still large, 
$d(c_1,c_2) \approx 7~a$.

Near the maximum multiplicity, before the percolation sets in, for a suitable
ultraviolet cut-off $\lambda_{\rm cut}$ the number of $q$-clusters is of the same 
order as the number of (anti-)selfdual $R$-clusters (domains) 
defined with a corresponding
number of modes.  Moreover, it turns out, that the respective isosurface plots 
are closely correlated. We consider this coincidence as an argument for the 
importance of locally (anti-)selfdual infrared gauge fields as pert
of the vacuum structure at a cut-off scale of $200$ MeV.
The correlation of positions and sizes of both types of clusters is shown 
in Fig.~\ref{fig:compare-isosurfaces} where one timeslice is visualized by 
means of $R(x)$ on the left side and of $q_{\lambda_{\rm cut}}(x)$ with  
$\lambda_{\rm cut}=200$ MeV (which actually corresponds to 9 pairs of 
non-zero modes for this configuration) on the right side.
The same kind of similarity exists in all timeslices.

Summarizing, in Fig.~\ref{fig:2dcuts_with_R} we show the various ways to present 
the topological 
structure in a two-dimensional section. The upper panels present the 
profiles of various densities, 
$\tilde{q}_{\rm IR}(x)$ based on the infrared field strength, 
the mode-truncated density $q_{\lambda_{\rm cut}}(x)$ 
and the all-scale density $q(x)$.
The lower panels show how the positive and negative part of $R(x)$ 
restricted to the interval $0.95 < |R(x)| < 1$ highlights the regions 
of selfduality and antiselfduality. The plots illustrate this for the 
same $Q=0$ example configuration as visualized many times.
			    
\section{Discussion and Conclusions}
\label{sec:conclusions}

We have reported on several attempts to elucidate the structure of
the (quenched) QCD vacuum using overlap fermions as a probe. 
In particular, we have looked at the localization properties
of the eigenmodes and of the topological charge density which is built 
up by all modes. Furthermore, we have studied the chiral properties of 
the eigenmodes of the overlap Dirac operator and tried to relate this
to the topological charge density. Finally we have searched for 
(anti-)selfdual domains of the ultraviolet filtered field strength tensor.

As a side-result of this work, we have obtained the topological susceptibility
for the ensembles in use (from the index of each configuration) and the 
average spectral density. The analysis of the latter
has resulted in an effective chiral condensate $\Sigma_{\rm eff}$ for
each lattice size. The final analysis of this in the light of quenched chiral
perturbation theory, the comparison with random matrix theory, in particular 
under the aspect of spectral correlations, will be subject of a separate 
publication. The emphasis of the present paper was put on the chiral and 
topological {\it structural properties}.

While average properties as the dimensionality of eigenmodes (for zero modes
and non-zero modes in special intervals of eigenvalues $\lambda$) could be 
obtained by the scaling properties of the average inverse participation ratio 
(and its generalization based on higher moments of the scalar density) in 
agreement with critical spectral correlations (not described in this paper), 
the space-time structure of single modes and of the topological charge density 
could be studied by employing a cluster algorithm and, subsequently, a random 
walk algorithm. While the cluster algorithm locates and separates the clusters, 
depending on a cut ($q_{\rm cut}$ or $p_{\rm cut}$) applied to the quantity 
that is being considered, the random walkers help to specify the space-time 
dimension of the cluster by means of their return probability to the cluster 
center from which they start.

A necessary condition for a pion to be able to propagate through a given
lattice vacuum is that the scalar density of the modes forms percolating 
clusters, such that any two separate points on the lattice must be 
``connected'' (in the sense defined in Section~\ref{sec:2.5} with non-zero 
measure when integrated over all configurations). Similar properties might 
be established for other quantities, like the topological charge density. 

In the specific case of the topological charge density one finds at low 
enough density only two connected clusters of opposite charge, which are 
infinitely thin in the co-direction and form a multi-layered compound 
that covers all space-time. This result can hardly be reconciled with 
the picture of a dilute gas of semi-classical, finite-sized, in four 
dimensions coherently extended excitations (instantons or other classical
solutions). Since the topological charge density correlator {\it in the 
vacuum} must be negative for all non-zero distances (on the lattice at 
least beyond a distance of $O(a)$ set by the non-ultralocal density 
definition in terms of the overlap operator), this requires a local fractal
dimension $d^{*} < 4$ for the topological density. The simultaneous presence 
of infrared (long-range) and ultraviolet ($O(a)$) features motivates to call 
the overlap topological density ``all-scale density'', as long as no 
mode-truncation is applied. This picture is known from the work of
Horvath {\it et al.}, but it does not exclude the presence of even lower
dimensional structure inside the two clusters. Here we complete the picture 
by pointing out (see Table V in the Appendix) that percolating clusters
of topological charge already appear with lower dimensions $d^{*} = 1 - 2$ at 
higher $q_{\rm cut} \approx 0.25 q_{\rm max}$.   

Similar things are observed for single modes of the overlap operator. 
Percolation sets in 
at dimensions $d^{*} \approx 2$ and $p_{\rm cut} \approx 0.20 p_{\rm max}$
for zero modes and the first few non-zero modes, whereas higher modes 
percolate already at higher $p_{\rm cut}/p_{\rm max}$.

\begin{figure}[t]
\begin{center}
\hspace*{-1.0cm}\epsfig{file=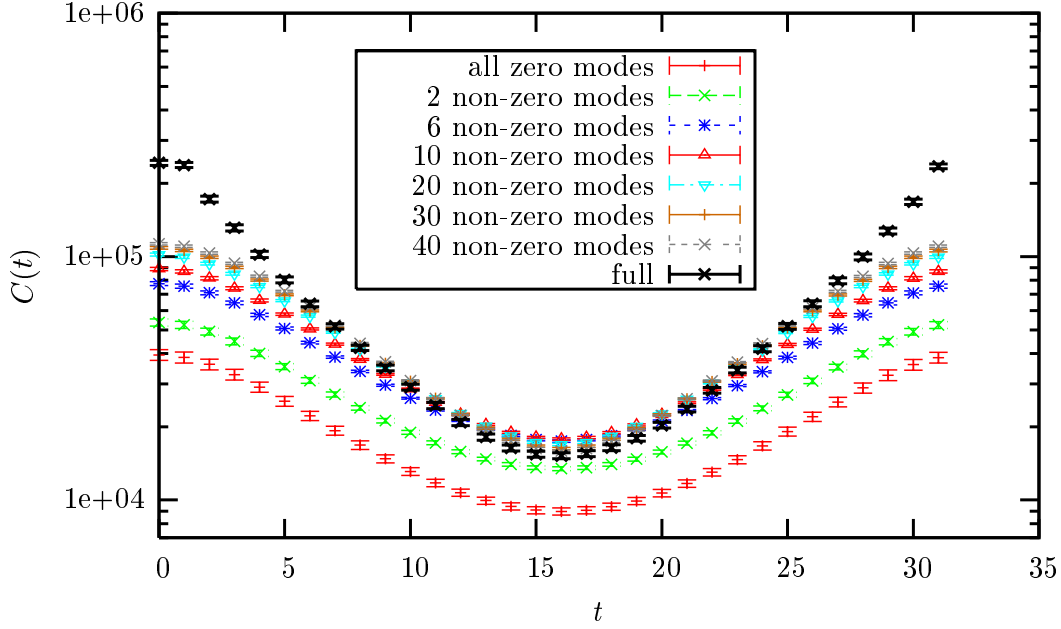,width=14cm, clip=}
\end{center}
\caption{The full pion propagator compared with the 
contribution of the zero modes alone and the cumulative approximation 
by various numbers of non-zero modes (see legend box) is shown in analogy 
to Fig.~\ref{fig:pionpropagator}.
For this calculation 250 configurations have been used from the 
$16^3\times32$ lattice ensemble generated at $\beta=8.45$.}
\label{fig:saturation_1}
\end{figure}

In the Introduction (in Fig.~\ref{fig:pionpropagator}) we have seen that 
the long-distance part of the pion propagator is well saturated by the lowest 
40 eigenmodes of the overlap operator. In more detail, 
Fig.~\ref{fig:saturation_1}  
shows the pion correlator separately for an increasing number of modes, from 
the zero modes alone up to the additional inclusion of 40 non-zero modes. 

\begin{figure}[t]
\begin{center}
\hspace*{-1.0cm}\epsfig{file=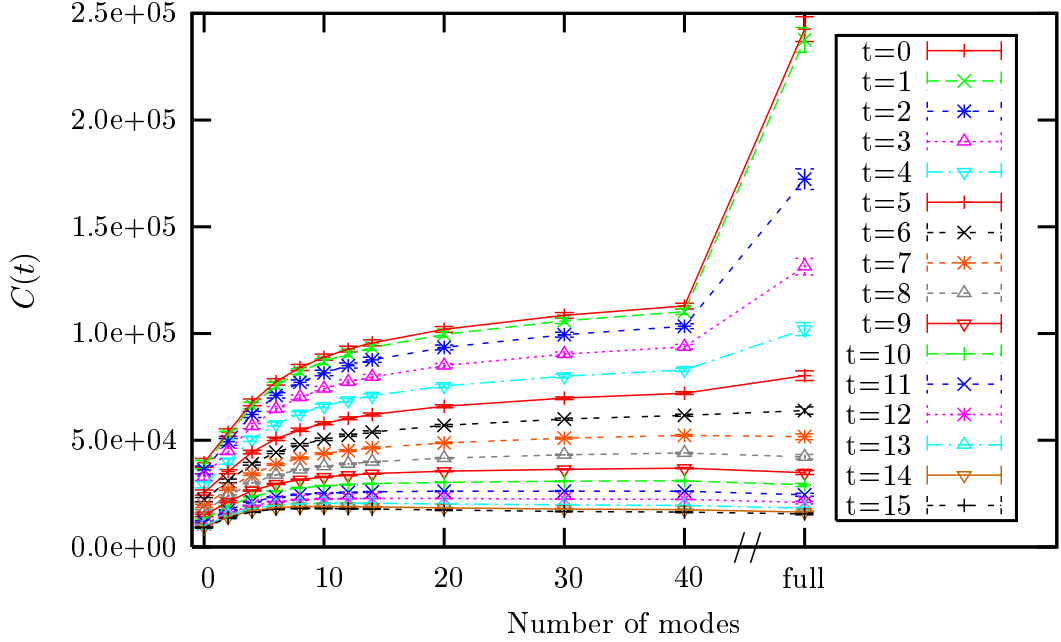,width=14cm, clip=}
\end{center}
\caption{The dependence of the pion propagator for various $t$ 
(see legend box) on the number of accumulated non-zero modes is presented 
for the $16^3\times32$ lattice ensemble generated at $\beta=8.45$.}
\label{fig:saturation_2}
\end{figure}

\begin{figure}[t]
\begin{center}
\hspace*{-1.0cm}\epsfig{file=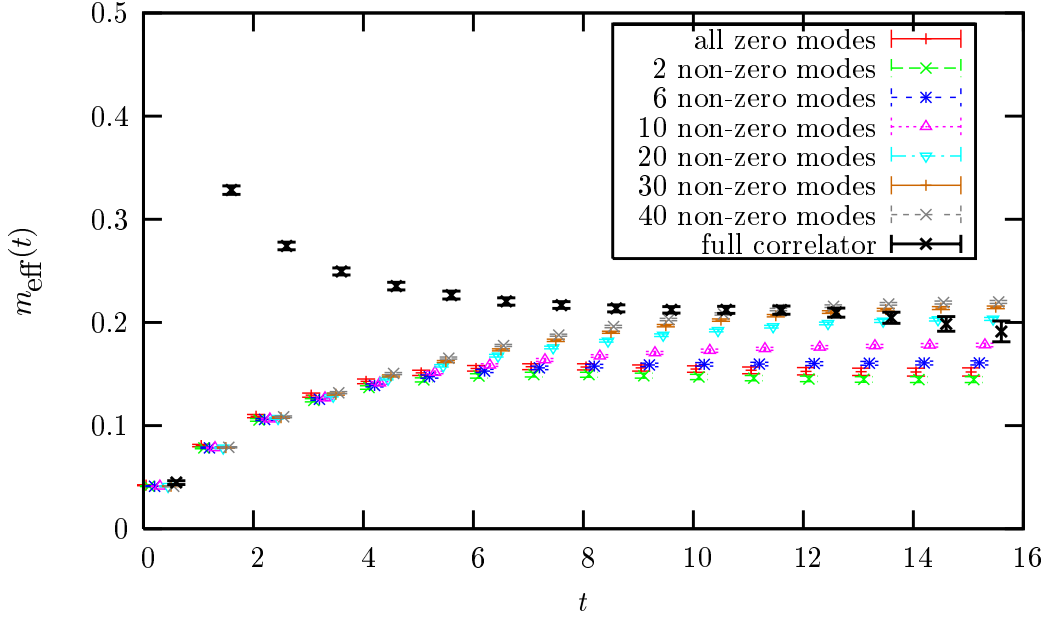,width=14cm, clip=}
\end{center}
\caption{The effective-mass plot {\it vs.} $t$  
for the $16^3\times32$ lattice ensemble generated at $\beta=8.45$.
This is shown for the full pion propagator and for various numbers 
(see legend box) of non-zero modes included in the approximation.} 
\label{fig:saturation_3}
\end{figure}

In Fig.~\ref{fig:saturation_2} we show the convergence to the full
propagator separately for various time differences $t$ and in
Fig.~\ref{fig:saturation_3} the effective pion mass separately for 
the full propagator and different
mode-truncated approximations to it, also as a function of $t$.
We see that the pion propagates 
with approximately 70 \% of its physical mass even when the quark propagator 
is stripped down to the zero modes. 

If we wanted to define the mobility edge 
by the lowest mode at which propagation sets in, we would have to
conclude that the mobility edge is at $\lambda=0$, and that the (quenched) 
QCD vacuum is critical in the whole lower part ($\lambda < 200 {\rm MeV}$)
of the spectrum, very analogous to certain 2D or 3D disordered condensed matter 
systems~\cite{Markos:2006}. 
The level compressibility (not described in 
this paper) supports this analogy. 

What changes in the vacuum picture if we would restrict our analysis 
of the vacuum structure to the lowest part of the spectrum? 
If the topological density is ultraviolet filtered by 
truncating its definition to the lowest modes, the emerging picture is 
completely different from what we have outlined above. The topological charge 
density correlator would now be resembling the presence of clusters of finite
extent being subject to some weak but non-trivial intercluster correlations,
increasing with the cut-off $\lambda_{\rm cut}$. 
This picture is basically 
reproduced by the cluster analysis and the random walker algorithm. 

In the Appendix we summarize our main results concerning the dimensionality 
of the all-scale and of the mode-truncated (ultraviolet filtered) topological 
charge density (Tables V and VI) and of the scalar density of zero modes
and selected non-zero modes (Tables VII and VIII). Here we also test the
reliability of the ``covering sphere method'' introduced in 
subsection \ref{sec:4.5}, that attempts to estimate the effective dimensionality
by the growth of the integrated quantity under consideration (the topological
charge and scalar density) with the radius of a sphere covering a cluster.
Although this latter method clearly defines a certain cut-off that separates a linear 
growth from a nonlinear one in both cases, the estimated dimensionalities 
do not coincide with the random walker method. This shows that the density 
inside the clusters is very inhomogeneous.

Concerning the all-scale topological density, it is remarkable that at 
low values of the density (at the lowest cut-off $q_{\rm cut}$ defining
the clusters) the dimensionality is practically independent of the lattice
cut-off $a$ and becomes almost equal three. The covering sphere method finds it
somewhat bigger than three. At higher values of the density smaller 
dimensionalities are found. There the covering sphere method gives an 
effective dimension equal to one, {\it i.e.} the charge inside a cluster
grows linearly with the size. For the mode-truncated density at lowest 
values of the density (at a low cut-off defining the clusters) 
the clusters' dimensionality surprisingly is also close to three. 
How the infrared clusters (with a fixed $\lambda_{\rm cut}$) start to 
percolate we can describe in the following way: 
at high values of the density they are well-separated and localized to zero 
dimensional balls or one dimensional rods of finite thickness, before they 
start to percolate at somewhat lower density with an effective dimension 
still relatively low, $d^{*}= 1.5 - 2$. Finally, at even lower values of the
density, they become extended in essentially three dimensions with an 
$a$-independent thickness of $O(\Lambda_{\rm QCD})$. 

A nonvanishing local chirality $X_{\rm \lambda}(x)$ is assumed by individual 
lower-lying non-zero modes in regions of their preferential localization 
({\it i.e.} at high scalar density $p(x)$). This effect could be demonstrated 
to be correlated to the ultraviolet filtered topological density 
$q_{\rm \lambda}$ over a distance comparable to the correlation length of
the density itself. Here we can only say that this ultraviolet filtered density is 
resulting from the correlated behavior of {\it all lowest-lying} modes. 
On the other hand, this type of correlation of $X_{\rm \lambda}$ with the 
topological charge density is a natural consequence in the 
{\it instanton model} where the topological charge is simply carried by the 
semiclassical background and the non-zero modes are superpositions of the
(anti-)instantons' zero modes.

The all-scale topological density does not follow this pattern. It deserves
its name at low cut-off $q_{\rm cut}$ where it combines long-range coherence 
on one hand and lower-dimensional, {\it i.e.} singular structure on the other. 
When the cut-off is increased, percolation breaks down and an increasing 
number of isolated clusters appears with lower and 
lower dimensions. In this sense the all-scale topological density is 
multifractal, too. It should be noted that the maximal number of these
clusters seems to depend only on the lattice size in lattice units.
In a given physical volume it is diverging with the lattice spacing 
$a \to 0$.

Applying the same methods to the scalar density of individual modes 
(from zero modes up to the highest analyzed modes) we also have found that at very 
low cut-off $p_{\rm cut}$
they do not extend isotropically but show a lower-dimensional structure. 
At the lowest cut-off the effective dimension ranges from two-dimensional 
for zero modes and the very first non-zero modes to three-dimensional for 
the modes up to the 120th non-zero mode. This means that all the modes that 
are sufficient to saturate the pion propagator share this property of 
low-dimensionality. This 
is in agreement with the critical level statistics (being ``non-metallic'' 
following the condensed-matter terminology). 

Summarizing we can state that at sufficiently low density ($q(x)$ or $p(x)$) we
find consistently a dimensionality $d^{*} \approx 3$ to characterize most 
considered densities.
The only exception are the very lowest modes which show a two-dimensional
structure with respect to the growth of the integrated norm. This special
feature is corroborating the average properties that have been concluded 
from the scaling behavior of the IPR. 

It is remarkable that the mode truncated topological charge density 
percolates already at relatively low (volume) packing fraction, when its 
domains have a finite extension, 
forming coherent clusters of finite thickness $O(\Lambda_{\rm CD})$. This is
impossible for the all-scale topological density. This kind of clustering 
of the ultraviolet filtered topological charge is corroborated by the 
search  for (anti-)selfdual domains, employing the ultraviolet filtered 
field strength tensor. The latter can be defined by means of the overlap 
operator, too, employing filtering in the form of truncating the spectral 
representation. 

In conclusion: the vacuum picture exhibited by the modes that are sufficient 
to build up the pion propagator resembles 
a dilute liquid of separated (anti-)selfdual excitations if looked at it at 
higher cut-off 
$q_{\rm cut}$ of the topological charge density and a dilute,
percolating network at lower cut-off. The less-than-three dimensional
pinning-down of the very lowest modes at very high scalar density $p(x)$ 
(and of the 
topological density at very high values of $q_{\lambda}(x)$) 
might be related to singular (confining) 
excitations like monopoles or vortices. 
This possibility is presently under investigation.
In particular, the two-dimensional structure of the zero modes 
(with pointlike peaks on top of the latter) intriguingly suggests that 
they are pinned to vortex surfaces and vortex 
intersections~\cite{Gattnar:2004gx}. 
In this sense we may conjecture that the lowest part of the fermion 
spectrum is both realizing chiral symmetry breaking and tracing the agents 
of confinement. This would be close in spirit to recent attempts in the 
literature to extract the dynamics of the Polyakov loop from the spectral 
properties of the Dirac 
operator~\cite{Gattringer:2006ci,Bruckmann:2006kx,Synatschke:2007bz}.

It will be our next task to extend these studies to finite temperature 
and full QCD in order to disclose the changes of topological structure 
under the influence of dynamical quarks and those accompanying the chiral 
phase transition.

\clearpage

\section{Appendix: Tables of  dimensions }

\vspace*{2cm}

\begin{center}
\begin{table}[h]
\begin{tabular*}{13cm}{|c@{\extracolsep{\fill}}|cccccc|}\hline
 $q_{\mbox{cut}}/q_{\mbox{max}}$  &0.00&0.10&0.20&0.30&0.40&0.50\\\hline
 $12^3\times 24$ all-scale &\textbf{2.7(1)}&\textbf{1.5(1)}&\textbf{1.1(1)}&0.7(1)&0.4(1)&0.1(1)\\
 $12^3\times 24$ $\lambda_{\mbox{cut}}\le$ 200 MeV &\textbf{3.1(1)}&\textbf{1.8(1)}&\textbf{1.5(1)}&1.0(1)&0.7(1)&0.2(1)\\
 $12^3\times 24$ $\lambda_{\mbox{cut}}\le$ 800 MeV &\textbf{3.0(1)}&\textbf{1.8(1)}&\textbf{1.5(1)}&\textbf{1.0(1)}&0.7(1)&0.2(1)\\\hline
 $16^3\times 32$  all-scale  &\textbf{2.6(2)}&\textbf{1.5(1)}&\textbf{1.1(1)}&0.9(1)&0.5(1)&0.1(1)\\
 $16^3\times 32$ $\lambda_{\mbox{cut}}\le$ 200 MeV &\textbf{3.0(1)}&\textbf{2.1(2)}&\textbf{1.8(1)}&1.2(1)&0.8(1)&0.3(1)\\
 $16^3\times 32$ $\lambda_{\mbox{cut}}\le$ 634 MeV &\textbf{3.0(1)}&\textbf{1.9(1)}&\textbf{1.7(1)}&1.2(1)&0.7(1)&0.3(1)\\\hline 
\end{tabular*}
\vspace{0.3cm}
\caption{Effective dimension $d^*$ of the leading cluster of the topological charge 
density based on the random walker approach. 
Dimensions referring to the percolating regime are printed in bold face. 
The data is averaged over the  $12^3\times 24$ ($16^3\times 32$) lattice ensemble 
at $\beta=8.10$ ($\beta=8.45)$.}
\end{table}

\begin{table}[h]
\begin{tabular*}{13cm}{|c@{\extracolsep{\fill}}|cccccc|}\hline
 $q_{\mbox{cut}}/q_{\mbox{max}}$                      &0.00&0.10&0.20&0.30&0.40&0.50\\\hline
 $12^3\times 24$ all-scale &\textbf{3.3(1)}&\textbf{3.1(1)}&1&\hphantom{00}1\hphantom{00}&\hphantom{00}1\hphantom{00}&\hphantom{00}1\hphantom{00}\\
 $12^3\times 24$ $\lambda_{\mbox{cut}}\le$ 200 MeV    &\textbf{2.5(2)}&1&1&1&1&1\\
 $12^3\times 24$ $\lambda_{\mbox{cut}}\le$ 800 MeV    &\textbf{2.9(1)}&\textbf{1.8(1)}&1&1&1&1\\\hline
 $16^3\times 32$  all-scale  &\textbf{3.3(1)}&\textbf{3.3(1)}&\textbf{2.5(1)}&1 &1&1\\
 $16^3\times 32$ $\lambda_{\mbox{cut}}\le$ 200 MeV    &\textbf{1.9(1)}&1&1&1&1&1\\ 
 $16^3\times 32$ $\lambda_{\mbox{cut}}\le$ 634 MeV    &\textbf{2.3(1)}&1&1&1&1&1\\\hline
\end{tabular*}
\vspace{0.3cm}
\caption{Effective dimension of the leading cluster of the topological charge 
density based on the covering sphere method. The data is averaged over the  
$12^3\times 24$ ($16^3\times 32$) lattice ensemble at $\beta=8.10$ ($\beta=8.45)$.}
\end{table}

\clearpage

\begin{table}[h]
\begin{tabular*}{13cm}{|c@{\extracolsep{\fill}}|ccccc|}\hline
 $p_{\mbox{cut}}/p_{\mbox{max}}$  &0.00&0.10&0.20&0.30&0.40\\\hline
zeromodes & \textbf{3.6(2)} &\textbf{2.8(2)}&\textbf{2.0(2)}&1.4(1)&1.1(1)\\
1st non-zero mode &\textbf{3.5(1)}&\textbf{2.9(2)}&\textbf{1.9(2)}&1.4(1)&1.1(1)\\
10th non-zero mode &\textbf{3.6(2)}&\textbf{3.0(2)}&\textbf{2.2(2)}&\textbf{1.6(1)}&\textbf{1.1(1)}\\
30th non-zero mode&\textbf{3.5(1)}&\textbf{3.2(2)}&\textbf{2.6(1)}&\textbf{2.0(1)}&\textbf{1.4(1)}\\
50th non-zero mode&\textbf{3.6(2)}&\textbf{3.3(1)}&\textbf{2.8(1)}&\textbf{2.1(1)}&1.5(1)\\
70th non-zero mode &\textbf{3.6(2)}&\textbf{3.4(1)}&\textbf{2.9(1)}&\textbf{2.2(1)}&\textbf{1.6(1)}\\
90th non-zero mode&\textbf{3.5(2)}&\textbf{3.3(2)}&\textbf{3.0(1)}&\textbf{2.4(1)}&\textbf{1.7(1)}\\
120th non-zero mode &\textbf{3.6(2)}&\textbf{3.4(2)}&\textbf{3.1(1)}&\textbf{2.5(1)}&\textbf{1.8(1)}\\\hline

\end{tabular*}
\vspace{0.3cm}
\caption{Effective dimension $d^*$ of the leading cluster of the scalar density 
based on the random walker approach. 
The data is averaged over the $16^3\times32$ lattice ensemble generated at 
$\beta=8.45$. 
Dimensions referring to the percolating regime are printed in bold face.} 
\end{table}


\begin{table}[h]
\begin{tabular*}{13cm}{|c@{\extracolsep{\fill}}|ccccc|}\hline
 $p_{\mbox{cut}}/p_{\mbox{max}}$  &0.00    &0.10&0.20 &0.30 &0.40\\\hline
zeromodes                      &\textbf{1.9(1)}&\textbf{1}&\textbf{1}&1&1\\
1st non-zero mode              &\textbf{2.1(1)}&\textbf{1}&\textbf{1}&1&1\\
10th non-zero mode             &\textbf{2.5(1)}&\textbf{1}&\textbf{1}&\textbf{1}&\textbf{1}\\
30th non-zero mode             &\textbf{2.9(1)}&\textbf{2.3(1)}&\textbf{1}&\textbf{1}&\textbf{1}\\
50th non-zero mode             &\textbf{2.9(1)}&\textbf{2.4(1)}&\textbf{1}&\textbf{1}&1\\
70th non-zero mode             &\textbf{3.0(1)}&\textbf{2.7(1)}&\textbf{1.9(1)}&\textbf{\hphantom{00}1\hphantom{00}}&\textbf{\hphantom{00}1\hphantom{00}}\\
90th non-zero mode             &\textbf{3.1(1)}&\textbf{2.6(1)}&\textbf{1.9(1)}&\textbf{1}&\textbf{1}\\
120th non-zero mode            &\textbf{3.1(1)}&\textbf{2.9(1)}&\textbf{2.3(1)}&\textbf{1}&\textbf{1}\\\hline

\end{tabular*}
\vspace{0.3cm}

\caption{Effective dimension of the leading cluster of the scalar density based 
on the covering sphere method. 
The data is averaged over the $16^3\times32$ lattice ensemble generated at 
$\beta=8.45$.} 
\end{table}
\end{center}

\clearpage

\section*{Acknowledgement}
The numerical calculations have been performed on the IBM p690 at HLRN
(Berlin) and NIC (J\"ulich), as well as the PC farm at DESY (Zeuthen), 
the LRZ M\"unchen and the Sektion Physik of LMU M\"unchen. We thank all 
these institutions for technical support. 
This work has been financially supported in part by the EU Integrated 
Infrastructure Initiative Hadron Physics (I3HP) under contract 
RII3-CT-2004-506078 and by the DFG under contract FOR 465 (Forschergruppe 
Gitter-Hadronen-Ph\"anomenologie).
E.-M.~I. is grateful to P. van Baal, F. Bruckmann, C. Gattringer, 
M. M\"uller-Preussker, A. Sch\"afer and S. Solbrig for inspiring discussions 
and collaboration on related projects.
V.~W. thanks the Yukawa Institute for Theoretical Physics at Kyoto University
for the opportunity to participate in the YITP workshop on "Actions and symmetries 
in lattice gauge theory" (YITP-W-05-25) and to present part of the results.
The discussions there were useful to complete this work. 
E.-M.~I. thanks the organizers of the Program INT-06-1 at the Institute for
Nuclear Theory of the University of Washington, in particular E. Shuryak, 
for the opportunity to discuss these results, and the DoE for support of his
participation.

\bibliography{overlap_paper230407.bib}

\begin{thebibliography}{85}
\expandafter\ifx\csname natexlab\endcsname\relax\def\natexlab#1{#1}\fi
\expandafter\ifx\csname bibnamefont\endcsname\relax
  \def\bibnamefont#1{#1}\fi
\expandafter\ifx\csname bibfnamefont\endcsname\relax
  \def\bibfnamefont#1{#1}\fi
\expandafter\ifx\csname citenamefont\endcsname\relax
  \def\citenamefont#1{#1}\fi
\expandafter\ifx\csname url\endcsname\relax
  \def\url#1{\texttt{#1}}\fi
\expandafter\ifx\csname urlprefix\endcsname\relax\def\urlprefix{URL }\fi
\providecommand{\bibinfo}[2]{#2}
\providecommand{\eprint}[2][]{\url{#2}}

\bibitem[{\citenamefont{Neuberger}(1998{\natexlab{a}})}]{Neuberger:1997fp}
\bibinfo{author}{\bibfnamefont{H.}~\bibnamefont{Neuberger}},
  \bibinfo{journal}{Phys. Lett.} \textbf{\bibinfo{volume}{B417}},
  \bibinfo{pages}{141} (\bibinfo{year}{1998}{\natexlab{a}}),
  \eprint{hep-lat/9707022}.

\bibitem[{\citenamefont{Neuberger}(1998{\natexlab{b}})}]{Neuberger:1998wv}
\bibinfo{author}{\bibfnamefont{H.}~\bibnamefont{Neuberger}},
  \bibinfo{journal}{Phys. Lett.} \textbf{\bibinfo{volume}{B427}},
  \bibinfo{pages}{353} (\bibinfo{year}{1998}{\natexlab{b}}),
  \eprint{hep-lat/9801031}.

\bibitem[{\citenamefont{Luscher}(1998)}]{Luscher:1998pq}
\bibinfo{author}{\bibfnamefont{M.}~\bibnamefont{L\"uscher}},
  \bibinfo{journal}{Phys. Lett.} \textbf{\bibinfo{volume}{B428}},
  \bibinfo{pages}{342} (\bibinfo{year}{1998}), \eprint{hep-lat/9802011}.

\bibitem[{\citenamefont{Hasenfratz et~al.}(1998)\citenamefont{Hasenfratz,
  Laliena, and Niedermayer}}]{Hasenfratz:1998ri}
\bibinfo{author}{\bibfnamefont{P.}~\bibnamefont{Hasenfratz}},
  \bibinfo{author}{\bibfnamefont{V.}~\bibnamefont{Laliena}}, \bibnamefont{and}
  \bibinfo{author}{\bibfnamefont{F.}~\bibnamefont{Niedermayer}},
  \bibinfo{journal}{Phys. Lett.} \textbf{\bibinfo{volume}{B427}},
  \bibinfo{pages}{125} (\bibinfo{year}{1998}), \eprint{hep-lat/9801021}.

\bibitem[{\citenamefont{Niedermayer}(1999)}]{Niedermayer:1998bi}
\bibinfo{author}{\bibfnamefont{F.}~\bibnamefont{Niedermayer}},
  \bibinfo{journal}{Nucl. Phys. Proc. Suppl.} \textbf{\bibinfo{volume}{73}},
  \bibinfo{pages}{105} (\bibinfo{year}{1999}), \eprint{hep-lat/9810026}.

\bibitem[{\citenamefont{Koma et~al.}(2006)}]{Koma:2005sw}
\bibinfo{author}{\bibfnamefont{Y.}~\bibnamefont{Koma}} \bibnamefont{et~al.},
  \bibinfo{journal}{PoS} \textbf{\bibinfo{volume}{LAT2005}},
  \bibinfo{pages}{300} (\bibinfo{year}{2006}), \eprint{hep-lat/0509164}.

\bibitem[{\citenamefont{Ilgenfritz
  et~al.}(2006{\natexlab{a}})}]{Ilgenfritz:2005hh}
\bibinfo{author}{\bibfnamefont{E.~M.} \bibnamefont{Ilgenfritz}}
  \bibnamefont{et~al.}, \bibinfo{journal}{Nucl. Phys. Proc. Suppl.}
  \textbf{\bibinfo{volume}{153}}, \bibinfo{pages}{328}
  (\bibinfo{year}{2006}{\natexlab{a}}), \eprint{hep-lat/0512005}.

\bibitem[{\citenamefont{Weinberg et~al.}(2006{\natexlab{a}})}]{Weinberg:2006ju}
\bibinfo{author}{\bibfnamefont{V.}~\bibnamefont{Weinberg}}
  \bibnamefont{et~al.}, \bibinfo{journal}{PoS}
  \textbf{\bibinfo{volume}{LAT2006}}, \bibinfo{pages}{078}
  (\bibinfo{year}{2006}{\natexlab{a}}), \eprint{hep-lat/0610087}.

\bibitem[{\citenamefont{Ilgenfritz
  et~al.}(2006{\natexlab{b}})}]{Ilgenfritz:2006gc}
\bibinfo{author}{\bibfnamefont{E.~M.} \bibnamefont{Ilgenfritz}}
  \bibnamefont{et~al.} (\bibinfo{year}{2006}{\natexlab{b}}),
  \eprint{hep-lat/0611007}.

\bibitem[{\citenamefont{Weinberg et~al.}(2006{\natexlab{b}})}]{Weinberg:2005dh}
\bibinfo{author}{\bibfnamefont{V.}~\bibnamefont{Weinberg}}
  \bibnamefont{et~al.}, \bibinfo{journal}{PoS}
  \textbf{\bibinfo{volume}{LAT2005}}, \bibinfo{pages}{171}
  (\bibinfo{year}{2006}{\natexlab{b}}), \eprint{hep-lat/0510056}.

\bibitem[{\citenamefont{DeGrand and Hasenfratz}(2001)}]{DeGrand:2000gq}
\bibinfo{author}{\bibfnamefont{T.~A.} \bibnamefont{DeGrand}} \bibnamefont{and}
  \bibinfo{author}{\bibfnamefont{A.}~\bibnamefont{Hasenfratz}},
  \bibinfo{journal}{Phys. Rev.} \textbf{\bibinfo{volume}{D64}},
  \bibinfo{pages}{034512} (\bibinfo{year}{2001}), \eprint{hep-lat/0012021}.

\bibitem[{\citenamefont{DeGrand}(2001)}]{DeGrand:2001tm}
\bibinfo{author}{\bibfnamefont{T.~A.} \bibnamefont{DeGrand}},
  \bibinfo{journal}{Phys. Rev.} \textbf{\bibinfo{volume}{D64}},
  \bibinfo{pages}{094508} (\bibinfo{year}{2001}), \eprint{hep-lat/0106001}.

\bibitem[{\citenamefont{Ivanenko and Negele}(1998)}]{Ivanenko:1997nb}
\bibinfo{author}{\bibfnamefont{T.~L.} \bibnamefont{Ivanenko}} \bibnamefont{and}
  \bibinfo{author}{\bibfnamefont{J.~W.} \bibnamefont{Negele}},
  \bibinfo{journal}{Nucl. Phys. Proc. Suppl.} \textbf{\bibinfo{volume}{63}},
  \bibinfo{pages}{504} (\bibinfo{year}{1998}), \eprint{hep-lat/9709130}.

\bibitem[{\citenamefont{Neff et~al.}(2001)\citenamefont{Neff, Eicker, Lippert,
  Negele, and Schilling}}]{Neff:2001zr}
\bibinfo{author}{\bibfnamefont{H.}~\bibnamefont{Neff}},
  \bibinfo{author}{\bibfnamefont{N.}~\bibnamefont{Eicker}},
  \bibinfo{author}{\bibfnamefont{T.}~\bibnamefont{Lippert}},
  \bibinfo{author}{\bibfnamefont{J.~W.} \bibnamefont{Negele}},
  \bibnamefont{and}
  \bibinfo{author}{\bibfnamefont{K.}~\bibnamefont{Schilling}},
  \bibinfo{journal}{Phys. Rev.} \textbf{\bibinfo{volume}{D64}},
  \bibinfo{pages}{114509} (\bibinfo{year}{2001}), \eprint{hep-lat/0106016}.

\bibitem[{\citenamefont{Haymaker}(1999)}]{Haymaker:1998cw}
\bibinfo{author}{\bibfnamefont{R.~W.} \bibnamefont{Haymaker}},
  \bibinfo{journal}{Phys. Rept.} \textbf{\bibinfo{volume}{315}},
  \bibinfo{pages}{153} (\bibinfo{year}{1999}), \eprint{hep-lat/9809094}.

\bibitem[{\citenamefont{Greensite}(2003)}]{Greensite:2003bk}
\bibinfo{author}{\bibfnamefont{J.}~\bibnamefont{Greensite}},
  \bibinfo{journal}{Prog. Part. Nucl. Phys.} \textbf{\bibinfo{volume}{51}},
  \bibinfo{pages}{1} (\bibinfo{year}{2003}), \eprint{hep-lat/0301023}.

\bibitem[{\citenamefont{Di~Giacomo}(2005)}]{DiGiacomo:2005yq}
\bibinfo{author}{\bibfnamefont{A.}~\bibnamefont{Di~Giacomo}},
  \bibinfo{journal}{Acta Phys. Polon.} \textbf{\bibinfo{volume}{B36}},
  \bibinfo{pages}{3723} (\bibinfo{year}{2005}), \eprint{hep-lat/0510065}.

\bibitem[{\citenamefont{Wetterich}(2005)}]{Wetterich:2004kg}
\bibinfo{author}{\bibfnamefont{C.}~\bibnamefont{Wetterich}},
  \bibinfo{journal}{AIP Conf. Proc.} \textbf{\bibinfo{volume}{739}},
  \bibinfo{pages}{123} (\bibinfo{year}{2005}), \eprint{hep-ph/0410057}.

\bibitem[{\citenamefont{Schafer and Shuryak}(1998)}]{Schafer:1996wv}
\bibinfo{author}{\bibfnamefont{T.}~\bibnamefont{Sch\"afer}} \bibnamefont{and}
  \bibinfo{author}{\bibfnamefont{E.~V.} \bibnamefont{Shuryak}},
  \bibinfo{journal}{Rev. Mod. Phys.} \textbf{\bibinfo{volume}{70}},
  \bibinfo{pages}{323} (\bibinfo{year}{1998}), \eprint{hep-ph/9610451}.

\bibitem[{\citenamefont{Gerhold et~al.}(2007)\citenamefont{Gerhold, Ilgenfritz,
  and M\"uller-Preussker}}]{Gerhold:2006sk}
\bibinfo{author}{\bibfnamefont{P.}~\bibnamefont{Gerhold}},
  \bibinfo{author}{\bibfnamefont{E.~M.} \bibnamefont{Ilgenfritz}},
  \bibnamefont{and}
  \bibinfo{author}{\bibfnamefont{M.}~\bibnamefont{M\"uller-Preussker}},
  \bibinfo{journal}{Nucl. Phys.} \textbf{\bibinfo{volume}{B760}},
  \bibinfo{pages}{1} (\bibinfo{year}{2007}), \eprint{hep-ph/0607315}.

\bibitem[{\citenamefont{Ilgenfritz et~al.}(1986)\citenamefont{Ilgenfritz,
  Laursen, Schierholz, M\"uller-Preussker, and Schiller}}]{Ilgenfritz:1985dz}
\bibinfo{author}{\bibfnamefont{E.-M.} \bibnamefont{Ilgenfritz}},
  \bibinfo{author}{\bibfnamefont{M.~L.} \bibnamefont{Laursen}},
  \bibinfo{author}{\bibfnamefont{G.}~\bibnamefont{Schierholz}},
  \bibinfo{author}{\bibfnamefont{M.}~\bibnamefont{M\"uller-Preussker}},
  \bibnamefont{and} \bibinfo{author}{\bibfnamefont{H.}~\bibnamefont{Schiller}},
  \bibinfo{journal}{Nucl. Phys.} \textbf{\bibinfo{volume}{B268}},
  \bibinfo{pages}{693} (\bibinfo{year}{1986}).

\bibitem[{\citenamefont{Teper}(1986)}]{Teper:1985ek}
\bibinfo{author}{\bibfnamefont{M.}~\bibnamefont{Teper}},
  \bibinfo{journal}{Phys. Lett.} \textbf{\bibinfo{volume}{B171}},
  \bibinfo{pages}{86} (\bibinfo{year}{1986}).

\bibitem[{\citenamefont{Smith and Teper}(1998)}]{Smith:1998wt}
\bibinfo{author}{\bibfnamefont{D.~A.} \bibnamefont{Smith}} \bibnamefont{and}
  \bibinfo{author}{\bibfnamefont{M.~J.} \bibnamefont{Teper}}
  (\bibinfo{collaboration}{UKQCD}), \bibinfo{journal}{Phys. Rev.}
  \textbf{\bibinfo{volume}{D58}}, \bibinfo{pages}{014505}
  (\bibinfo{year}{1998}), \eprint{hep-lat/9801008}.

\bibitem[{\citenamefont{Garcia~Perez et~al.}(1999)\citenamefont{Garcia~Perez,
  Philipsen, and Stamatescu}}]{GarciaPerez:1998ru}
\bibinfo{author}{\bibfnamefont{M.}~\bibnamefont{Garcia~Perez}},
  \bibinfo{author}{\bibfnamefont{O.}~\bibnamefont{Philipsen}},
  \bibnamefont{and} \bibinfo{author}{\bibfnamefont{I.-O.}
  \bibnamefont{Stamatescu}}, \bibinfo{journal}{Nucl. Phys.}
  \textbf{\bibinfo{volume}{B551}}, \bibinfo{pages}{293} (\bibinfo{year}{1999}),
  \eprint{hep-lat/9812006}.

\bibitem[{\citenamefont{DeGrand
  et~al.}(1996{\natexlab{a}})\citenamefont{DeGrand, Hasenfratz, and
  Zhu}}]{DeGrand:1996ih}
\bibinfo{author}{\bibfnamefont{T.~A.} \bibnamefont{DeGrand}},
  \bibinfo{author}{\bibfnamefont{A.}~\bibnamefont{Hasenfratz}},
  \bibnamefont{and} \bibinfo{author}{\bibfnamefont{D.-c.} \bibnamefont{Zhu}},
  \bibinfo{journal}{Nucl. Phys.} \textbf{\bibinfo{volume}{B475}},
  \bibinfo{pages}{321} (\bibinfo{year}{1996}{\natexlab{a}}),
  \eprint{hep-lat/9603015}.

\bibitem[{\citenamefont{DeGrand
  et~al.}(1996{\natexlab{b}})\citenamefont{DeGrand, Hasenfratz, and
  Zhu}}]{DeGrand:1996zb}
\bibinfo{author}{\bibfnamefont{T.~A.} \bibnamefont{DeGrand}},
  \bibinfo{author}{\bibfnamefont{A.}~\bibnamefont{Hasenfratz}},
  \bibnamefont{and} \bibinfo{author}{\bibfnamefont{D.-c.} \bibnamefont{Zhu}},
  \bibinfo{journal}{Nucl. Phys.} \textbf{\bibinfo{volume}{B478}},
  \bibinfo{pages}{349} (\bibinfo{year}{1996}{\natexlab{b}}),
  \eprint{hep-lat/9604018}.

\bibitem[{\citenamefont{Feurstein et~al.}(1998)\citenamefont{Feurstein,
  Ilgenfritz, M\"uller-Preussker, and Thurner}}]{Feurstein:1996cf}
\bibinfo{author}{\bibfnamefont{M.}~\bibnamefont{Feurstein}},
  \bibinfo{author}{\bibfnamefont{E.-M.} \bibnamefont{Ilgenfritz}},
  \bibinfo{author}{\bibfnamefont{M.}~\bibnamefont{M\"uller-Preussker}},
  \bibnamefont{and} \bibinfo{author}{\bibfnamefont{S.}~\bibnamefont{Thurner}},
  \bibinfo{journal}{Nucl. Phys.} \textbf{\bibinfo{volume}{B511}},
  \bibinfo{pages}{421} (\bibinfo{year}{1998}), \eprint{hep-lat/9611024}.

\bibitem[{\citenamefont{DeGrand et~al.}(1997)\citenamefont{DeGrand, Hasenfratz,
  and Kovacs}}]{DeGrand:1997gu}
\bibinfo{author}{\bibfnamefont{T.~A.} \bibnamefont{DeGrand}},
  \bibinfo{author}{\bibfnamefont{A.}~\bibnamefont{Hasenfratz}},
  \bibnamefont{and} \bibinfo{author}{\bibfnamefont{T.~G.}
  \bibnamefont{Kovacs}}, \bibinfo{journal}{Nucl. Phys.}
  \textbf{\bibinfo{volume}{B505}}, \bibinfo{pages}{417} (\bibinfo{year}{1997}),
  \eprint{hep-lat/9705009}.

\bibitem[{\citenamefont{DeGrand et~al.}(1998)\citenamefont{DeGrand, Hasenfratz,
  and Kovacs}}]{DeGrand:1997ss}
\bibinfo{author}{\bibfnamefont{T.~A.} \bibnamefont{DeGrand}},
  \bibinfo{author}{\bibfnamefont{A.}~\bibnamefont{Hasenfratz}},
  \bibnamefont{and} \bibinfo{author}{\bibfnamefont{T.~G.}
  \bibnamefont{Kovacs}}, \bibinfo{journal}{Nucl. Phys.}
  \textbf{\bibinfo{volume}{B520}}, \bibinfo{pages}{301} (\bibinfo{year}{1998}),
  \eprint{hep-lat/9711032}.

\bibitem[{\citenamefont{Negele}(1999)}]{Negele:1998ev}
\bibinfo{author}{\bibfnamefont{J.~W.} \bibnamefont{Negele}},
  \bibinfo{journal}{Nucl. Phys. Proc. Suppl.} \textbf{\bibinfo{volume}{73}},
  \bibinfo{pages}{92} (\bibinfo{year}{1999}), \eprint{hep-lat/9810053}.

\bibitem[{\citenamefont{Horvath et~al.}(2003{\natexlab{a}})}]{Horvath:2002yn}
\bibinfo{author}{\bibfnamefont{I.}~\bibnamefont{Horvath}} \bibnamefont{et~al.},
  \bibinfo{journal}{Phys. Rev.} \textbf{\bibinfo{volume}{D67}},
  \bibinfo{pages}{011501} (\bibinfo{year}{2003}{\natexlab{a}}),
  \eprint{hep-lat/0203027}.

\bibitem[{\citenamefont{Bilson-Thompson
  et~al.}(2003)\citenamefont{Bilson-Thompson, Leinweber, and
  Williams}}]{Bilson-Thompson:2002jk}
\bibinfo{author}{\bibfnamefont{S.~O.} \bibnamefont{Bilson-Thompson}},
  \bibinfo{author}{\bibfnamefont{D.~B.} \bibnamefont{Leinweber}},
  \bibnamefont{and} \bibinfo{author}{\bibfnamefont{A.~G.}
  \bibnamefont{Williams}}, \bibinfo{journal}{Ann. Phys.}
  \textbf{\bibinfo{volume}{304}}, \bibinfo{pages}{1} (\bibinfo{year}{2003}),
  \eprint{hep-lat/0203008}.

\bibitem[{\citenamefont{Horvath et~al.}(2003{\natexlab{b}})}]{Horvath:2003yj}
\bibinfo{author}{\bibfnamefont{I.}~\bibnamefont{Horvath}} \bibnamefont{et~al.},
  \bibinfo{journal}{Phys. Rev.} \textbf{\bibinfo{volume}{D68}},
  \bibinfo{pages}{114505} (\bibinfo{year}{2003}{\natexlab{b}}),
  \eprint{hep-lat/0302009}.

\bibitem[{\citenamefont{Horvath et~al.}(2005{\natexlab{a}})}]{Horvath:2005rv}
\bibinfo{author}{\bibfnamefont{I.}~\bibnamefont{Horvath}} \bibnamefont{et~al.},
  \bibinfo{journal}{Phys. Lett.} \textbf{\bibinfo{volume}{B612}},
  \bibinfo{pages}{21} (\bibinfo{year}{2005}{\natexlab{a}}),
  \eprint{hep-lat/0501025}.

\bibitem[{\citenamefont{Witten}(1979)}]{Witten:1978bc}
\bibinfo{author}{\bibfnamefont{E.}~\bibnamefont{Witten}},
  \bibinfo{journal}{Nucl. Phys.} \textbf{\bibinfo{volume}{B149}},
  \bibinfo{pages}{285} (\bibinfo{year}{1979}).

\bibitem[{\citenamefont{Horvath
  et~al.}(2002{\natexlab{a}})\citenamefont{Horvath, Isgur, McCune, and
  Thacker}}]{Horvath:2001ir}
\bibinfo{author}{\bibfnamefont{I.}~\bibnamefont{Horvath}},
  \bibinfo{author}{\bibfnamefont{N.}~\bibnamefont{Isgur}},
  \bibinfo{author}{\bibfnamefont{J.}~\bibnamefont{McCune}}, \bibnamefont{and}
  \bibinfo{author}{\bibfnamefont{H.~B.} \bibnamefont{Thacker}},
  \bibinfo{journal}{Phys. Rev.} \textbf{\bibinfo{volume}{D65}},
  \bibinfo{pages}{014502} (\bibinfo{year}{2002}{\natexlab{a}}),
  \eprint{hep-lat/0102003}.

\bibitem[{\citenamefont{Horvath et~al.}(2002{\natexlab{b}})}]{Horvath:2002gk}
\bibinfo{author}{\bibfnamefont{I.}~\bibnamefont{Horvath}} \bibnamefont{et~al.},
  \bibinfo{journal}{Phys. Rev.} \textbf{\bibinfo{volume}{D66}},
  \bibinfo{pages}{034501} (\bibinfo{year}{2002}{\natexlab{b}}),
  \eprint{hep-lat/0201008}.

\bibitem[{\citenamefont{Gattringer}(2002)}]{Gattringer:2002gn}
\bibinfo{author}{\bibfnamefont{C.}~\bibnamefont{Gattringer}},
  \bibinfo{journal}{Phys. Rev. Lett.} \textbf{\bibinfo{volume}{88}},
  \bibinfo{pages}{221601} (\bibinfo{year}{2002}), \eprint{hep-lat/0202002}.

\bibitem[{\citenamefont{Liu}(2006)}]{Liu:2006wa}
\bibinfo{author}{\bibfnamefont{K.~F.} \bibnamefont{Liu}}
  (\bibinfo{year}{2006}), \eprint{hep-lat/0609033}.

\bibitem[{\citenamefont{Liu et~al.}(2007)\citenamefont{Liu, Alexandru, and
  Horvath}}]{Liu:2007hq}
\bibinfo{author}{\bibfnamefont{K.~F.} \bibnamefont{Liu}},
  \bibinfo{author}{\bibfnamefont{A.}~\bibnamefont{Alexandru}},
  \bibnamefont{and} \bibinfo{author}{\bibfnamefont{I.}~\bibnamefont{Horvath}}
  (\bibinfo{year}{2007}), \eprint{hep-lat/0703010}.

\bibitem[{\citenamefont{DeGrand and Schaefer}(2004)}]{DeGrand:2004qw}
\bibinfo{author}{\bibfnamefont{T.~A.} \bibnamefont{DeGrand}} \bibnamefont{and}
  \bibinfo{author}{\bibfnamefont{S.}~\bibnamefont{Schaefer}},
  \bibinfo{journal}{Comput. Phys. Commun.} \textbf{\bibinfo{volume}{159}},
  \bibinfo{pages}{185} (\bibinfo{year}{2004}), \eprint{hep-lat/0401011}.

\bibitem[{\citenamefont{Giusti and Necco}(2006)}]{Giusti:2005sx}
\bibinfo{author}{\bibfnamefont{L.}~\bibnamefont{Giusti}} \bibnamefont{and}
  \bibinfo{author}{\bibfnamefont{S.}~\bibnamefont{Necco}},
  \bibinfo{journal}{PoS} \textbf{\bibinfo{volume}{LAT2005}},
  \bibinfo{pages}{132} (\bibinfo{year}{2006}), \eprint{hep-lat/0510011}.

\bibitem[{\citenamefont{Galletly et~al.}(2006)}]{Galletly:2006hq}
\bibinfo{author}{\bibfnamefont{D.}~\bibnamefont{Galletly}} \bibnamefont{et~al.}
  (\bibinfo{year}{2006}), \eprint{hep-lat/0607024}.

\bibitem[{\citenamefont{Aubin et~al.}(2005)}]{Aubin:2004mp}
\bibinfo{author}{\bibfnamefont{C.}~\bibnamefont{Aubin}} \bibnamefont{et~al.}
  (\bibinfo{collaboration}{MILC}), \bibinfo{journal}{Nucl. Phys. Proc. Suppl.}
  \textbf{\bibinfo{volume}{140}}, \bibinfo{pages}{626} (\bibinfo{year}{2005}),
  \eprint{hep-lat/0410024}.

\bibitem[{\citenamefont{Bernard et~al.}(2006)}]{Bernard:2005nv}
\bibinfo{author}{\bibfnamefont{C.}~\bibnamefont{Bernard}} \bibnamefont{et~al.},
  \bibinfo{journal}{PoS} \textbf{\bibinfo{volume}{LAT2005}},
  \bibinfo{pages}{299} (\bibinfo{year}{2006}), \eprint{hep-lat/0510025}.

\bibitem[{\citenamefont{Gubarev et~al.}(2005)\citenamefont{Gubarev, Morozov,
  Polikarpov, and Zakharov}}]{Gubarev:2005az}
\bibinfo{author}{\bibfnamefont{F.~V.} \bibnamefont{Gubarev}},
  \bibinfo{author}{\bibfnamefont{S.~M.} \bibnamefont{Morozov}},
  \bibinfo{author}{\bibfnamefont{M.~I.} \bibnamefont{Polikarpov}},
  \bibnamefont{and} \bibinfo{author}{\bibfnamefont{V.~I.}
  \bibnamefont{Zakharov}}, \bibinfo{journal}{JETP Lett.}
  \textbf{\bibinfo{volume}{82}}, \bibinfo{pages}{343} (\bibinfo{year}{2005}),
  \eprint{hep-lat/0505016}.

\bibitem[{\citenamefont{Polikarpov
  et~al.}(2006{\natexlab{a}})\citenamefont{Polikarpov, Gubarev, Morozov, and
  Zakharov}}]{Polikarpov:2005ey}
\bibinfo{author}{\bibfnamefont{M.~I.} \bibnamefont{Polikarpov}},
  \bibinfo{author}{\bibfnamefont{F.~V.} \bibnamefont{Gubarev}},
  \bibinfo{author}{\bibfnamefont{S.~M.} \bibnamefont{Morozov}},
  \bibnamefont{and} \bibinfo{author}{\bibfnamefont{V.~I.}
  \bibnamefont{Zakharov}}, \bibinfo{journal}{PoS}
  \textbf{\bibinfo{volume}{LAT2005}}, \bibinfo{pages}{143}
  (\bibinfo{year}{2006}{\natexlab{a}}), \eprint{hep-lat/0510098}.

\bibitem[{\citenamefont{Polikarpov
  et~al.}(2006{\natexlab{b}})\citenamefont{Polikarpov, Gubarev, Morozov, and
  Syritsyn}}]{Polikarpov:2006ki}
\bibinfo{author}{\bibfnamefont{M.~I.} \bibnamefont{Polikarpov}},
  \bibinfo{author}{\bibfnamefont{F.~V.} \bibnamefont{Gubarev}},
  \bibinfo{author}{\bibfnamefont{S.~M.} \bibnamefont{Morozov}},
  \bibnamefont{and} \bibinfo{author}{\bibfnamefont{S.~N.}
  \bibnamefont{Syritsyn}}, \bibinfo{journal}{Nucl. Phys. Proc. Suppl.}
  \textbf{\bibinfo{volume}{153}}, \bibinfo{pages}{221}
  (\bibinfo{year}{2006}{\natexlab{b}}).

\bibitem[{\citenamefont{DeGrand and Hasenfratz}(2002)}]{DeGrand:2001pj}
\bibinfo{author}{\bibfnamefont{T.~A.} \bibnamefont{DeGrand}} \bibnamefont{and}
  \bibinfo{author}{\bibfnamefont{A.}~\bibnamefont{Hasenfratz}},
  \bibinfo{journal}{Phys. Rev.} \textbf{\bibinfo{volume}{D65}},
  \bibinfo{pages}{014503} (\bibinfo{year}{2002}), \eprint{hep-lat/0103002}.

\bibitem[{\citenamefont{Edwards and Heller}(2002)}]{Edwards:2001nd}
\bibinfo{author}{\bibfnamefont{R.~G.} \bibnamefont{Edwards}} \bibnamefont{and}
  \bibinfo{author}{\bibfnamefont{U.~M.} \bibnamefont{Heller}},
  \bibinfo{journal}{Phys. Rev.} \textbf{\bibinfo{volume}{D65}},
  \bibinfo{pages}{014505} (\bibinfo{year}{2002}), \eprint{hep-lat/0105004}.

\bibitem[{\citenamefont{de~Forcrand}(2006)}]{deForcrand:2006my}
\bibinfo{author}{\bibfnamefont{P.}~\bibnamefont{de~Forcrand}}
  (\bibinfo{year}{2006}), \eprint{hep-lat/0611034}.

\bibitem[{\citenamefont{Pugh and Teper}(1989)}]{Pugh:1989ek}
\bibinfo{author}{\bibfnamefont{D.~J.~R.} \bibnamefont{Pugh}} \bibnamefont{and}
  \bibinfo{author}{\bibfnamefont{M.}~\bibnamefont{Teper}},
  \bibinfo{journal}{Phys. Lett.} \textbf{\bibinfo{volume}{B224}},
  \bibinfo{pages}{159} (\bibinfo{year}{1989}).

\bibitem[{\citenamefont{Zakharov}(2005)}]{Zakharov:2004jh}
\bibinfo{author}{\bibfnamefont{V.~I.} \bibnamefont{Zakharov}},
  \bibinfo{journal}{Phys. Atom. Nucl.} \textbf{\bibinfo{volume}{68}},
  \bibinfo{pages}{573} (\bibinfo{year}{2005}), \eprint{hep-ph/0410034}.

\bibitem[{\citenamefont{Zakharov}(2006{\natexlab{a}})}]{Zakharov:2006vt}
\bibinfo{author}{\bibfnamefont{V.~I.} \bibnamefont{Zakharov}}
  (\bibinfo{year}{2006}{\natexlab{a}}), \eprint{hep-ph/0602141}.

\bibitem[{\citenamefont{Zakharov}(2006{\natexlab{b}})}]{Zakharov:2006te}
\bibinfo{author}{\bibfnamefont{V.~I.} \bibnamefont{Zakharov}}
  (\bibinfo{year}{2006}{\natexlab{b}}), \eprint{hep-ph/0612341}.

\bibitem[{\citenamefont{Zakharov}(2006{\natexlab{c}})}]{Zakharov:2006tf}
\bibinfo{author}{\bibfnamefont{V.~I.} \bibnamefont{Zakharov}}
  (\bibinfo{year}{2006}{\natexlab{c}}), \eprint{hep-ph/0612342}.

\bibitem[{\citenamefont{Garcia-Garcia and
  Osborn}(2006{\natexlab{a}})}]{Garcia-Garcia:2005dp}
\bibinfo{author}{\bibfnamefont{A.~M.} \bibnamefont{Garcia-Garcia}}
  \bibnamefont{and} \bibinfo{author}{\bibfnamefont{J.~C.}
  \bibnamefont{Osborn}}, \bibinfo{journal}{PoS}
  \textbf{\bibinfo{volume}{LAT2005}}, \bibinfo{pages}{265}
  (\bibinfo{year}{2006}{\natexlab{a}}), \eprint{hep-lat/0509118}.

\bibitem[{\citenamefont{Garcia-Garcia and
  Osborn}(2006{\natexlab{b}})}]{Garcia-Garcia:2005vj}
\bibinfo{author}{\bibfnamefont{A.~M.} \bibnamefont{Garcia-Garcia}}
  \bibnamefont{and} \bibinfo{author}{\bibfnamefont{J.~C.}
  \bibnamefont{Osborn}}, \bibinfo{journal}{Nucl. Phys.}
  \textbf{\bibinfo{volume}{A770}}, \bibinfo{pages}{141}
  (\bibinfo{year}{2006}{\natexlab{b}}), \eprint{hep-lat/0512025}.

\bibitem[{\citenamefont{Garcia-Garcia and
  Osborn}(2006{\natexlab{c}})}]{Garcia-Garcia:2006gr}
\bibinfo{author}{\bibfnamefont{A.~M.} \bibnamefont{Garcia-Garcia}}
  \bibnamefont{and} \bibinfo{author}{\bibfnamefont{J.~C.} \bibnamefont{Osborn}}
  (\bibinfo{year}{2006}{\natexlab{c}}), \eprint{hep-lat/0611019}.

\bibitem[{\citenamefont{Damgaard}(2001)}]{Damgaard:2001xr}
\bibinfo{author}{\bibfnamefont{P.~H.} \bibnamefont{Damgaard}},
  \bibinfo{journal}{Nucl. Phys.} \textbf{\bibinfo{volume}{B608}},
  \bibinfo{pages}{162} (\bibinfo{year}{2001}), \eprint{hep-lat/0105010}.

\bibitem[{\citenamefont{Horvath et~al.}(2005{\natexlab{b}})}]{Horvath:2005cv}
\bibinfo{author}{\bibfnamefont{I.}~\bibnamefont{Horvath}} \bibnamefont{et~al.},
  \bibinfo{journal}{Phys. Lett.} \textbf{\bibinfo{volume}{B617}},
  \bibinfo{pages}{49} (\bibinfo{year}{2005}{\natexlab{b}}),
  \eprint{hep-lat/0504005}.

\bibitem[{\citenamefont{Gockeler et~al.}(1989)\citenamefont{Gockeler, Kronfeld,
  Laursen, Schierholz, and Wiese}}]{Gockeler:1989qg}
\bibinfo{author}{\bibfnamefont{M.}~\bibnamefont{G\"ockeler}},
  \bibinfo{author}{\bibfnamefont{A.~S.} \bibnamefont{Kronfeld}},
  \bibinfo{author}{\bibfnamefont{M.~L.} \bibnamefont{Laursen}},
  \bibinfo{author}{\bibfnamefont{G.}~\bibnamefont{Schierholz}},
  \bibnamefont{and} \bibinfo{author}{\bibfnamefont{U.~J.} \bibnamefont{Wiese}},
  \bibinfo{journal}{Phys. Lett.} \textbf{\bibinfo{volume}{B233}},
  \bibinfo{pages}{192} (\bibinfo{year}{1989}).

\bibitem[{\citenamefont{Luscher and
  Weisz}(1985{\natexlab{a}})}]{Luscher:1984xn}
\bibinfo{author}{\bibfnamefont{M.}~\bibnamefont{L\"uscher}} \bibnamefont{and}
  \bibinfo{author}{\bibfnamefont{P.}~\bibnamefont{Weisz}},
  \bibinfo{journal}{Commun. Math. Phys.} \textbf{\bibinfo{volume}{97}},
  \bibinfo{pages}{59} (\bibinfo{year}{1985}{\natexlab{a}}).

\bibitem[{\citenamefont{Luscher and
  Weisz}(1985{\natexlab{b}})}]{Luscher:1985zq}
\bibinfo{author}{\bibfnamefont{M.}~\bibnamefont{L\"uscher}} \bibnamefont{and}
  \bibinfo{author}{\bibfnamefont{P.}~\bibnamefont{Weisz}},
  \bibinfo{journal}{Phys. Lett.} \textbf{\bibinfo{volume}{B158}},
  \bibinfo{pages}{250} (\bibinfo{year}{1985}{\natexlab{b}}).

\bibitem[{\citenamefont{Lepage and Mackenzie}(1993)}]{Lepage:1992xa}
\bibinfo{author}{\bibfnamefont{G.~P.} \bibnamefont{Lepage}} \bibnamefont{and}
  \bibinfo{author}{\bibfnamefont{P.~B.} \bibnamefont{Mackenzie}},
  \bibinfo{journal}{Phys. Rev.} \textbf{\bibinfo{volume}{D48}},
  \bibinfo{pages}{2250} (\bibinfo{year}{1993}), \eprint{hep-lat/9209022}.

\bibitem[{\citenamefont{Alford et~al.}(1995)\citenamefont{Alford, Dimm, Lepage,
  Hockney, and Mackenzie}}]{Alford:1995hw}
\bibinfo{author}{\bibfnamefont{M.~G.} \bibnamefont{Alford}},
  \bibinfo{author}{\bibfnamefont{W.}~\bibnamefont{Dimm}},
  \bibinfo{author}{\bibfnamefont{G.~P.} \bibnamefont{Lepage}},
  \bibinfo{author}{\bibfnamefont{G.}~\bibnamefont{Hockney}}, \bibnamefont{and}
  \bibinfo{author}{\bibfnamefont{P.~B.} \bibnamefont{Mackenzie}},
  \bibinfo{journal}{Phys. Lett.} \textbf{\bibinfo{volume}{B361}},
  \bibinfo{pages}{87} (\bibinfo{year}{1995}), \eprint{hep-lat/9507010}.

\bibitem[{\citenamefont{Snippe}(1997)}]{Snippe:1997ru}
\bibinfo{author}{\bibfnamefont{J.}~\bibnamefont{Snippe}},
  \bibinfo{journal}{Nucl. Phys.} \textbf{\bibinfo{volume}{B498}},
  \bibinfo{pages}{347} (\bibinfo{year}{1997}), \eprint{hep-lat/9701002}.

\bibitem[{\citenamefont{Gattringer et~al.}(2002)\citenamefont{Gattringer,
  Hoffmann, and Schaefer}}]{Gattringer:2001jf}
\bibinfo{author}{\bibfnamefont{C.}~\bibnamefont{Gattringer}},
  \bibinfo{author}{\bibfnamefont{R.}~\bibnamefont{Hoffmann}}, \bibnamefont{and}
  \bibinfo{author}{\bibfnamefont{S.}~\bibnamefont{Schaefer}},
  \bibinfo{journal}{Phys. Rev.} \textbf{\bibinfo{volume}{D65}},
  \bibinfo{pages}{094503} (\bibinfo{year}{2002}), \eprint{hep-lat/0112024}.

\bibitem[{\citenamefont{Capitani et~al.}(1999)\citenamefont{Capitani, Gockeler,
  Horsley, Rakow, and Schierholz}}]{Capitani:1999uz}
\bibinfo{author}{\bibfnamefont{S.}~\bibnamefont{Capitani}},
  \bibinfo{author}{\bibfnamefont{M.}~\bibnamefont{G\"ockeler}},
  \bibinfo{author}{\bibfnamefont{R.}~\bibnamefont{Horsley}},
  \bibinfo{author}{\bibfnamefont{P.~E.~L.} \bibnamefont{Rakow}},
  \bibnamefont{and}
  \bibinfo{author}{\bibfnamefont{G.}~\bibnamefont{Schierholz}},
  \bibinfo{journal}{Phys. Lett.} \textbf{\bibinfo{volume}{B468}},
  \bibinfo{pages}{150} (\bibinfo{year}{1999}), \eprint{hep-lat/9908029}.

\bibitem[{\citenamefont{Hernandez et~al.}(1999)\citenamefont{Hernandez, Jansen,
  and Luscher}}]{Hernandez:1998et}
\bibinfo{author}{\bibfnamefont{P.}~\bibnamefont{Hernandez}},
  \bibinfo{author}{\bibfnamefont{K.}~\bibnamefont{Jansen}}, \bibnamefont{and}
  \bibinfo{author}{\bibfnamefont{M.}~\bibnamefont{L\"uscher}},
  \bibinfo{journal}{Nucl. Phys.} \textbf{\bibinfo{volume}{B552}},
  \bibinfo{pages}{363} (\bibinfo{year}{1999}), \eprint{hep-lat/9808010}.

\bibitem[{\citenamefont{Giusti et~al.}(2003)\citenamefont{Giusti, Hoelbling,
  Luscher, and Wittig}}]{Giusti:2002sm}
\bibinfo{author}{\bibfnamefont{L.}~\bibnamefont{Giusti}},
  \bibinfo{author}{\bibfnamefont{C.}~\bibnamefont{Hoelbling}},
  \bibinfo{author}{\bibfnamefont{M.}~\bibnamefont{L\"uscher}}, \bibnamefont{and}
  \bibinfo{author}{\bibfnamefont{H.}~\bibnamefont{Wittig}},
  \bibinfo{journal}{Comput. Phys. Commun.} \textbf{\bibinfo{volume}{153}},
  \bibinfo{pages}{31} (\bibinfo{year}{2003}), \eprint{hep-lat/0212012}.


\bibitem[{\citenamefont{Chiu et~al.}(1998)\citenamefont{Chiu, Wang, and Zenkin}}]{Chiu:1998aa}
\bibinfo{author}{\bibfnamefont{T.-W.} \bibnamefont{Chiu}},
 \bibinfo{author}{\bibfnamefont{C.-W.} \bibnamefont{Wang}}, \bibnamefont{and}
 \bibinfo{author}{\bibfnamefont{S.~V.} \bibnamefont{Zenkin}},
 \bibinfo{journal}{Phys. Lett.} \textbf{\bibinfo{volume}{B438}},
 \bibinfo{pages}{321} (\bibinfo{year}{1998}), \eprint{hep-lat/9806031}.

\bibitem[{\citenamefont{Chiu and Zenkin}(1999)}]{Chiu:1998gp}
 \bibinfo{author}{\bibfnamefont{T.-W.} \bibnamefont{Chiu}} \bibnamefont{and}
 \bibinfo{author}{\bibfnamefont{S.~V.} \bibnamefont{Zenkin}},
 \bibinfo{journal}{Phys. Rev.} \textbf{\bibinfo{volume}{D59}},
 \bibinfo{pages}{074501} (\bibinfo{year}{1999}), \eprint{hep-lat/9806019}.
		
\bibitem[{\citenamefont{Del~Debbio et~al.}(2005)\citenamefont{Del~Debbio,
  Giusti, and Pica}}]{DelDebbio:2004ns}
\bibinfo{author}{\bibfnamefont{L.}~\bibnamefont{Del~Debbio}},
  \bibinfo{author}{\bibfnamefont{L.}~\bibnamefont{Giusti}}, \bibnamefont{and}
  \bibinfo{author}{\bibfnamefont{C.}~\bibnamefont{Pica}},
  \bibinfo{journal}{Phys. Rev. Lett.} \textbf{\bibinfo{volume}{94}},
  \bibinfo{pages}{032003} (\bibinfo{year}{2005}), \eprint{hep-th/0407052}.

\bibitem[{\citenamefont{Banks and Casher}(1980)}]{Banks:1979yr}
\bibinfo{author}{\bibfnamefont{T.}~\bibnamefont{Banks}} \bibnamefont{and}
  \bibinfo{author}{\bibfnamefont{A.}~\bibnamefont{Casher}},
  \bibinfo{journal}{Nucl. Phys.} \textbf{\bibinfo{volume}{B169}},
  \bibinfo{pages}{103} (\bibinfo{year}{1980}).

\bibitem[{\citenamefont{Wilke et~al.}(1998)\citenamefont{Wilke, Guhr, and
  Wettig}}]{Wilke:1997gf}
\bibinfo{author}{\bibfnamefont{T.}~\bibnamefont{Wilke}},
  \bibinfo{author}{\bibfnamefont{T.}~\bibnamefont{Guhr}}, \bibnamefont{and}
  \bibinfo{author}{\bibfnamefont{T.}~\bibnamefont{Wettig}},
  \bibinfo{journal}{Phys. Rev.} \textbf{\bibinfo{volume}{D57}},
  \bibinfo{pages}{6486} (\bibinfo{year}{1998}), \eprint{hep-th/9711057}.

\bibitem[{\citenamefont{Kravtsov and Muttalib}()}]{Kravtsov:1997}
\bibinfo{author}{\bibfnamefont{V.~E.} \bibnamefont{Kravtsov}} \bibnamefont{and}
  \bibinfo{author}{\bibfnamefont{K.~A.} \bibnamefont{Muttalib}},
  \bibinfo{journal}{Phys. Rev. Lett.} \textbf{\bibinfo{volume}{79}},
  \bibinfo{pages}{1913} (\bibinfo{year}{1997}), \eprint{cond-mat/9703167}.

\bibitem[{\citenamefont{Kravtsov}()}]{Kravtsov:1996}
\bibinfo{author}{\bibfnamefont{V.~E.} \bibnamefont{Kravtsov}},
  \eprint{cond-mat/9603166}.

\bibitem[{\citenamefont{Diakonov and Petrov}(1986)}]{Diakonov:1985eg}
\bibinfo{author}{\bibfnamefont{D.}~\bibnamefont{Diakonov}} \bibnamefont{and}
  \bibinfo{author}{\bibfnamefont{V.~Y.} \bibnamefont{Petrov}},
  \bibinfo{journal}{Nucl. Phys.} \textbf{\bibinfo{volume}{B272}},
  \bibinfo{pages}{457} (\bibinfo{year}{1986}).

\bibitem[{\citenamefont{Greensite et~al.}(2005)\citenamefont{Greensite,
  Olejnik, Polikarpov, Syritsyn, and Zakharov}}]{Greensite:2005yu}
\bibinfo{author}{\bibfnamefont{J.}~\bibnamefont{Greensite}},
  \bibinfo{author}{\bibfnamefont{S.}~\bibnamefont{Olejnik}},
  \bibinfo{author}{\bibfnamefont{M.}~\bibnamefont{Polikarpov}},
  \bibinfo{author}{\bibfnamefont{S.}~\bibnamefont{Syritsyn}}, \bibnamefont{and}
  \bibinfo{author}{\bibfnamefont{V.}~\bibnamefont{Zakharov}},
  \bibinfo{journal}{Phys. Rev.} \textbf{\bibinfo{volume}{D71}},
  \bibinfo{pages}{114507} (\bibinfo{year}{2005}), \eprint{hep-lat/0504008}.

\bibitem[{\citenamefont{Seiler}(2002)}]{Seiler:2001je}
\bibinfo{author}{\bibfnamefont{E.}~\bibnamefont{Seiler}},
  \bibinfo{journal}{Phys. Lett.} \textbf{\bibinfo{volume}{B525}},
  \bibinfo{pages}{355} (\bibinfo{year}{2002}), \eprint{hep-th/0111125}.

\bibitem[{\citenamefont{Hasenfratz}(2000)}]{Hasenfratz:1999ng}
\bibinfo{author}{\bibfnamefont{A.}~\bibnamefont{Hasenfratz}},
  \bibinfo{journal}{Phys. Lett.} \textbf{\bibinfo{volume}{B476}},
  \bibinfo{pages}{188} (\bibinfo{year}{2000}), \eprint{hep-lat/9912053}.

\bibitem[{\citenamefont{Ilgenfritz and de~Forcrand}()}]{EMIunpublished}
\bibinfo{author}{\bibfnamefont{E.~M.} \bibnamefont{Ilgenfritz}}
  \bibnamefont{and}
  \bibinfo{author}{\bibfnamefont{P.}~\bibnamefont{de~Forcrand}} (unpublished).

\bibitem[{\citenamefont{Gattringer et~al.}(2001)\citenamefont{Gattringer, Hip,
  and Lang}}]{Gattringer:2000qu}
\bibinfo{author}{\bibfnamefont{C.}~\bibnamefont{Gattringer}},
  \bibinfo{author}{\bibfnamefont{I.}~\bibnamefont{Hip}}, \bibnamefont{and}
  \bibinfo{author}{\bibfnamefont{C.~B.} \bibnamefont{Lang}},
  \bibinfo{journal}{Nucl. Phys.} \textbf{\bibinfo{volume}{B597}},
  \bibinfo{pages}{451} (\bibinfo{year}{2001}), \eprint{hep-lat/0007042}.

\bibitem[{\citenamefont{Gattnar et~al.}(2005)}]{Gattnar:2004gx}
\bibinfo{author}{\bibfnamefont{J.}~\bibnamefont{Gattnar}} \bibnamefont{et~al.},
  \bibinfo{journal}{Nucl. Phys.} \textbf{\bibinfo{volume}{B716}},
  \bibinfo{pages}{105} (\bibinfo{year}{2005}), \eprint{hep-lat/0412032}.

\bibitem[{\citenamefont{Gattringer}(2006)}]{Gattringer:2006ci}
\bibinfo{author}{\bibfnamefont{C.}~\bibnamefont{Gattringer}},
  \bibinfo{journal}{Phys. Rev. Lett.} \textbf{\bibinfo{volume}{97}},
  \bibinfo{pages}{032003} (\bibinfo{year}{2006}), \eprint{hep-lat/0605018}.

\bibitem[{\citenamefont{Markos}(2006)}]{Markos:2006}
\bibinfo{author}{\bibfnamefont{P.}~\bibnamefont{Markos}},
  \bibinfo{journal}{acta phys. slovaca} \textbf{\bibinfo{volume}{56}},
  \bibinfo{pages}{561} (\bibinfo{year}{2006}), \eprint{cond-mat/0609580}.

\bibitem[{\citenamefont{Bruckmann et~al.}(2007)\citenamefont{Bruckmann,
  Gattringer, and Hagen}}]{Bruckmann:2006kx}
\bibinfo{author}{\bibfnamefont{F.}~\bibnamefont{Bruckmann}},
  \bibinfo{author}{\bibfnamefont{C.}~\bibnamefont{Gattringer}},
  \bibnamefont{and} \bibinfo{author}{\bibfnamefont{C.}~\bibnamefont{Hagen}},
  \bibinfo{journal}{Phys. Lett.} \textbf{\bibinfo{volume}{B647}},
  \bibinfo{pages}{56} (\bibinfo{year}{2007}), \eprint{hep-lat/0612020}.

\bibitem[{\citenamefont{Synatschke et~al.}(2007)\citenamefont{Synatschke, Wipf,
  and Wozar}}]{Synatschke:2007bz}
\bibinfo{author}{\bibfnamefont{F.}~\bibnamefont{Synatschke}},
  \bibinfo{author}{\bibfnamefont{A.}~\bibnamefont{Wipf}}, \bibnamefont{and}
  \bibinfo{author}{\bibfnamefont{C.}~\bibnamefont{Wozar}}
  (\bibinfo{year}{2007}), \eprint{hep-lat/0703018}.

\end{thebibliography}

\end{document}